\documentclass[12pt]{article}
\usepackage{epsfig}
\usepackage{amsmath}
\usepackage{amssymb}
\usepackage{graphicx}
\usepackage{subfigure}
\usepackage{array}
\usepackage{cite}
\usepackage{color}
\usepackage{authblk}
\usepackage{epstopdf}

\usepackage[top=1.in,left=1.in,right=1.in,bottom=1.in]{geometry}

\normalsize
\def\ba{\begin{array}}
\def\ea{\end{array}}
\def\beq{\begin{equation}}
\def\eeq{\end{equation}}
\def\beqs{\begin{equation*}}
\def\eeqs{\end{equation*}}
\def\bea{\begin{eqnarray}}
\def\eea{\end{eqnarray}}
\def\beas{\begin{eqnarray*}}
\def\eeas{\end{eqnarray*}}
\def\bi{\begin{itemize}}
\def\ei{\end{itemize}}

\def\a{\alpha}
\def\g{\gamma}
\def\d{\delta}
\def\e{\varepsilon}

\def\k{\kappa}
\def\l{\lambda}
\def\m{\mu}

\def\r{\rho}

\def\t{\tau}

\def\o{\omega}

\def\cA{{\cal A}}

\def\hC{\mathbb{C}}

\def\hN{\mathbb{N}}

\def\hR{\mathbb{R}}

\def\hZ{\mathbb{Z}}

\def\({\textnormal{(}}
\def\){\textnormal{)}}
\def\[{[\neg[}
\def\]{]\neg]}

\def\q{\quad}
\def\qq{\qquad}

\def\neg{\negthinspace}

\def\Ra {\mathop{\Rightarrow }}

\def\wt{\widetilde}

\def\b1{{\bf 1}}

\begin{document}
\title{Breathers in a locally resonant granular chain with precompression}
\author[1]{{\normalsize Lifeng Liu}}
\author[2]{{\normalsize Guillaume James}}
\author[3,4]{{\normalsize Panayotis Kevrekidis}}
\author[1,5]{{\normalsize Anna Vainchtein}}
\affil[1]{{\small\emph{Department of Mathematics\\University of Pittsburgh\\Pittsburgh\\Pennsylvania 15260\\USA}}}
\affil[2]{{\small\emph{INRIA Grenoble - Rh\^one-Alpes, Bipop Team-Project,
Inovall\'ee, 655 Avenue de l'Europe, 38334 Saint Ismier Cedex, France.}}}
\affil[3]{{\small\emph{Department of Mathematics and Statistics\\University of Massachusetts\\Amherst\\Massachusetts 01003\\USA}}}
\affil[4]{{\small\emph{Theoretical Division and Center for Nonlinear Studies\\Los Alamos National Laboratory\\Los Alamos\\New Mexico 87545\\USA}}}
\affil[5]{{\small\emph{Corresponding author: aav4@pitt.edu (email), 412 624 8309 (phone), 412 624 8397 (fax)}}}
\maketitle

\begin{abstract}
We study a locally resonant granular material in the form of a precompressed
Hertzian chain with linear internal resonators. Using an asymptotic reduction,
we derive an effective nonlinear Schr\"odinger (NLS) modulation equation.
This, in turn, leads us to provide analytical evidence,
subsequently corroborated numerically, for the existence of
two distinct types of discrete breathers related to acoustic or optical modes:
(a) traveling bright breathers
with a strain profile exponentially vanishing at infinity and
(b) stationary and traveling dark breathers, exponentially localized, time-periodic states mounted
on top of a non-vanishing background.
The stability and bifurcation structure
of numerically computed exact stationary dark breathers is also examined.
Stationary bright breathers cannot be identified using the NLS equation,
which is defocusing at the upper edges of the phonon bands and
becomes linear at the lower edge of the optical band.
\end{abstract}

\noindent {\bf Keywords:} locally resonant granular material; precompression; discrete breather; modulation equation; stability

\section{Introduction}\label{sec:intro}
Granular crystals are tightly packed arrays of solid particles that deform elastically upon contact via nonlinear Hertzian interactions \cite{Nest01,Sen200821,Coste97}. The dynamics of these systems ranges from \emph{weakly nonlinear}, when the initial overlap of the neighboring particles due to the static precompression is much larger than their relative displacement, to the \emph{strongly nonlinear} regime characterized by relatively small or zero precompression. This provides an ideal setting for exploring nonlinear waves, including traveling \cite{Nest01,Sen200821,Coste97} and shock waves \cite{herbo,molinari}.

A particularly interesting class of nonlinear excitations exhibited by these materials are the so-called \emph{discrete breathers} \cite{Boechler10,Chong13,chong14,GJ11,JKC13,jobmelo,Hoog12,Theo09,Theo10}, i.e., time-periodic and exponentially
 localized in space
oscillations that are also encountered in a wide variety of other nonlinear systems (see \cite{Aubry1997201,Flach08} and references therein). There are two distinct types of breathers. \emph{Bright breathers} have
a profile (of strain in the case of granular systems) exponentially decaying to zero at infinity and are known to exist in granular materials with defects \cite{jobmelo, Theo09}, heterogeneous granular chains
such as dimers or trimers \cite{Boechler10, Hoog12, Theo10} and Hertzian chains with a harmonic onsite potential modeling Newton's cradle
or granular chains embedded in a matrix \cite{GJ11,JKC13,hasan15}.
\emph{Dark breathers}, on the other hand, are spatially
modulated standing waves with amplitude that is constant at infinity and vanishes at the center. They have been recently identified and analyzed in a homogeneous granular chain with precompression \cite{Chong13}, and their existence was experimentally verified in damped, driven granular chains in~\cite{chong14}.

In this work we consider both types of discrete breathers in a locally resonant granular chain characterized by very rich nonlinear dynamics \cite{kimyang15, haitao}. This novel granular metamaterial has tunable band gaps and can be potentially used in engineering applications involving shock absorption and vibration mitigation. The system consists of a regular granular chain with additional degrees of freedom due to attached linear resonators. Its recent experimental
implementations include mass-in-mass granular chains with internal linear
resonators placed inside the primary beads \cite{Bonanomi14},
mass-with-mass chains with external ring resonators attached to the
beads~\cite{Gantzounis13} (see also \cite{PGK12}) and woodpile phononic crystals consisting of vertically stacked slender cylindrical rods in orthogonal contact \cite{Kim14}. Under certain assumptions, each of these experimental setups can be modeled by a Hertzian chain with a secondary mass attached to each primary bead by a linearly elastic spring, with the ratio of the secondary and primary masses being the main control parameter. In a recent work \cite{LJKV15}, we studied the strongly nonlinear dynamics of this system in the absence of precompression.
Through a combination of asymptotic analysis and numerical computations, we provided evidence for the existence of exact dark breathers in the locally resonant granular chain and investigated their stability and bifurcation structure. In addition, we studied small-amplitude periodic traveling waves and identified the conditions under which the system has long-lived (but not exact) bright breathers.

Here we turn our attention to the locally resonant granular chain under nonzero precompression. In the non-resonant limit (regular granular chain, zero mass ratio), such a system belongs to the general class of Fermi-Pasta-Ulam (FPU) lattice models
(e.g. see \cite{FPU55, FW94, FP99, FP02, Iooss00, IJ05, Schneider10, James01, James03, FP04, Kevrekidis11} and references therein),
with a dispersion relation for plane wave solutions of the linearized problem possessing only acoustic spectrum. At finite mass ratio, the dispersion relation has both acoustic and optical branches. In this respect the problem is somewhat reminiscent of diatomic FPU chains, although the optical branch is quite different in our case. In the small-amplitude limit the dynamics of the system is weakly nonlinear. This dynamical regime has been well studied for the FPU problem, as has its generalized version with an additional onsite potential \cite{FP99, FP02, Iooss00, James01, James03, FP04, Gian04, Gian06, Flach08, Schneider10}. In particular, the established conditions for bifurcation of discrete breathers for this class of problems
\cite{Flach08, James01, James03} rule out the existence of bright breathers in the
homogeneous non-resonant granular chain under precompression,
the limiting case of our problem when the mass ratio is zero and the dispersion relation has only an acoustic branch. In this case, dark breathers were identified in \cite{Chong13} as the \emph{only} possible type of intrinsically localized mode. The \emph{defocusing} nonlinear Schr\"odinger equation (NLS), which has tanh-type solutions, is derived in \cite{Chong13} as the modulation equation for waves with frequencies near the edge of the linear acoustic spectrum and used to construct initial conditions for numerical computation and analysis of the dark breathers. In another limiting case, when the mass ratio goes to infinity and the secondary masses have zero initial conditions, the system approaches the Newton's cradle model with precompression, a problem with a purely optical dispersion relation. In this case, traveling bright breathers were investigated in \cite{JKC13} via the analysis of the corresponding \emph{focusing} NLS, which admits sech-type solutions.

To explore the weakly nonlinear dynamics at finite mass ratio, we use a multiscale asymptotic method
(see \cite{Gian04,Gian06,Schneider10} and references therein) and derive the classical NLS equation,
yielding closed-form solutions of sech-type and tanh-type in the focusing and defocusing cases, respectively.
In particular, we show that parameters (mass ratio and precompression) can modify
the number of focusing regions in the acoustic and optical bands, a phenomenon which
does not occur in classical homogeneous and diatomic granular chains \cite{Chong13,Theo10}.
This property is particularly interesting for applications because precompression is easy to tune experimentally.
Another special feature of the resonant granular chain is that the cubic NLS coefficient vanishes
at the zero wavenumber corresponding to the lower edge of the optical band. Since the NLS equation
is defocusing at the upper edges of the optical and acoustic bands, the NLS equation cannot be used
to approximate stationary bright breathers in the present context.

Having identified focusing and defocusing parameter regimes,
we first investigate how well the focusing NLS equation approximates moving bright breather solutions of the original system.
Provided that certain resonances are avoided,
we find that the focusing NLS equation successfully approximates small-amplitude optical bright breathers at various mass ratios and wave numbers. This very good correspondence is established by integrating the lattice differential equation starting from
the NLS approximation, which leads to robust motion of the bright breather over long times.
We also demonstrate that bright breathers can be generated in the resonant granular
chain initially at rest, and driven from a boundary at a frequency within the focusing region of the optical band
(see \cite{Boechler10,hasan15} for related works).
In addition, we analyze discrepancies between numerical solutions and NLS approximations
which can be observed at some other wave numbers for the optical branch and in the acoustic case. In particular, in some cases we observe formation and robust propagation of \emph{nanoptera}, bright breathers that emit small-amplitude oscillations behind them.

Following the approach in \cite{Chong13}, we also consider the defocusing NLS at the edges of both optical and acoustic branches
that correspond to wave number equal to $\pi$, and use the static solutions of the modulation equation to construct the approximate standing dark breather solutions.
A continuation procedure based on a Newton-type method with initial conditions
built from the approximation ansatz is employed to compute
numerically exact stationary dark breathers for a wide range of frequencies and at
different mass ratios. Interestingly, the resulting branches of solutions also include large-amplitude dark breathers, whose dynamics is \emph{strongly nonlinear}.

We examine numerically the stability of the exact dark breathers of both weakly and strongly nonlinear types,
using both a Floquet analysis and direct numerical simulations.
Our results suggest that small-amplitude weakly nonlinear dark breather solutions with frequencies close to the linear frequencies of the system are stable, in analogy
to what was found for the homogeneous granular
chain in~\cite{Chong13}. As the amplitude of the dark breather solution becomes relatively large compared to the amount of precompression, the solution starts to exhibit a very strong modulational instability of the background, eventually leading to its complete destruction and the emergence of chaotic dynamics.
However, when a real instability of the solution is dominant in the Floquet spectrum,
it may give rise to steady propagation of a dark breather at large enough time.
Interestingly, we observe that such types of propagating dark breathers can also form spontaneously,
as a result of the instability of certain acoustic periodic traveling waves.
We also show that the mass ratio plays a substantial role in oscillatory instabilities of the background of the dark breathers. In contrast, the value of the mass ratio has a less significant effect on real instability modes for both the strongly and weakly nonlinear solutions.

The paper is organized as follows. Sec.~\ref{sec:model} introduces the model and the dispersion relation for plane waves is derived.
In Sec.~\ref{nlsl}, we derive the modulation equation of NLS type (with more technical details included in the Appendix),
recall basic features of the focusing and defocusing regimes of NLS, and localize these different regimes
in the parameter space of the original lattice.
In Sec.~\ref{sec:focusing} we investigate the existence of moving bright breathers for the original system at different mass ratios,
and test the validity of the NLS approximation.
In Sec.~\ref{sec:defocusing} we analyze the existence and stability of stationary dark breathers and
discuss the excitation of traveling dark breathers by different means.
Concluding remarks can be found in Sec.~\ref{sec:end_weak}.

\section{The model}\label{sec:model}

We consider a granular chain of identical beads of mass $m_1$ precompressed by the static load $F_0$. To obtain a locally resonant granular chain, we attach a secondary mass $m_2$ attached to each primary bead via a linear spring of stiffness $K>0$. The primary and secondary masses are constrained to move in the horizontal direction with displacements $\tilde{u}_n(\tilde{t})$ and $\tilde{v}_n(\tilde{t})$, respectively. In what follows, we assume that the deformation of the neighboring primary masses is confined to a sufficiently small region near the contact point and varies slowly enough on the time scale of interest. We also assume that the effects of plasticity and dissipation can be neglected. Under these assumptions, the interaction of the $n$th and $(n+1)$th primary beads is governed by the static Hertzian contact law \cite{Coste97,Nest01,Sen200821}
corresponding to the force $F=\mathcal{A}(\tilde{\d_0} + \tilde{u}_n - \tilde{u}_{n+1})_{+}^{\a}$, where $(x)_+ = \text{max}\{x,0\}$, $\mathcal{A}>0$ is the Hertzian constant determined by material properties of the beads and the
radius of the contact curvature, $\a$ is the Hertzian nonlinear exponent which is equal to $3/2$ for spherical beads, $\tilde{\d_0} = (F_0/\cA)^{1/\a}$ is the equilibrium overlap of the adjacent primary masses due to the precompression.
The equations of motion are then given by
\beq
\begin{split}
m_1\dfrac{d^2 \tilde{u}_{n}}{d\tilde{t}^2} &= \mathcal{A}(\tilde{\d_0} + \tilde{u}_{n-1} - \tilde{u}_n )_{+}^{\a} - \mathcal{A}(\tilde{\d_0} + \tilde{u}_n - \tilde{u}_{n+1})_{+}^{\a}
- K(\tilde{u}_n - \tilde{v}_n),\\
m_2\dfrac{d^2 \tilde{v}_{n}}{d\tilde{t}^2}&= K(\tilde{u}_{n} - \tilde{v}_{n}).
\label{eq:dyn}
\end{split}
\eeq
Let $R$ be a given characteristic length scale. We introduce dimensionless variables
$$
u_n=\dfrac{\tilde{u}_n}{R}, \quad v_n=\dfrac{\tilde{v}_n}{R}, \quad t=\tilde{t}\sqrt{\dfrac{R^{\a-1}\mathcal{A}}{m_1}},
$$
as well as three dimensionless parameters
$$
\r=\dfrac{m_2}{m_1}, \quad \k=\dfrac{K}{\mathcal{A}R^{\a-1}}, \quad \d_0=\dfrac{\tilde{\d}_0}{R},
$$
where $\r$ is the ratio of two masses and $\d_0$ the dimensionless overlap due to precompression.
When $R$ corresponds to the radius of spherical primary beads, the parameter
$\k$ measures the relative strength of the linearly elastic spring and Hertzian interactions
for compressions at the scale of $R$. Another relevant rescaling consists in fixing
$R = (K/\mathcal{A})^{1/(\alpha -1)}$, so that $\kappa = 1$.

\begin{figure}
\centerline{\includegraphics[width=0.6\textwidth]{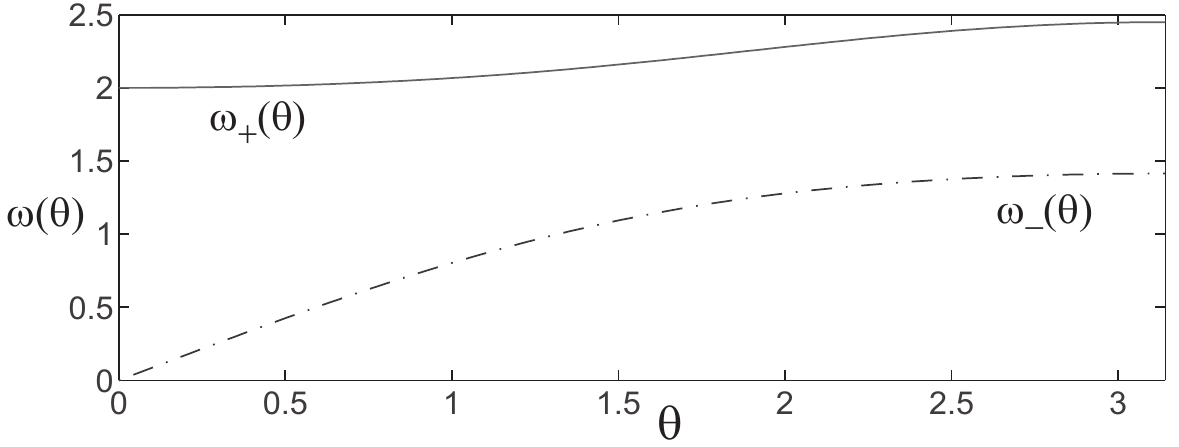}}
\caption{\footnotesize The optical (solid) and acoustic (dashed) branches of the dispersion relation \eqref{disp1d}. Due to the even symmetry about $\theta =0$, only $[0,\pi]$ interval is shown. Here $\d_{0} = 4/9$, $\k =1$ and $\r = 1/3$.}
\label{fig:w_vs_q}
\end{figure}

Rewriting the system \eqref{eq:dyn} in terms of the dimensionless variables and parameters, we obtain
\beq
\begin{split}
\ddot u_{n} &= V^{'}(\d^+u_n) - V^{'}(\d^-u_n) - \k(u_n - v_n)\\
\r\ddot v_{n}&= \k(u_{n} - v_{n}),
\label{eq:Hertz}
\end{split}
\eeq
where $\ddot u_{n}$ and $\ddot v_{n}$ are the accelerations of the primary and secondary masses, respectively, $\d^+u_n = u_{n+1} - u_n$ and $\d^-u_n = u_{n} - u_{n-1}$ denote the shift operators, and
$V(r)$ is the interaction potential in the form
\beq\label{eq:Vr}
V(r) = \frac{1}{\a + 1}(\d_0 - r)^{\a+1}_{+} +\d_0^{\a}r - \frac{1}{\a+1}\d_0^{\a+1}
\eeq
that satisfies $V(0) = V'(0) = 0$. In the weakly nonlinear regime when $r \ll \d_0$, it is relevant to consider the Taylor expansion
\beq
\label{expv}
V(r) = K_2\frac{r^2}{2} + K_3\frac{r^3}{3} + K_4\frac{r^4}{4} + O(|r|^5),
\eeq
where $K_2 = \a\d_0^{\a -1}$, $K_3 = -\frac{1}{2}\a(\a-1)\d_0^{\a-2}$ and $K_4 = \frac{1}{6}\a(\a-1)(\a-2)\d_0^{\a-3}$. Linearizing the system \eqref{eq:Hertz} about the equilibrium state, we obtain
\beq
\ddot u_n = K_2(u_{n+1} -2u_n + u_{n-1}) - \k(u_n - v_n), \q \r\ddot v_n = -\k(v_n - u_n).
\eeq
The linear system has nontrivial plane wave solutions in the form
\beqs
u_n(t) = Ae^{i(n\theta - \o t)}, \q v_n(t) = Be^{i(n\theta - \o t)}
\eeqs
with wave number $\theta \in (-\pi, \pi]$, frequency $\o$ and amplitudes $A$ and $B$, provided that the matrix
\bea\label{M}
\mathbf{M}
= \left( \begin{array}{cc}
\o^2 - D - \k & \k\\
\k / \r & \o^2 - \k / \r \end{array} \right)
\eea
with $D= D(\theta) = 4K_2 \sin^2(\theta/2)$
has a vanishing determinant. This yields the dispersion relation
\bea\label{disp1d}
\o^2 = \o_{\pm}^2(\theta) = \frac{D + \k + \k/\r \pm \sqrt{(D + \k + \k/\r)^2 - 4D\k/\r}}{2}.
\eea
The relation has two branches: optical, $\o_{+}(\theta)$, and acoustic, $\o_{-}(\theta)$, as shown in Fig.~\ref{fig:w_vs_q}. One can show that $\o_{+}(\theta)$ and $\o_{-}(\theta)$ are increasing functions of $\theta$ in $[0, \pi]$ and that $\o_{-}(\pi) < \o_{+}(0)$, implying the existence of a gap between the two branches.

Note that in the limit $\r\rightarrow 0$, the model reduces to the one for a regular (non-resonant) homogeneous granular chain with precompression, which is governed by \cite{Coste97}
\beq
\ddot u_{n} =(\d_0 + u_{n-1} - u_{n})^{\a}_{+} - (\d_0 + u_{n} - u_{n+1})^{\a}_{+}.
\eeq
In this case the dispersion relation for the linearized problem only has the acoustic branch
\beq
\o^2(\theta) = 4K_2\sin^2(\theta/2).
\eeq
In the opposite limit of $\rho \rightarrow \infty$ and for zero initial conditions for $v_n(t)$, the system approaches a precompressed granular chain with quadratic onsite potential, i.e., a Newton's cradle model with precompression \cite{JKC13}, described by
\beq
\ddot u_n + \k u_n = (\d_0 + u_{n-1} - u_n)^{\a}_+ - (\d_0 + u_{n} - u_{n+1})^{\a}_+.
\eeq
In this limit, the dispersion relation is purely optical and given by
\beq
\o^2(\theta) = 4K_2\sin^2(\theta/2) + \k.
\eeq
\section{\label{nlsl}Nonlinear Schr\"{o}dinger limit}
\subsection{Derivation of the nonlinear Schr\"{o}dinger equation}\label{sec:derivation_NLS}
To study the weakly nonlinear dynamics of the locally resonant chain, we begin by deriving the modulation equations for the plane-wave mode $E(t,n) := e^{i(n\theta - \o t)}$, with $\theta \in (-\pi, \pi]$ and $\omega = \omega_+$ or $\omega_-$ defined in (\ref{disp1d}).
Using a small parameter $\e>0$, we introduce the slow time $\t = \e^2 t$ and the macroscopic traveling wave coordinate $\xi = \e(n - ct)$,
where $c=c_\pm :=\o_{\pm}^{'}(\theta)$ is the group velocity. We seek solutions of \eqref{eq:Hertz} in the form of fast oscillating, small-amplitude patterns modulated by envelopes that vary slowly in space and time:
\beq
\begin{split}
&u_n(t) = u_n^{A}(t) + O(\e^2) = \e \{A_{1,0}(\t, \xi) + A_{1,1}(\t, \xi)E(t,n) + c.c.\} + O(\e^2)\\
&v_n(t) = v_n^{B}(t) + O(\e^2) = \e \{B_{1,0}(\t,\xi) + B_{1,1}(\t, \xi)E(t,n) + c.c.\} + O(\e^2),
\label{ansatz_uv}
\end{split}
\eeq
where $c.c.$ denotes complex conjugate. More precisely, following \cite{Gian04,Gian06} (see also \cite{Schneider10}), we substitute the multiple-scale ansatz
\beq\label{eq:ansatz}
u_n(t) = \sum_{k\in \hN_1} \e^k\sum_{j=-k}^{k} A_{k,j}(\t, \xi)E^{j}(t,n), \qq
v_n(t) = \sum_{k\in \hN_1} \e^k\sum_{j=-k}^{k} B_{k,j}(\t, \xi)E^{j}(t,n),
\eeq
where $\hN_1$ is the set of natural numbers $k \in \hN, k \ge 1$, $A_{k,j},  B_{k,j}\in \hC$, $A_{k,-j} = \bar A_{k,j}$ and $B_{k,-j} = \bar B_{k,j}$, into \eqref{eq:Hertz}. As shown in Appendix~\ref{App:mod}, this leads to the coupled modulation equations
\bea
&&i\partial_{\t}A_{1,1} + \beta\partial_{\xi}A_{1,0}A_{1,1} + \frac{1}{2}\o''\partial_{\xi}^2A_{1,1} - h|A_{1,1}|^2A_{1,1}=0 ,\label{eq:mod1} \\
&&[c^2(1+\r)-K_{2}]\partial_{\xi}^{2}A_{1,0} = 8K_3\sin^2{(\theta/2)}\bar A_{1,1}\partial_{\xi}A_{1,1} + c.c.\label{eq:mod2}
\eea
and the identities
\begin{equation}
\label{b10b11}
B_{1,0}=A_{1,0}, \ \ \  B_{1,1} =  \frac{\k}{\k - \r\o^2} \, A_{1,1}
\end{equation}
(one can check that $\k - \r\o_\pm^2 \neq 0$).
In equations (\ref{eq:mod1})-(\ref{eq:mod2}), we assume two \emph{non-resonance conditions},
\bea\label{eq:nonresonance1}
(4\o^2 - 4K_2\sin^2\theta-\k)(\k-4\o^2\r) + \k^2 \neq 0
\eea
and
\beq\label{eq:nonresonance2}
c^2(1 + \r) - K_2 \neq 0,
\eeq
and set
\bea\label{eq:ppOmega}
\beta = -4K_3\sin^2{(\theta/2)}\g, \qq \o'' = \left\{K_2\cos\theta - c^2\left[1 - \frac{3\o^2\r^2\k^2+\r\k^3}{(\r\o^2-\k)^3}\right]\right\}\g
\eea
and
\bea\label{eq:h}
h= \left\{\frac{16K_3^2\sin^2\theta(1-\cos\theta)^2(\k-4\o^2\r)}{(4\o^2-4K_2\sin^2\theta-\k)(\k-4\o^2\r)+\k^2}+6K_4(1-\cos\theta)^2\right\}\g,
\eea
where
\beq\label{eq:gamma}
\gamma = \frac{(\r\o^2 - \k)^2}{\o[\r\k^2 + (\r\o^2-\k)^2]}.
\eeq
Note that $h$ is non-singular when \eqref{eq:nonresonance1} holds.
Observe also that \eqref{eq:mod2} can be rewritten as
\bea\label{eq:mod21}
\partial_{\xi}^{2}A_{1,0} = \l \partial_{\xi}|A_{1,1}|^2,
\eea
where
\beq
\label{eq:lambda}
\l = \frac{8K_3\sin^2{(\theta/2)}}{c^2(1 + \r) - K_2}
\eeq
is well-defined under \eqref{eq:nonresonance2}.
Integrating both sides of \eqref{eq:mod21} with respect to $\xi$ yields
\beq\label{eq:parA10}
\partial_{\xi} A_{1,0} = \l |A_{1,1}|^2 + f(\t),
\eeq
where $f$ is an arbitrary real-valued function. Substituting $\partial_{\xi}A_{1,0}$ into \eqref{eq:mod1} then gives
\beq\label{eq:mod1New}
i\partial_{\t} A_{1,1} + \beta f(\t)A_{1,1} + \frac{1}{2} \o''\partial_{\xi}^2A_{1,1} + (\l\beta - h)|A_{1,1}|^2 A_{1,1} = 0.
\eeq
Let
\beq\label{eq:A11}
A(\xi, \t) = A_{1,1}(\xi, \t) e^{-i\beta F(\t)},
\eeq
where $F(\t)$ is the antiderivative of $f(\t)$.
From \eqref{eq:mod1New} it then follows that $A(\xi, \t)$ satisfies the classical time-dependent nonlinear Schr\"{o}dinger (NLS) equation
\beq\label{eq:NLS}
i\partial_{\t}A + \frac{1}{2}\o''\partial_{\xi}^2A - \wt h |A|^2A = 0,
\eeq
where we set
\beq
\wt h = h - \l\beta.
\label{eq:h_tilde}
\eeq
Solutions of (\ref{eq:NLS}) provide approximate solutions
to the original lattice system \eqref{eq:Hertz}
through identities
\eqref{ansatz_uv}, \eqref{eq:A11}, \eqref{eq:parA10} and \eqref{b10b11}.

\subsection{Bright and dark breathers solutions}

Equation (\ref{eq:NLS}) has solutions in the form
\beq\label{eq:A}
A(\xi,\t) = \wt A(\xi)e^{i\m\t},
\eeq
where $\m \in \hR$ and $\wt A(\xi)$ is a real-valued function satisfying the stationary NLS equation
\beq\label{eq:NLS_stat}
-\mu\wt A + \frac{1}{2}\o''\partial_{\xi}^2\wt A - \wt h \wt A^3 = 0 .
\eeq
The \emph{focusing case} of the NLS equation \eqref{eq:NLS} occurs for $\o''\wt h <0$. In this case \eqref{eq:NLS_stat} admits sech-type solution
\beq\label{sechA}
\wt A(\xi) = \sqrt{\frac{-2 \m}{\wt h}} \text{sech}\biggl(\sqrt{\frac{2 \m}{\o''}}\xi\biggr)
\eeq
for $\mu$ such that $\m \o'' >0$ and $\m \wt h <0$.
The solutions of (\ref{eq:NLS}) given by \eqref{sechA}-\eqref{eq:A} are denoted as {\em bright breather solutions}.
Using \eqref{eq:A11} and \eqref{eq:A}, we then obtain
\beq\label{sechA11}
A_{1,1}(\xi, \t) = \sqrt{\frac{-2 \m}{\wt h}} \text{sech}\left(\sqrt{\frac{2 \m}{\o''}}\xi\right)e^{i[\m\t + \beta F(\t)]},
\eeq
and integrating \eqref{eq:parA10} yields
\beq\label{sechA10}
A_{1,0}(\xi, \t) =\frac{\l\sqrt{2\o''\m}}{|\wt h|}\text{tanh}\left(\sqrt{\frac{2 \m}{\o''}}\xi\right)+f(\t)\xi + C(\t).
\eeq
In the focusing case,
the spatially homogeneous solutions $A_{\theta , r}(\tau )=r\, e^{-i\, \tilde{h}\, r^2 \tau}$ of (\ref{eq:NLS})
are unstable, with typical perturbations giving rise to localized waves close to bright breathers \cite{Osbo10}.

On the other hand, when $\o''\wt h>0$, equation \eqref{eq:NLS} becomes
\emph{defocusing}, and \eqref{eq:NLS_stat} has a  tanh-type solution
\beq\label{tanh}
\wt A(\xi) = \sqrt{\frac{-\m}{\wt h}} \text{tanh}\left(\sqrt{\frac{-\m}{\o''}}\xi\right)
\eeq
for $\m$ satisfying $\m\o'' <0$ and $\m\wt h < 0$.
Expressions \eqref{tanh}-\eqref{eq:A} define {\em dark breather solutions} of (\ref{eq:NLS}).
We then have
\beq\label{tanhA11}
A_{1,1}(\xi, \t) =  \sqrt{\frac{-\m}{\wt h}} \text{tanh}\left(\sqrt{\frac{-\m}{\o''}}\xi\right)e^{i[\m\t + \beta F(\t)]}.
\eeq
Integrating \eqref{eq:parA10}, we obtain
\beq\label{tanhA10}
A_{1,0}(\xi, \t) = -\frac{\l\sqrt{-\o''\m}}{|\wt h|}\text{tanh}\left(\sqrt{\frac{-\m}{\o''}}\xi\right)+\left(f(\t)-\frac{\l\m}{\wt h}\right)\xi + C(\t).
\eeq

\subsection{\label{focdef}Focusing and defocusing parameter regimes}

\begin{figure}[h]
\centerline{\includegraphics[width=\textwidth]{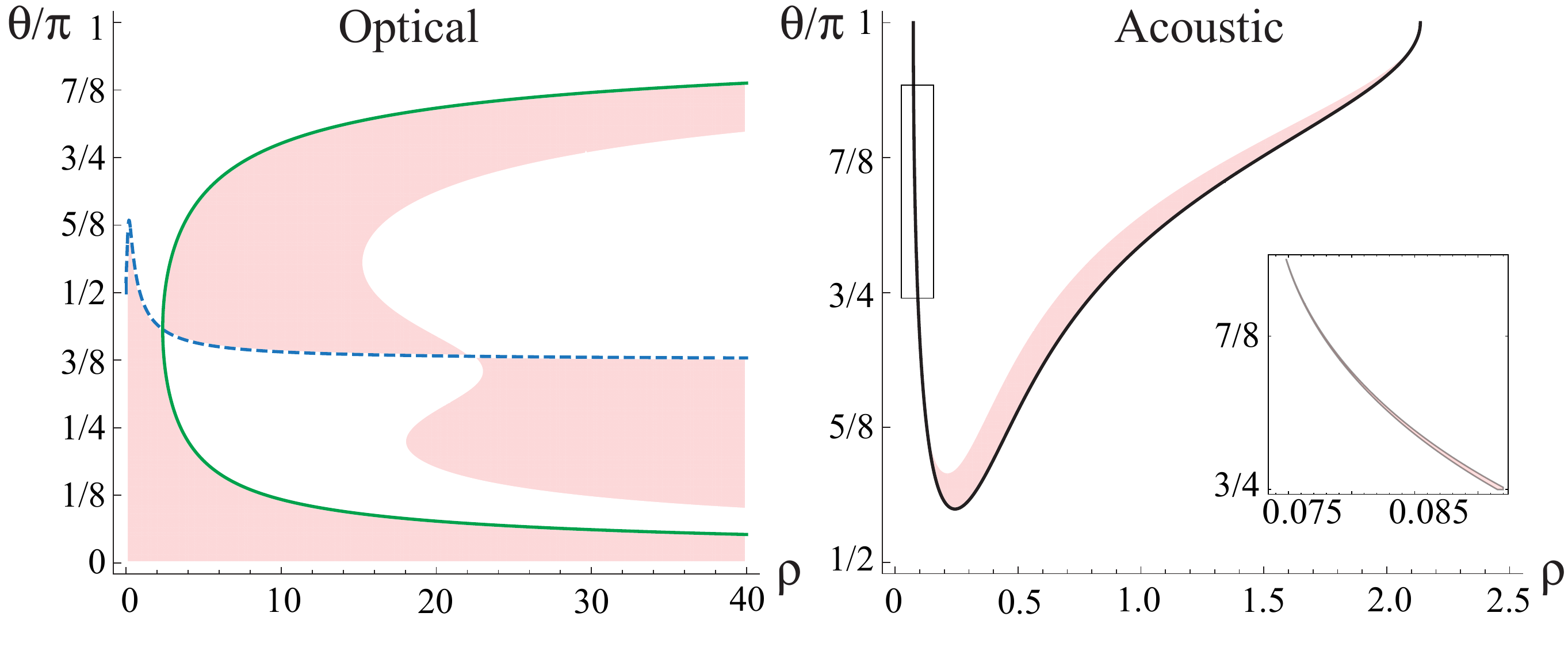}}
\caption{\footnotesize Plots of $\text{sign}(\o'' \wt h)$ for optical (left) and acoustic (right) branches. The focusing region is shaded in pink
(gray in the black and white version). Its boundaries for the optical case include $\o_{+}'' = 0$ (blue dashed line) and the curve along which the second non-resonance condition \eqref{eq:nonresonance2} is violated (greed solid line). Black solid line in the acoustic case (right plot) corresponds to the curve along which the first non-resonance condition \eqref{eq:nonresonance1} breaks down. The remaining boundaries separating focusing (pink/gray) and defocusing (white) regions corresponds to $\wt h = 0$. Inset zooms in on the region inside the rectangle. Here $\a = 3/2$, $\d_{0} = 4/9$, $\k =1$.}
\label{fig:region}
\end{figure}

Representative plots of $\text{sign}(\o'' \wt h)$ as a function of $\theta$ and $\rho$ for optical and acoustic branches are shown in Fig.~\ref{fig:region}, where we vary $\rho$ but fix the other parameters at the same values as in Fig.~\ref{fig:w_vs_q} ($\k = 1$, $\d_0 = 4/9$ and $\a = 3/2$).
Interestingly, it turns out that the number of focusing wavenumber intervals can change when varying $\rho$,
both for optical and acoustic modes. In particular, the topology of focusing regions is more complex for
optical modes, as shown in the left panel of Fig.~\ref{fig:region}.

Let us now study the origin of this phenomenon.
From \eqref{eq:ppOmega}, \eqref{eq:h}, \eqref{eq:lambda} and \eqref{eq:h_tilde}, one can see that $\o'' \wt h$ can change sign when $\omega''=0$, $\tilde{h}=0$ or when $\tilde{h}$ has a singularity which occurs when either of the two non-resonance conditions, \eqref{eq:nonresonance1} and \eqref{eq:nonresonance2}, is violated for $\theta \in (0,\pi)$.

For the optical branch of the dispersion relation, \eqref{eq:nonresonance1} always holds for the parameters we consider in Fig.~\ref{fig:region},
but the other non-resonance condition, \eqref{eq:nonresonance2}, breaks down at some wave numbers for sufficiently large $\rho$ (solid curve in the left plot of Fig.~\ref{fig:region}). In this case the sign of $\o_{+}''\wt h_{+}$ also changes for any $\rho>0$ at the inflection point of the dispersion relation ($\o_{+}'' = 0$, dashed curve) and, for large enough $\rho$, when the numerator of $\tilde{h}$ in \eqref{eq:h_tilde} vanishes. Thus, for sufficiently small $\rho$, $0<\rho<2.356$, we have the focusing regime (shaded in Fig.~\ref{fig:region}) for $\theta<\theta_c$, where $\o_{+}''(\theta_c)=0$, and the defocusing NLS otherwise. For larger $\rho$, focusing and defocusing regimes alternate, due to breakdown of \eqref{eq:nonresonance2}, the inflection point, and, for $\rho \gtrsim 15.335$, $\tilde{h}=0$. Interestingly, the focusing and defocusing regions ``flip" at $\rho \approx 2.356$ and $\rho \approx 22.298$, where the boundary curves intersect, with focusing regime below the $\o_{+}''=0$ curve at small $\rho$, above it for a range of wave numbers at intermediate mass ratios and below the curve again for a $\theta$-range at larger $\rho$.

We now turn to $\text{sign}(\o''_{-} \wt h_{-})$ and examine the acoustic dispersion branch shown in the right plot of Fig.~\ref{fig:region}. One can show that in this case the curvature of the dispersion curve is always negative, $\o''_{-} (\theta)< 0$,  for $\theta \in (0, \pi]$, yielding $0 \le c_{-}=\o'_{-}(\theta) < \o'_{-}(0) = \sqrt{K_2/(1 + \r)}$. As a result, the second non-resonance condition \eqref{eq:nonresonance2} always holds for the acoustic branch at nonzero wave numbers. However, the first non-resonance condition breaks along the solid curve, changing the sign of $\tilde{h}_{-}$ from negative to positive, and hence the sign of $\o''_{-} \wt h_{-}$ from positive to negative, at sufficiently large $\theta<\pi$ in the interval of mass ratios $0.074<\rho<2.137$. For each $\rho$ in this interval, $\text{sign}(\o''_{-} \wt h_{-})$ changes again to positive (defocusing regime) at slightly larger $\theta$ due to $\tilde{h}=0$. This yields a very narrow shaded focusing region, with the lower boundary for $\theta<\pi$ coinciding with the solid curve where \eqref{eq:nonresonance2} is violated. It should be noted that although this singularity curve includes points where $\theta=\pi$, the focusing region approaches this value at its ends but does not include it because $h(\theta)$ in \eqref{eq:h}, and therefore $\tilde{h}(\theta)$ in \eqref{eq:h_tilde}, does not have a singularity at $\theta=\pi$. Instead, as we approach each of the two end points of the focusing region where $\theta \rightarrow \pi$, the values of $\theta$ where $\tilde{h}_{-}$ is singular and where it vanishes for given $\rho$ both approach $\pi$, so that at $\theta=\pi$ we have the defocusing case, $\tilde{h}_{-}(\pi)\omega''_{-}(\pi)>0$ for $\rho>0$. The focusing region is particularly narrow at smaller mass ratios, $\rho \gtrsim 0.074$ (see the inset in Fig.~\ref{fig:region}). Outside this region one has the defocusing regime. This includes $0<\rho<0.074$, in agreement with the non-resonant chain studied in \cite{Chong13} ($\rho=0$), where only the defocusing case is possible.

\begin{figure}[h]
\centerline{\includegraphics[width=\textwidth]{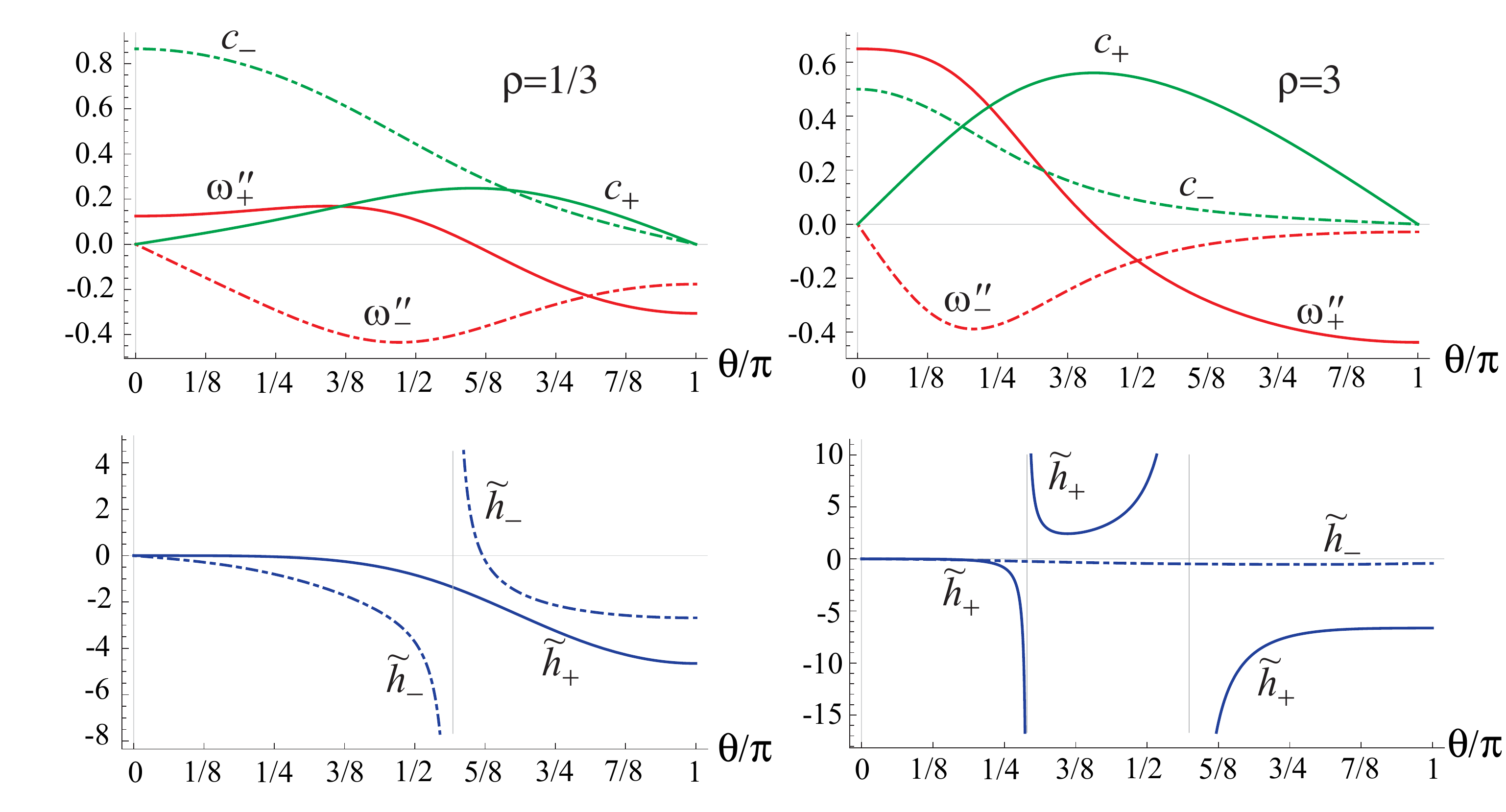}}
\caption{\footnotesize Representative plots of group velocity $c$ (green), curvature $\o''$ (red) and $\wt h$ (blue) as functions of the scaled wave number $\theta/\pi$ at different mass ratios for optical (solid) and acoustic (dashed) branches. Here $\a = 3/2$, $\d_{0} = 4/9$, $\k =1$.
Left column corresponds to $\rho = 1/3$ and right column to $\rho=3$.}
\label{fig:disp}
\end{figure}

To further illustrate these results, we show some plots of group velocity $c$, curvature $\o''$ and $\wt h$ as functions of the wave number $\theta$ in Fig.~\ref{fig:disp}.
One can see that at $\rho=1/3$ the optical branch has negative $\tilde{h}_{+}$ in $(0,\pi]$ with no singularities, and the single transition from focusing to defocusing regime occurs due to the inflection point at $\theta = \theta_{c} \approx 1.892$ in the dispersion curve ($\omega''_{+}(\theta_c)=0$). At $\rho=3$, however, $\tilde{h}_+$ changes sign twice via singularities (breakdown of \eqref{eq:nonresonance2}), and together with the inflection point this yields three transition points separating focusing and defocusing regions (see also the left plot in Fig.~\ref{fig:region}). Meanwhile, the acoustic branch, as already noted, has no inflection points in $(0,\pi]$, $\omega''_{-}<0$, and the transition from defocusing to focusing (for $\theta \in (1.784,1.949)$) and back at $\rho=1/3$ occurs through the change of sign of $\tilde{h}_-$ due to singularity (breakdown of \eqref{eq:nonresonance2}) and going through zero. At $\rho=3$ we have $\tilde{h}_{-}<0$ in $(0,\pi]$ without singularities, so the regime is defocusing.\\


To complete the above results, let us illustrate how the precompression $\delta_0$ in \eqref{eq:Vr}
influences the number of focusing regions for a given mass ratio $\rho$.
In what follows we fix $\kappa =1$ in \eqref{eq:Hertz}. Moreover,
we parameterize the optical and acoustic modes using their
frequency $\omega$ instead of wavenumber $\theta$. This parameterization is interesting from a practical point of view,
in order to analyze the response of granular chains to a periodic driving (see Sec.~\ref{mblm} for an example).

\begin{figure}[h]
\begin{center}
\includegraphics[scale=0.37]{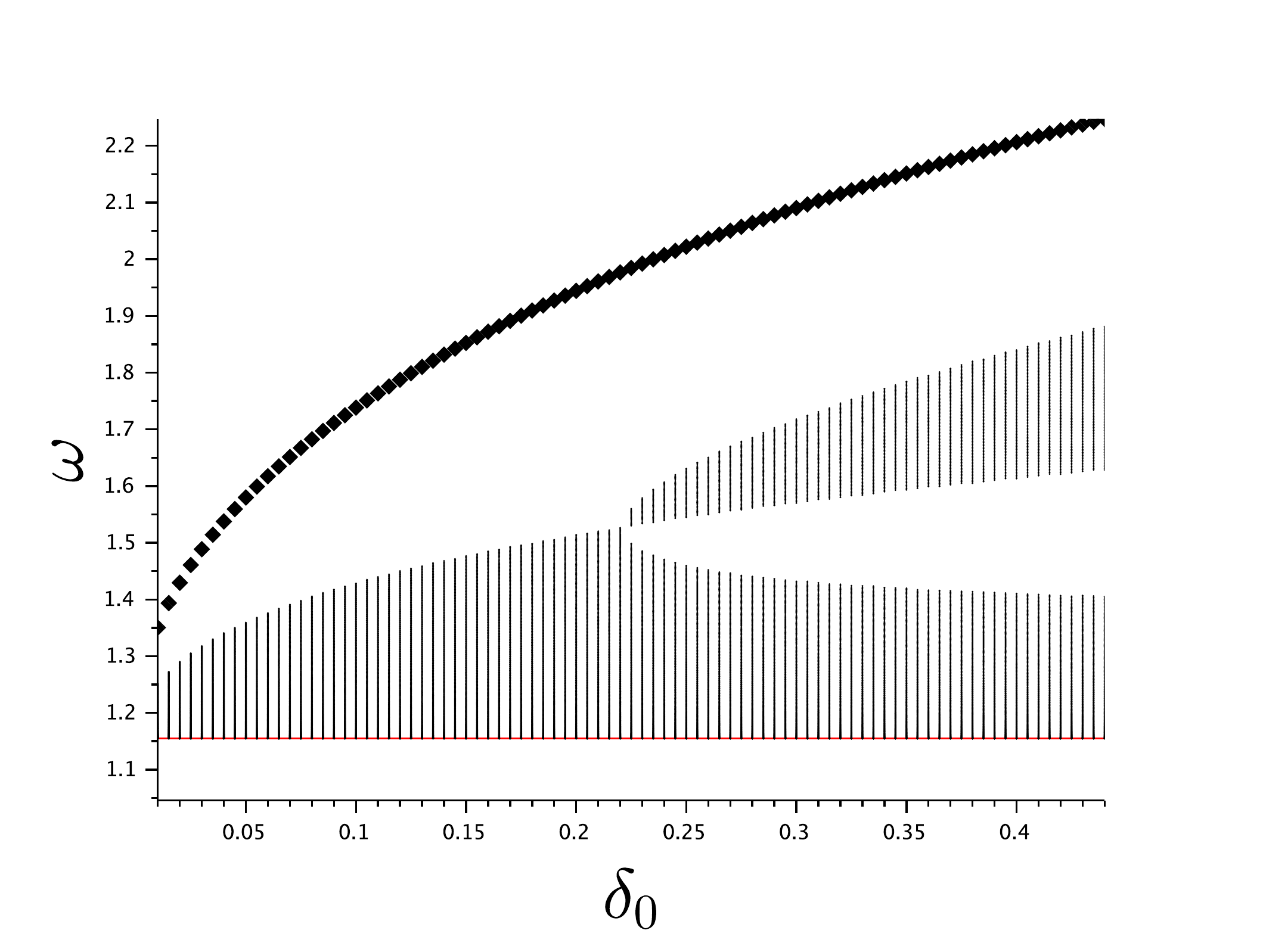}
\includegraphics[scale=0.37]{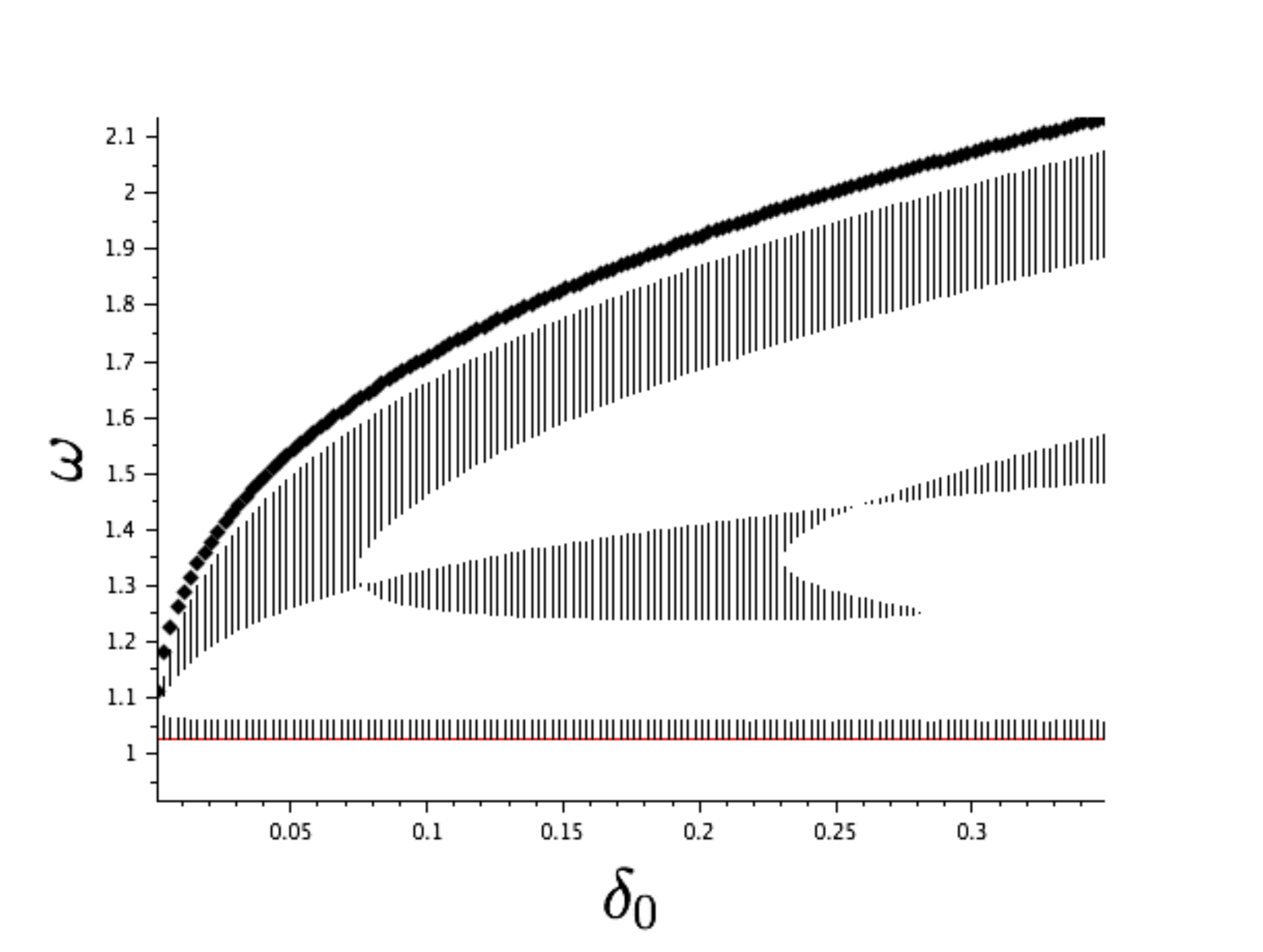}
\end{center}
\caption{\footnotesize
Optical band delimited by $\omega = \omega_{+}(\pi )$ (dotted curve) and
$\omega = \omega_{+}(0 )=\sqrt{\kappa (1+\rho^{-1})}$ (red line), versus precompression $\delta_0$.
We have fixed $\kappa =1$, $\rho=3$ (left plot) and $\rho = 17.8$ (right plot).
Hatched regions correspond to the focusing case $\o''\wt h <0$.}
\label{fig:opband}
\end{figure}

Fig.~\ref{fig:opband} illustrates how the focusing regions depend on $\delta_0$ in the case of optical modes.
For $\rho=3$, the left plot displays the optical band bounded by the graph of
$\delta_0 \mapsto \omega = {\omega_{+}}{(\pi)}$
and lying above the line $\omega = \omega_{+}(0)=\sqrt{\kappa (1+\rho^{-1})}$.
Within the optical band, the hatched region corresponds to the focusing case $\o_+''\wt h_+ <0$
and white regions to the defocusing case $\o_+''\wt h_+ >0$.
When $\delta_0$ exceeds a threshold (around $\delta_0 = 0.22$), the band of focusing modes
splits into two parts. The right plot describes the case of a larger mass ratio $\rho = 17.8$.
The above transition from one to two focusing bands still takes place, but the precompression
threshold is much lower, around $\delta_0 = 0.002$. When $\delta_0$ is further increased, the
number of focusing bands changes according to the transitions $2 \rightarrow 3 \rightarrow 4 \rightarrow 3$,
within the range of precompression previously examined for $\rho=3$.

\begin{figure}[h]
\begin{center}
\includegraphics[scale=0.37]{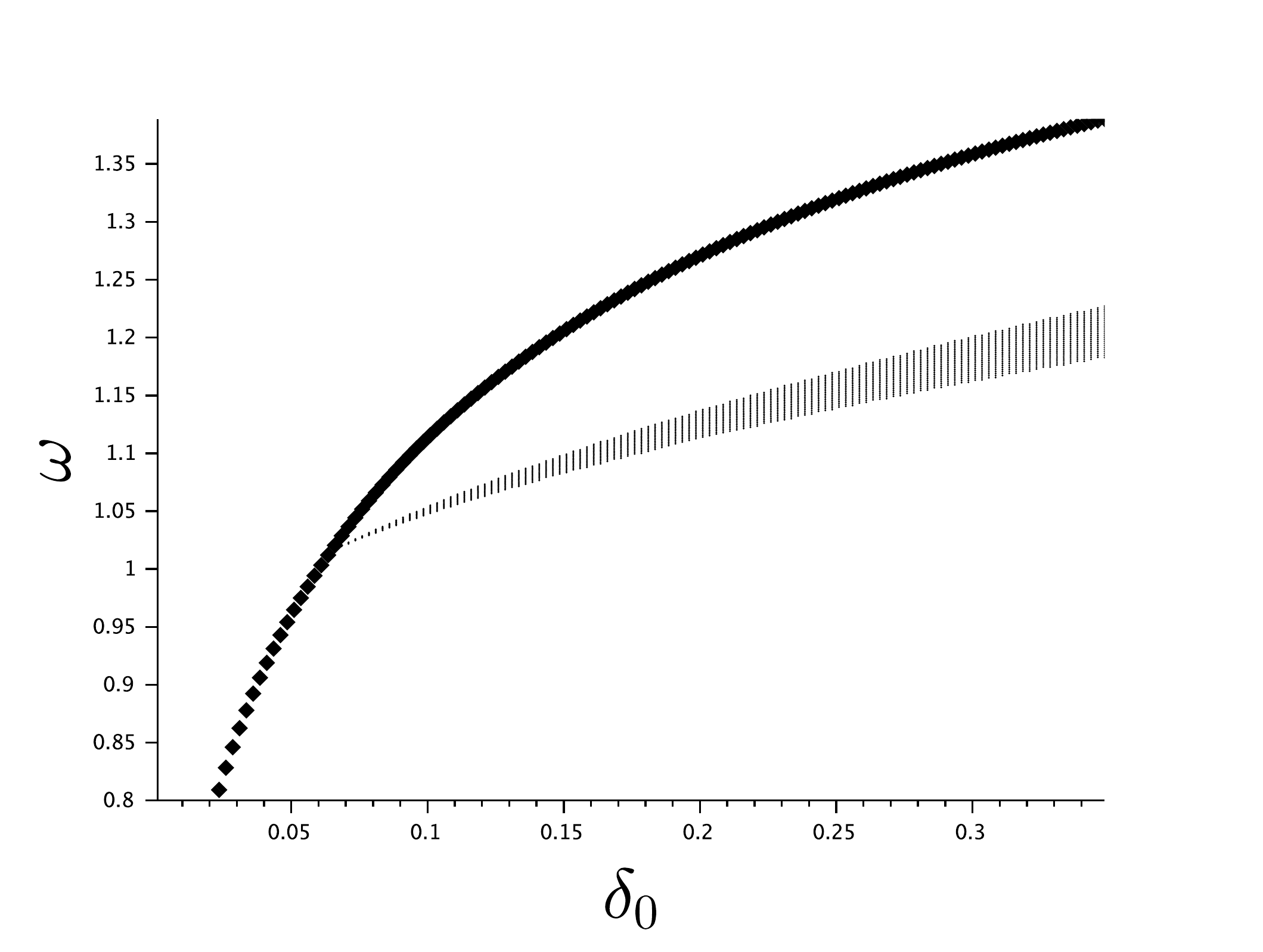}
\end{center}
\caption{\footnotesize
The upper curve (black diamonds) corresponds to the graph of $\delta_0 \mapsto \omega_{-}(\pi )$
delimiting the acoustic band $\omega \in [0 , \omega_{-}(\pi ) ]$,
where we have fixed $\kappa =1$ and $\rho = 0.32$.
The hatched region corresponds to the focusing case $\o''\wt h <0$.}
\label{fig:acband}
\end{figure}

The case of acoustic modes is illustrated by Fig.~\ref{fig:acband} for $\rho = 0.32$.
When precompression lies below a threshold (around $\delta_0 = 0.06$), all acoustic modes are defocusing, while
a thin band of focusing modes exists above this threshold.

\section{\label{sec:focusing} Moving bright breathers in the granular chain}

From direct numerical simulations of the granular chain,
we now investigate how well the dynamics governed by the focusing NLS equation
approximates the solutions of the original lattice.
Due to the existence of a kink component in the position variables $u_n$, $v_n$ for
bright breathers, it is convenient to rewrite \eqref{eq:Hertz} in terms of strain variables $x_n = \d^-u_n=u_n - u_{n-1}$ and $y_n = \d^-v_n=v_n - v_{n-1}$, which yields
\beq
\begin{split}
\ddot x_{n} &= V'(x_{n+1}) - 2V'(x_{n}) + V'(x_{n-1}) - \k(x_n - y_n),\\
\r\ddot y_{n}&= \k(x_{n} - y_{n}).
\label{eq:Hertzd0S}
\end{split}
\eeq
Numerical integration is performed on the lattice of size $2N + 1$ with zero-strain boundary conditions $x_{-N-1} = x_{N+1} = 0$ unless explicitly stated.
We use the same parameters as before ($\a = 3/2$, $\d_{0} = 4/9$, $\k =1$)
and simulate (\ref{eq:Hertzd0S}) for different mass ratio $\rho$ and initial conditions
described below.

\subsection{\label{aabp}Analytical approximation of breather profiles}

We recall that solutions of the NLS equation (\ref{eq:NLS})
define small amplitude approximate solutions \eqref{ansatz_uv} to the original lattice \eqref{eq:Hertz},
where $A(\xi ,\tau )$ describes the amplitude of a plane-wave mode $E(t,n) := e^{i(n\theta - \o t)}$.
Using \eqref{ansatz_uv}, \eqref{b10b11},
\eqref{sechA11} and \eqref{sechA10}, we find that the approximate breather solutions given by the
focusing NLS equation ($\o''\wt h <0$)
take the form
\beq
\begin{split}
&x_{n}^{A}(t) = \d^-u_n^{A}(t) = 2 M_2\d^-\{\text{sech}[a(n - n_0 -ct)]\cos{(n\theta - \o_{b}t + \beta F(\e^2t))}\}\\
&+ M_1\d^-\text{tanh}[a(n - n_0 - ct)] + \e^2 f(\e^2t),\\
&y_{n}^{A}(t) = \d^-v_n^{A}(t) = \frac{2\k M_2}{\k-\r\o^2}\d^-\{\text{sech}[a(n - n_0 -ct)]\cos{(n\theta - \o_{b}t + \beta F(\e^2 t))}\} \\
&+ M_1\d^-\text{tanh}[a(n - n_0 - ct)]+ \e^2 f(\e^2t),
\label{eq:focus_app}
\end{split}
\eeq
with arbitrary spatial translation by $n_0$. Here instead of $\mu \e^2$ we introduce the breather frequency $\o_b=\o-\mu\e^2$, a real number such that $|\o -\o_b| = O(\e^2)$, $(\o - \o_b)\o''>0$, $(\o - \o_b) \tilde{h}<0$, and set
\beqs
M_1 = \frac{\l\sqrt{2\o''(\o - \o_b)}}{|\wt h|}, \q M_2 = \sqrt{\frac{2(\o_b - \o)}{\wt h}} \q \text{and} \q  a = \sqrt{\frac{2 (\o - \o_b)}{\o''}}.
\eeqs
Recall that $f(\t)$ is a slow-time varying function, independent of $n$ and $F(\t)$ is its antiderivative. In what follows we simply set $F(\t)\equiv 0$ so that $f(\t) \equiv 0$.

To numerically integrate \eqref{eq:Hertzd0S}, we start with initial condition determined from the first order approximation \eqref{eq:focus_app} at $t = 0$, along with initial velocities given by
\beq
\begin{split}
&\dot x_n(0) =c\,a\left\{-M_{1}\d^-\text{sech}^2[a(n - n_0 )] +2 M_{2}\d^-\text{tanh}[a(n - n_0)]\text{sech}[a(n - n_0)]\cos{(n\theta )}\right\}\\
&+2\o_bM_2\d^-\text{sech}[a(n - n_0)]\sin{(n\theta ),} \\
&\dot y_n(0) =c\,a\left\{-M_{1}\d^-\text{sech}^2[a(n - n_0 )] +\frac{2\k M_2}{\k - \r\o^2}\d^-\text{tanh}[a(n - n_0)]\text{sech}[a(n - n_0)]\cos{(n\theta)}\right\}\\
&+\frac{2\o_b\k M_2}{\k - \r\o^2}\d^-\text{sech}[a(n - n_0)]\sin{(n\theta)} .
\end{split}
\label{eq:focus_ini}
\eeq
In numerical computations,
we fix $\o_b = \o + s\, \cdot 10^{-4}$ with $s = \text{sign}(\tilde{h})=\pm 1$. When $\tilde{h}$ and $\o''$ are
of order unity, the breather amplitude $M_2$ is then of order $10^{-2}$ and its width $a^{-1}$ close to $100$ particles.
Note however that the breather amplitude becomes larger in nearly degenerate cases where $\tilde{h}$
becomes small, or more strongly localized when $\o''$ is small.
The weakly nonlinear approximation (\ref{eq:focus_app}) is not expected to be accurate
in these more strongly nonlinear regimes, as will be numerically confirmed in Sec.~\ref{mmr}.

\subsection{\label{mmr}Moving bright breathers for moderate mass ratio}

We choose a moderate mass ratio $\r = 1/3$ and start by considering the optical case. Recall from the discussion in Sec.~\ref{sec:derivation_NLS} (see also the left plots in Fig.~\ref{fig:region} and Fig.~\ref{fig:disp}) that for this mass ratio the focusing regime takes place at $0<\theta<\theta_c$, where
$\theta_{c} \approx 1.892$ satisfies $\o''_{+}(\theta_c) = 0$.

\begin{figure}
\centerline{\includegraphics[width=\textwidth]{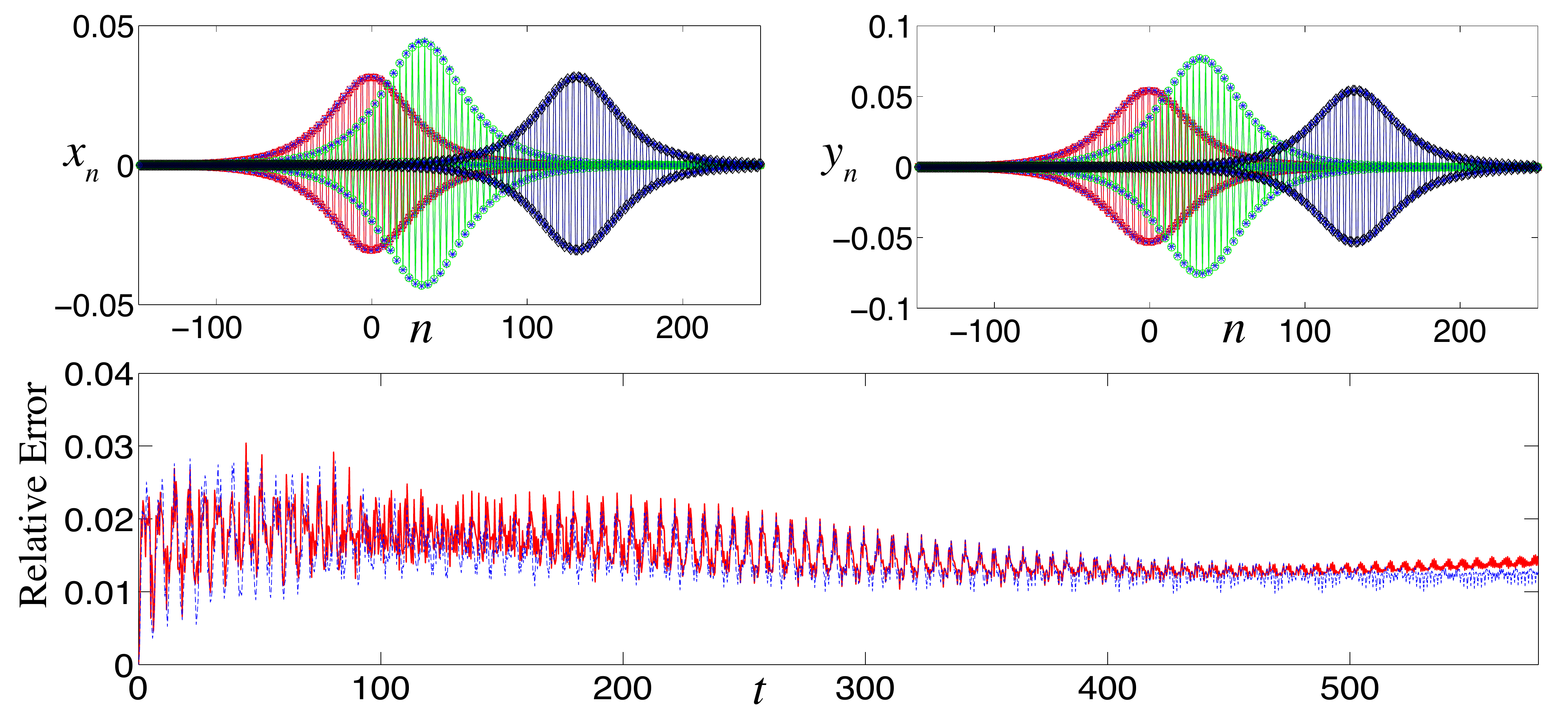}}
\caption{\footnotesize Top plots: snapshots of a moving optical bright breather solution $x_n$ and $y_n$ of the original system \eqref{eq:Hertz}, with initial data determined from \eqref{eq:focus_app}, \eqref{eq:focus_ini}. The breather moving from the middle of the chain to the right is shown here at $t =0$ (connected red squares), $t = 50.125\,T_b \approx 145$ (connected green circles) and $t = 200\,T_b \approx 578$ (connected black squares). The same plots compare the time snapshots of the approximate analytical
solution (connected blue stars) and the numerical evolution result at the same times. Bottom plot: relative errors $E_x(t)$ (solid red curve) and $E_y(t)$ (dashed blue curve). Here $\theta = \pi/2$, $\k = 1$, $\r = 1/3$, $\d_0 = 4/9$, $\o_b  = \o - 0.0001 = 2.1752$, $N = 500$ and
$n_0 = 0$.}
\label{fig:focus1}
\end{figure}

We first consider the wave number $\theta = \pi/2 < \theta_c$. As shown in Fig.~\ref{fig:focus1}, the agreement between the numerical
evolution of \eqref{eq:Hertzd0S} and the approximate analytical solution is excellent, even after a long time at $t = 200\,T_{b}$, where $T_{b} = 2\pi/\o_{b} = 2.8885$ is the period of the modulated wave. The relative errors of the approximation \eqref{eq:focus_app}, defined by $E_x(t) = ||x_n^A(t) - x_n(t)||_{\infty}/||x_{n}(0)||_{\infty}$ and $E_y(t) = ||y_n^A(t) - y_n(t)||_{\infty}/||y_{n}(0)||_{\infty}$, remain less than $3.5\%$ over the time of computation. In the simulation, we observe that the numerically exact solutions have a localized structure which moves to the right end at the speed approximately equal to the group velocity $c_+ \approx 0.23$.
Meanwhile, time variations of wave amplitude are well captured by the
NLS approximation \eqref{eq:focus_app}. Snapshots of these \emph{moving bright breathers} at different times are also shown in Fig.~\ref{fig:focus1}.

\begin{figure}
\centerline{\includegraphics[width=\textwidth]{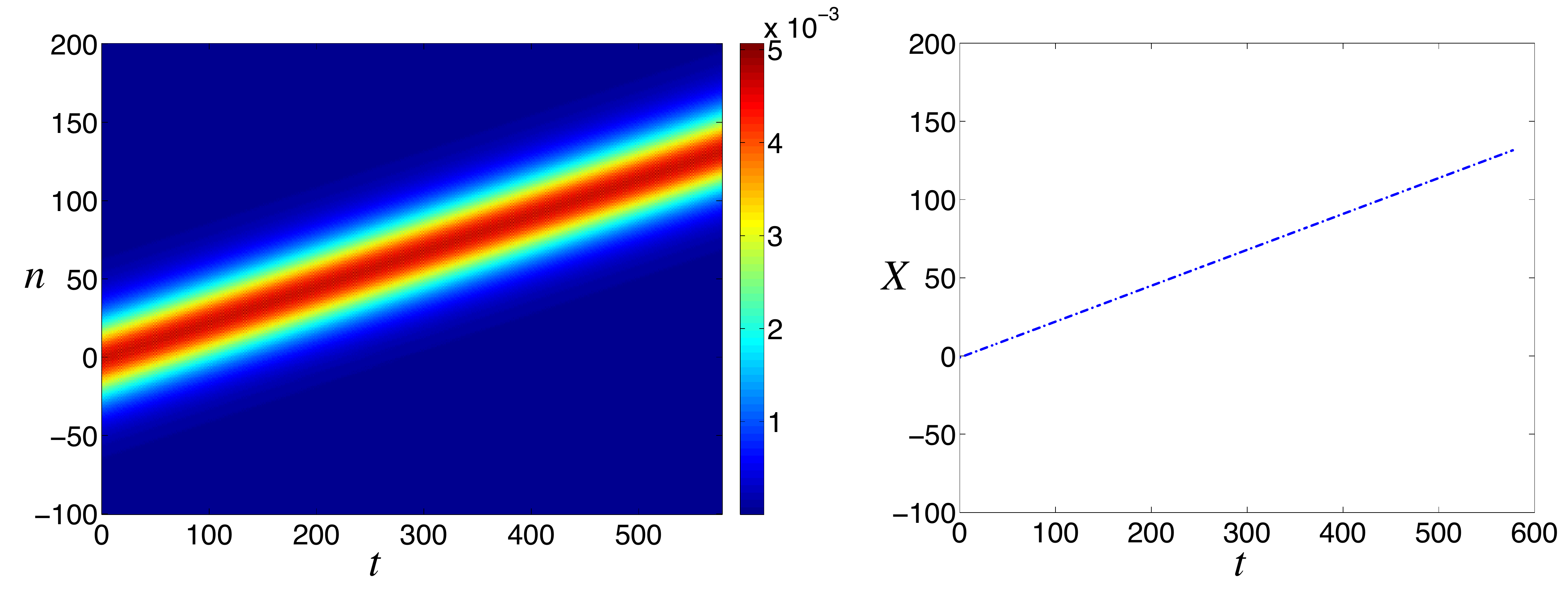}}
\caption{\footnotesize Left plot: energy density of a moving optical bright breather in the system \eqref{eq:Hertz}. Right plot: time evolution of the breather's energy center. Here all the parameters are the same as in Fig.~\ref{fig:focus1}.}
\label{fig:focus3}
\end{figure}

To further illustrate the strong mobility of the bright breather, we consider the energy density (energy stored at the $n$th site):
\beq\label{edensity}
e_n = \frac{1}{2}\dot u_n^2 + \frac{\r}{2}\dot v_n^2 + \frac{\k}{2}(u_n - v_n)^2 + \frac{1}{5}[(\d_0 - \d^-u_n)^{5/2}_{+} + (\d_0 -  \d^+u_n)^{5/2}_{+}] + \frac{1}{2}\d_0^{3/2}(u_{n+1} - u_{n-1}) - \frac{2}{5}\d_0^{5/2}.
\eeq
The left plot in Fig.~\ref{fig:focus3} shows the energy density in the system \eqref{eq:Hertz}, and the right plot displays the time evolution of the (local) energy center of mass, which is defined by
\beq\label{ectr}
X = \frac{\sum_{i = n' - m}^{n' + m}ie_i}{\sum_{i = n'-m}^{n'+m}e_i},
\eeq
with $n'$ being the location of the maximum energy density of the breather and $m>0$ an integer which accounts for the width of the breather (we set $m = 100$).

\begin{figure}
\centerline{\includegraphics[width=\textwidth]{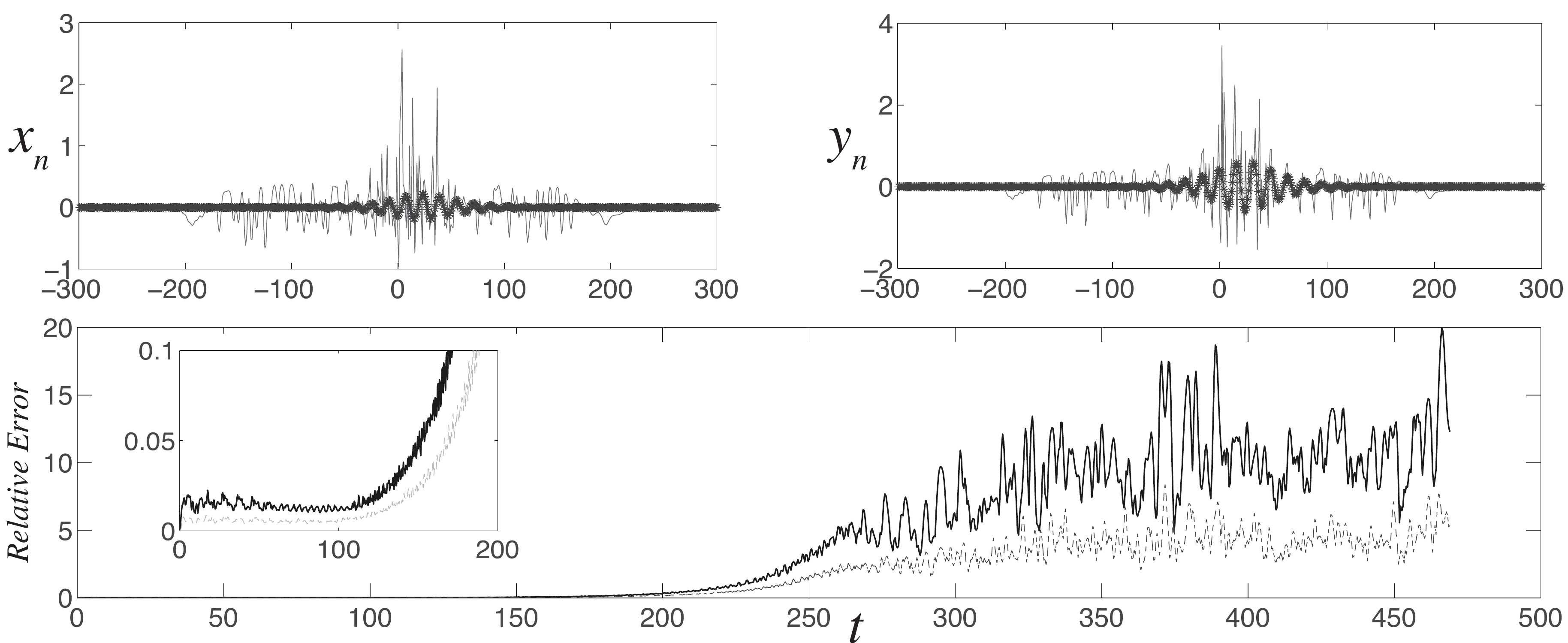}}
\caption{\footnotesize Top plots: comparison of time snapshot of approximate optical bright breather (connected stars) and numerical solution (grey curve) at $t=150\,T_b \approx 469$. Bottom plot: relative errors $E_x(t)$ (solid curve) and $E_y(t)$ (dashed curve). Inset: the relative error for $t \in[0, 200]$. Here $\theta = \pi/8$, $\o_b  = \o - 0.0001 = 2.0097$, and the other parameters are the same as in Fig.~\ref{fig:focus1}.}
\label{fig:focus4}
\end{figure}

\begin{figure}
\centerline{\includegraphics[width=\textwidth]{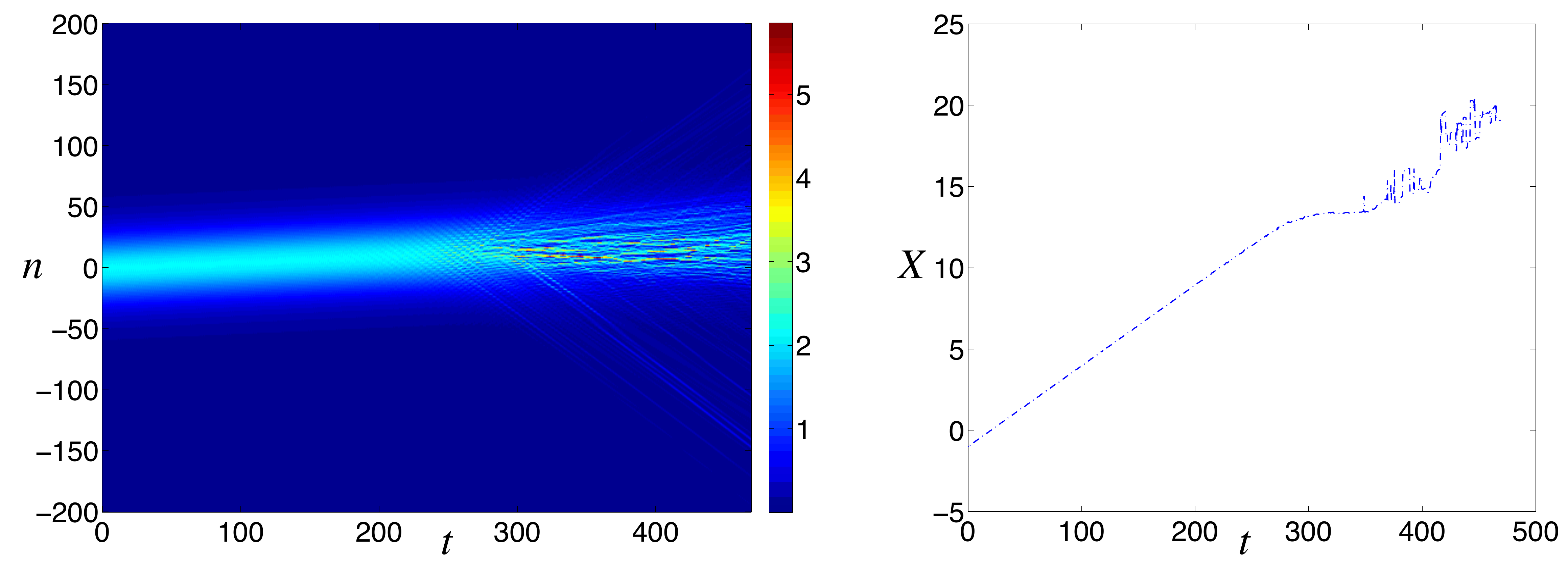}}
\caption{\footnotesize Left plot: energy density of a moving optical bright breather in system \eqref{eq:Hertz}. Right plot: time evolution of the breather's energy center. Clearly here, past a certain point in time, Eq.~(\ref{ectr})
used to identify the center incorporates the radiation emitted by the breather
and reflects the deformation of the structure illustrated in the left panel,
no longer accurately representing the breather center position.
Here $\theta = \pi/8$, $\o_b  = \o - 0.0001 = 2.0097$, and the other parameters are the same as in Fig.~\ref{fig:focus1}.}
\label{fig:focus5}
\end{figure}

In the second numerical run, we choose a smaller wave number $\theta = \pi/8$, while the other parameters remain the same. The linear frequency is now given by $\o = 2.0098$ which is fairly close to the $\theta = 0$ edge of the optical branch. As shown in Fig.~\ref{fig:disp}, the corresponding value of $|\wt h|$ decreases dramatically, so that the amplitude of the strain profiles increases. In fact, we now have $||x_n(0)||_{\infty} \approx 0.208 < \d_0$, which is of the same order as $
\d_0 = 0.444$. We perform the same numerical integration over
the time interval $[0, 150\,T_b]$, where $T_b = 2\pi/\o_b = 3.1265$ and $\o_b = \o - 0.0001= 2.0097$. As shown in the bottom plot of Fig.~\ref{fig:focus4}, the NLS approximation remains excellent at the early stage of the simulation, with relative errors less than $3\,\%$ when $t \le 150$. However, later the approximation error becomes large, and the energy spreads out towards both ends of the chain, as shown in Fig.~\ref{fig:focus5}. The ensuing waveform no longer preserves the structure of a breather.

An interesting question is whether there exist static bright breathers at the $\theta = 0$ edge of the optical branch. Recall that the order of the approximate solution \eqref{eq:focus_app} roughly depends on the magnitude of $M_1$ and $M_2$ and thus on $\wt h$ that appears in their denominators.  As shown in Fig. \ref{fig:disp}, the value of $\wt h$ becomes extremely small when $\theta$ is close to zero. Hence we need to choose $\o_b$ very close to $\o$ to ensure that the approximate solutions are within the small-amplitude regime. Meanwhile, the width $a^{-1}$
of the moving breather is $\sim |\o - \o_b|^{-1/2}$ when $\o \rightarrow \o_b$.
When $|\o - \o_b|$ is very small, we have to run simulations on an extremely long chain in order to observe the localized structure with a decaying tail. Therefore, it is numerically impractical to investigate static bright breathers as a limit of the moving ones using the focusing NLS approximation \eqref{eq:focus_app}. However, the multiscale analysis used to derive the modulation equations still holds for $\theta = 0$, yielding $\beta = h = c = 0$ so that \eqref{eq:mod1}-\eqref{eq:mod2} become the linear equations
\beq
i\partial_{\t}A_{1,1}  + \frac{1}{2}\o''\partial_{\xi}^2A_{1,1} =0 , \q \partial_{\xi}^{2}A_{1,0} = 0\label{eq:mod_theta0}
\eeq
where $\o'' = K_2\g = \k K_2/\o^3$.
Given the absence of localization at this order of the asymptotic expansion (\ref{ansatz_uv}),
we do not have any evidence of the existence of static bright breathers in this limit, contrary, e.g., to what is the case in the diatomic granular
chain case~\cite{Boechler10}.

\begin{figure}
\centerline{\includegraphics[width=\textwidth]{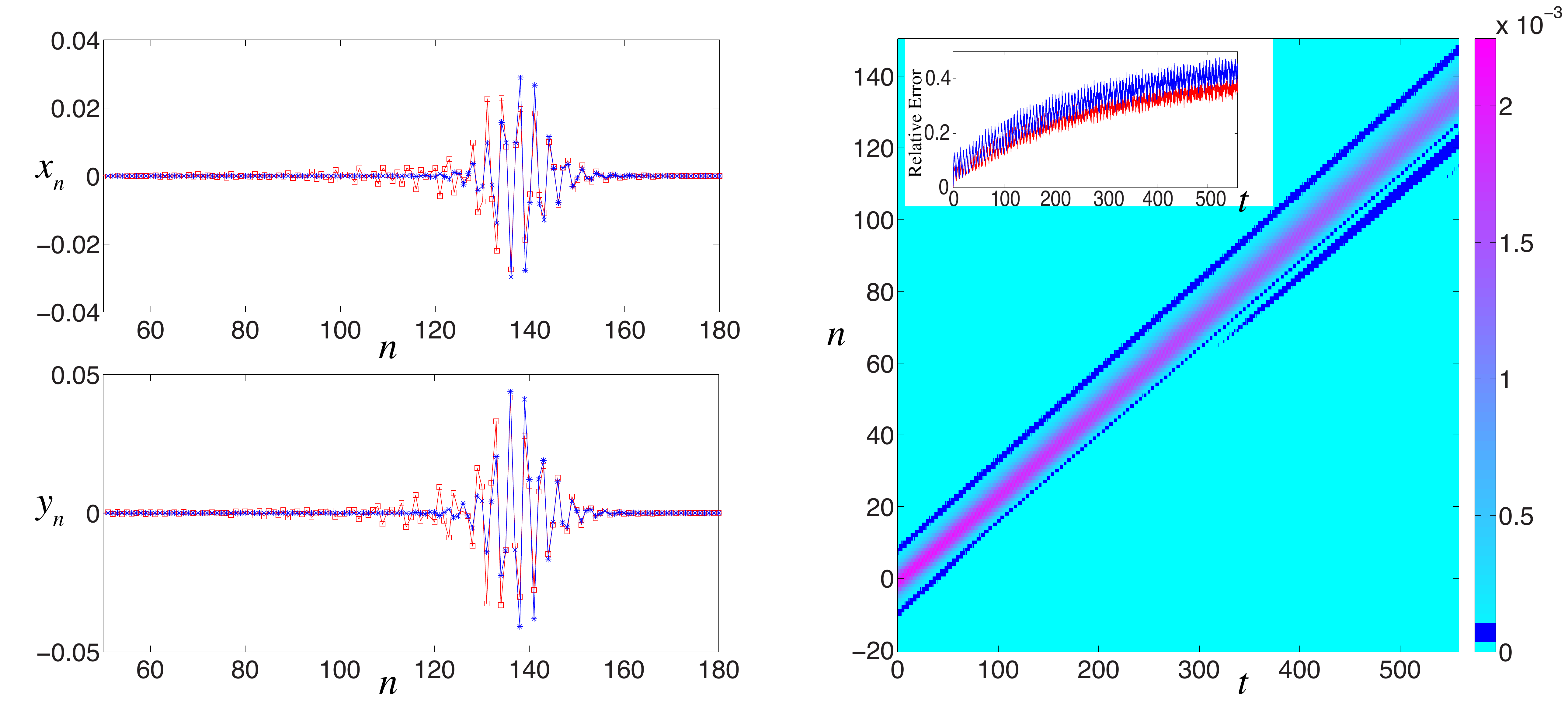}}
\caption{\footnotesize Left panels: comparison of the time snapshots of the approximate optical bright breather (connected blue stars) and the numerical evolution result (red squares) at $t=200\,T_b\approx 558.2$ for $\theta = 3\pi/5$, $\o_b  = \o - 0.0001 = 2.2512$, $N=1000$; the other parameters are the same as in Fig.~\ref{fig:focus1}. Right panels: energy density of the moving bright breather in the system \eqref{eq:Hertz}. The dark blue color marks a range of small energy densities in order to better show the growing size of the tail behind the breather. The inset depicts the relative errors $E_x(t)$ (red) and $E_y(t)$ (blue).}
\label{fig:focus6}
\end{figure}

We now consider wave numbers above $\pi/2$ but below $\theta_c$ at the same mass ratio $\rho = 1/3$. In the numerical simulation, we observe that the NLS approximation of the moving bright breathers remains excellent over a finite time interval until the wave number is close to $\theta_c \approx 1.892$, which, as we recall, marks the boundary between focusing and defocusing regions at our parameter values. To illustrate what happens below but very close to this boundary, we consider the wave number $\theta = 3\pi/5 \approx 1.885$ and perform the numerical integration over the time interval $[0, 200\,T_b]$, where $T_b = 2\pi/\o_b = 2.791$ and $\o_b = \o - 0.0001 = 2.2512$. As illustrated by the snapshots of the strain profiles of $x_n$ and $y_n$ at time $t = 200\,T_b \approx 558.2$, as well as by the space-time evolution of the energy density in Fig.~\ref{fig:focus6}, the resulting waveform mostly preserves its localized structure and moves to the right at the velocity approximately equal to $c_{+}$. However, we observe the growing trend of the relative error of the NLS approximation emerging from the very beginning of the simulation. In particular,
the approximation fails to capture the increasing size of the tail in the numerical solution, which suggests that the localized wave generated by the present initial condition
(approximation \eqref{eq:focus_app} at $t=0$) cannot be robustly sustained for long-time dynamical evolutions.
However, one can expect a better accuracy of the NLS ansatz \eqref{eq:focus_app} if $| \o_b -\o |$ is further reduced.

We next investigate bright breathers associated with the acoustic branch. From the discussion in Sec.~\ref{sec:derivation_NLS} (see also the right plot in Fig.~\ref{fig:region} and the left plots in Fig.~\ref{fig:disp}), we recall that in this case the focusing regime takes place only in a narrow interval of wave numbers for a range of small enough mass ratios. In the case $\rho=1/3$, with other parameters kept the same as before, this $\theta$-interval is $(1.7841,1.949)$.
%

For initial conditions with wave numbers $\theta$ in the lower part of the focusing $\theta$-interval,
the resulting waveform mostly preserves its localized structure over the time interval $[0, 100\,T_b]$
but we observe a growing trend of the relative error of the NLS approximation from the very beginning of the simulation.
For example, $\theta = 1.7845$ yields $T_b = 5.187$, and relative errors $E_x(t)$, $E_y(t)$ increase up to $0.1$ for $t \approx 100\,T_b$
(data not shown). Such values of $\theta$ are close to the singularity where the cubic coefficient $\wt h_{-}$ is very large
(see left bottom plot of Fig.~\ref{fig:disp}), which corresponds to a near-resonant situation with optical modes where
$\o_{+} (2\theta) \approx 2\o_{-}(\theta)$. In a neighborhood of this resonance,
the NLS approximation is expected to be less precise since it does not  account for the excitation of optical modes.

\begin{figure}
\centerline{\includegraphics[width=\textwidth]{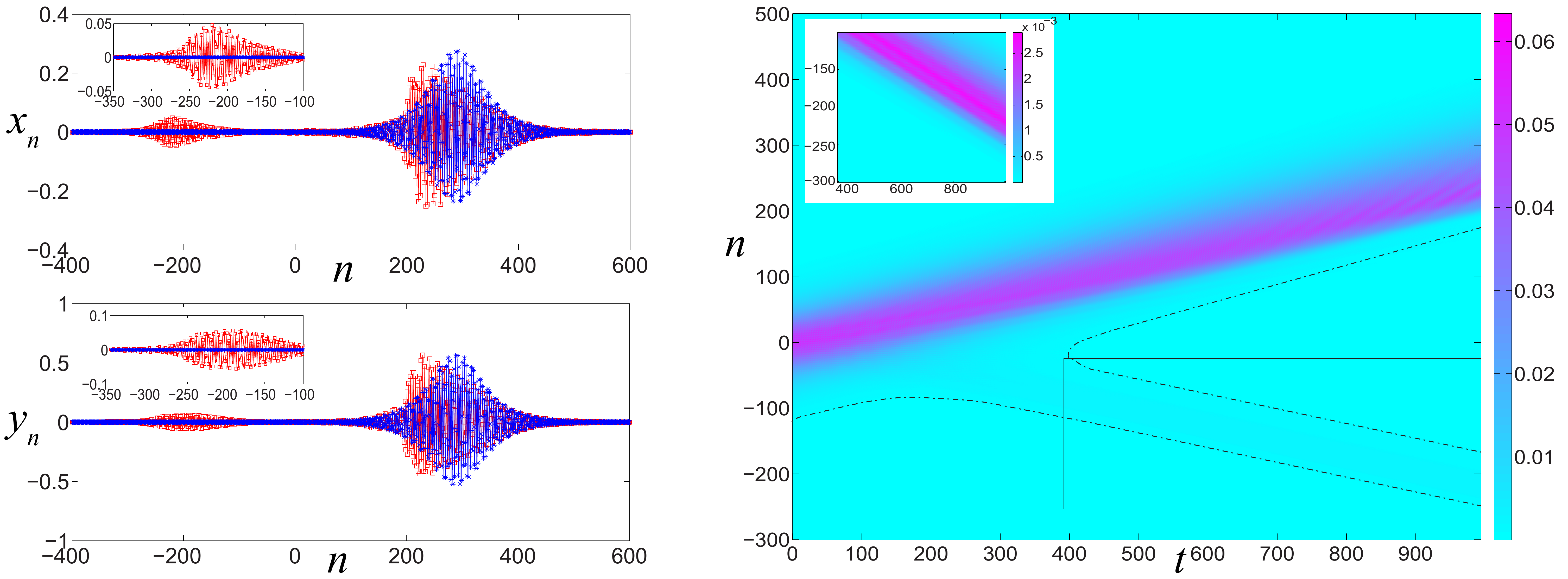}}
\caption{\footnotesize Left panels: comparison of the time snapshots of the approximate analytical solution (acoustic bright breather, connected blue stars) and the numerical evolution result (connected red squares) at $t = 200\,T_b \approx 994$ for $\theta = 1.948$, $\o_b  = \o + 0.0001 = 1.264$, $N=1000$; the other parameters are the same as in Fig.~\ref{fig:focus1}. Insets zoom in on the small-amplitude optical breather that eventually separates from the initial acoustic breather. Right panel: energy density of the numerical solution of the system \eqref{eq:Hertz} showing the energy density of the parent breather (darker color) and the small-amplitude breather detaching from it (faint lighter color bounded by the dash-dotted lines for better visibility), also shown in the inset that enlarges the region inside the rectangle.}
\label{fig:focus_acoustic2}
\end{figure}

At wave numbers near the upper bound of the focusing $\theta$-interval,
when $\wt h_{-}$ decreases and approaches zero, yielding a relatively large amplitude of $||x_n(0)||_{\infty}$, we observed another small-amplitude bright breather eventually detaching from the original waveform and moving to the left with constant velocity
near $-c_{+}$, suggesting that this second breather is an \emph{optical} one. See, for example, the results of the numerical simulation for $\theta = 1.948$ shown in Fig.~\ref{fig:focus_acoustic2}. Clearly, the NLS approximation does not capture this feature.
\begin{figure}
\centerline{\includegraphics[width=\textwidth]{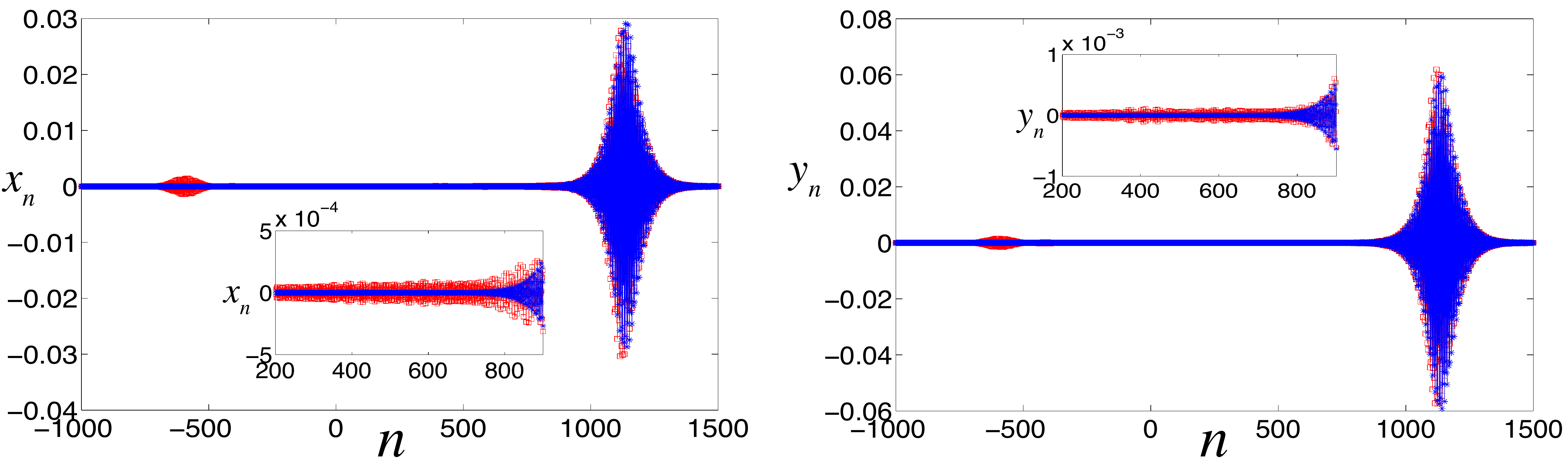}}
\caption{\footnotesize Comparison of the time snapshots of the approximate analytical solution (acoustic bright breather, connected blue stars) and the numerical evolution result (connected red squares) at $t = 700\,T_b \approx 3546.44$ for $\theta = 1.87$, $\o_b  = \o + 0.0001 = 1.2402$, $N=2000$; the other parameters are the same as in Fig.~\ref{fig:focus1}. Insets zoom in on the small-amplitude tail behind the primary acoustic breather. Similar tail forms behind the secondary optical breather propagating to the left.}
\label{fig:focus_nano}
\end{figure}

Finally, we consider $\theta=1.87$, which is in the middle of the focusing interval for the acoustic branch at $\rho=1/3$. Similar to the result shown in Fig.~\ref{fig:focus_acoustic2}, a small-amplitude optical breather eventually detaches from the parent breather and propagates to the left, although in this case the amplitudes of both parent and secondary breathers are much smaller, and the parent breather deviates significantly less from the initial NLS ansatz and robustly propagates through the lattice; see Fig.~\ref{fig:focus_nano}. Interestingly, we also observe small-amplitude oscillations emitted by both primary and secondary breathers, suggesting the existence of \emph{nanoptera} (bright breathers with small-amplitude oscillatory tails) in the lattice under consideration. To further investigate this phenomenon, we performed another simulation at $\theta=1.87$, where we solved \eqref{eq:Hertz} with emitting time-dependent boundary conditions obtained by evaluating the initial NLS ansatz at the ends of the computational domain. In this case no secondary breather was nucleated but robust propagation of an acoustic nanopteron was again observed.

\subsection{\label{mblm}Moving bright breathers for large mass ratio}

In the acoustic case, the NLS equation is defocusing at large enough $\rho$ and thus it does not admit
breather solutions. In what follows, we
investigate the effect of larger mass ratios on the optical bright breathers and their NLS approximation.\\


We first use the same approach as in Sec.~\ref{mmr}, integrating (\ref{eq:Hertzd0S})
from the initial conditions given in Sec.~\ref{aabp}.
We start with the mass ratio $\r = 10$, at which the focusing regime corresponds to two disjoint $\theta$-intervals
$(0, 0.3681)$ and $(1.2263, 2.4389)$.

At wave numbers in the upper part of each interval, we found that NLS provides an excellent approximation of the corresponding small-amplitude moving bright breather over a finite time.
Note that the divergence of $\tilde{h}$ for $\theta \approx 0.3681$ and $\theta \approx 2.4389$
corresponds to the case $c^2(1 + \r) - K_2 \approx 0$, where the group velocity of the carrier wave becomes close to the
maximal group velocity of acoustic modes. This phenomenon does not seem to affect the quality of the NLS approximation.

\begin{figure}[htp]
\centerline{\includegraphics[width=\textwidth]{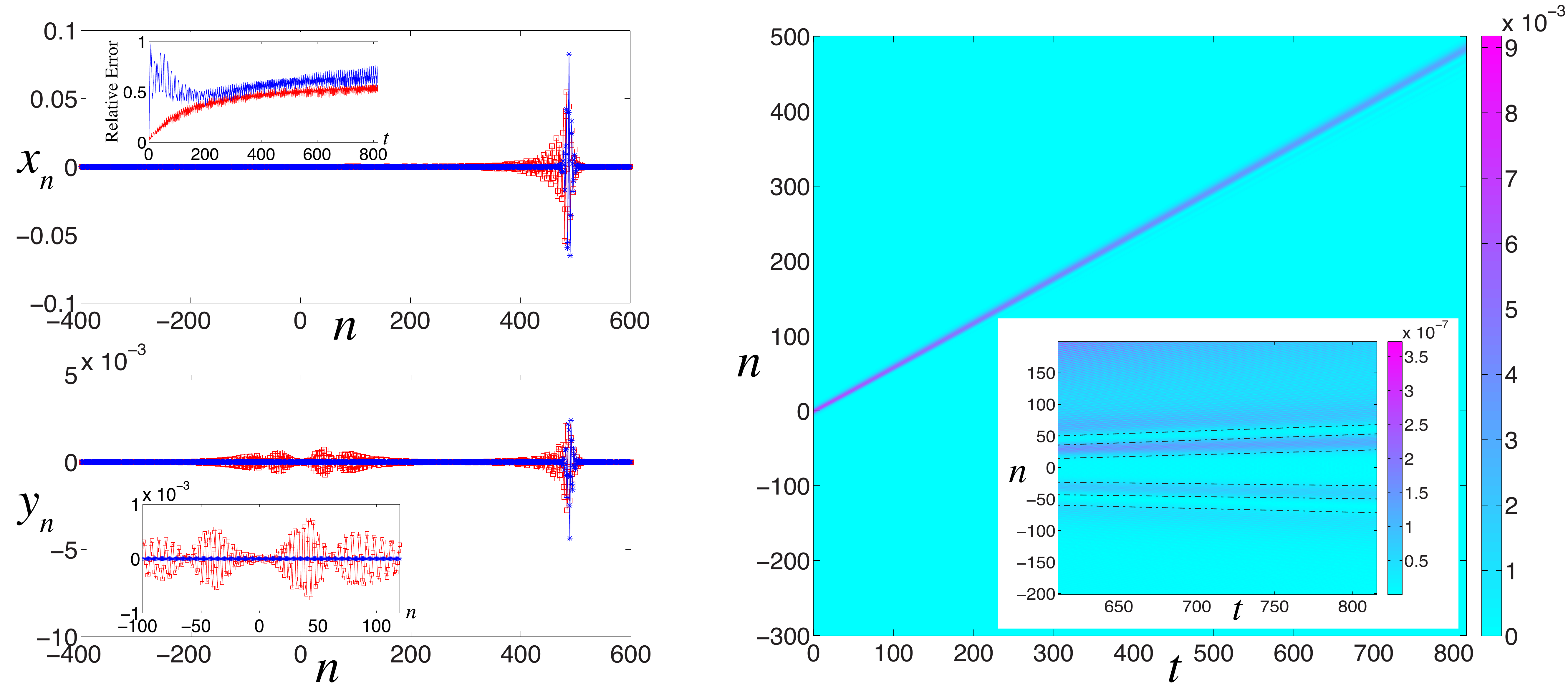}}
\caption{\footnotesize Left panels: comparison of the time snapshots of the approximate analytical solution (optical bright breather, connected blue stars) and the numerical evolution result (connected red squares) at $t = 200\,T_b \approx 815.28$ for $\rho=10$, $\theta = 1.23$, $\o_b  = \o + 0.0001 = 1.5414$, $N=1000$; the other parameters are the same as in Fig.~\ref{fig:focus1}. Relative errors $E_x(t)$ (red) and $E_y(t)$ (blue) are shown in the top left inset. Bottom left inset zooms in on the train
of acoustic waveforms that eventually separate from the initial optical breather. Right panel: energy density of the numerical solution of the system \eqref{eq:Hertz}. The inset zooms in on the train of acoustic breather-like structures (not visible in the main energy plot), with dash-dotted lines added for better visibility.}
\label{fig:focus_opt_r10}
\end{figure}

We observe that the accuracy of the NLS approximation deteriorates near the lower bounds of the two $\theta$-intervals.
Similarly to the results shown in Fig.~\ref{fig:focus4} and Fig.~\ref{fig:focus5} for the case $\r = 1/3$,
$\tilde{h}$ becomes small for $\theta \approx 0$ and the localized structure eventually breaks down.
At wave numbers near the lower bound of the $\theta$-interval $(1.2263, 2.4389)$, one has
$\o_{+}'' \approx 0$ and we observe not only the increasing size of the tail in the numerical solution, similar to the example shown in Fig.~\ref{fig:focus6} for the case $\r = 1/3$, but also the emergence of a train of bright-breather-like structures clearly visible in the $y_n$ variable that separate from the initial breather and slowly move in both directions with speed approximately equal to $c_{-}$, suggesting their acoustic nature. Meanwhile, the optical breather initiated by the NLS approximation propagates to the right with velocity close to $c_{+}$. See, for example, Fig.~\ref{fig:focus_opt_r10}, where the wave number is $\theta = 1.23$, and we have $c_{-} \approx 0.0754$ and $c_{+} \approx 0.6$.\\
%




To complete the above results,
we now illustrate the spontaneous formation of moving
bright breathers without resorting to well-prepared localized initial conditions.
A key mechanism for the formation of moving or static breathers in nonlinear lattices
is the modulational instability of periodic waves; see \cite{Flach08} for a review and
\cite{Boechler10,GJ11,JKC13,hasan15,LJKV15} for specific studies concerning granular chains.
In the weakly nonlinear regime, this phenomenon can be understood from the focusing NLS equation
($\omega^{\prime\prime}\, \tilde{h} <0$). In the focusing case,
the spatially homogeneous solutions $A_{\theta , r}(\tau )=r\, e^{-i\, \tilde{h}\, r^2 \tau}$ of (\ref{eq:NLS})
correspond formally to unstable periodic traveling waves of \eqref{eq:Hertz},
which self-localize under long-wave perturbations.
An interesting approach to explore modulational instabilities consists of
driving a bead with a particular modulationally unstable frequency (see e.g. the experiments in \cite{Boechler10}).
This approach is of practical relevance because
it is not straightforward to initialize desired periodic wave profiles throughout a granular chain
in experiments.

\begin{figure}[h]
\begin{center}
\includegraphics[scale=0.37]{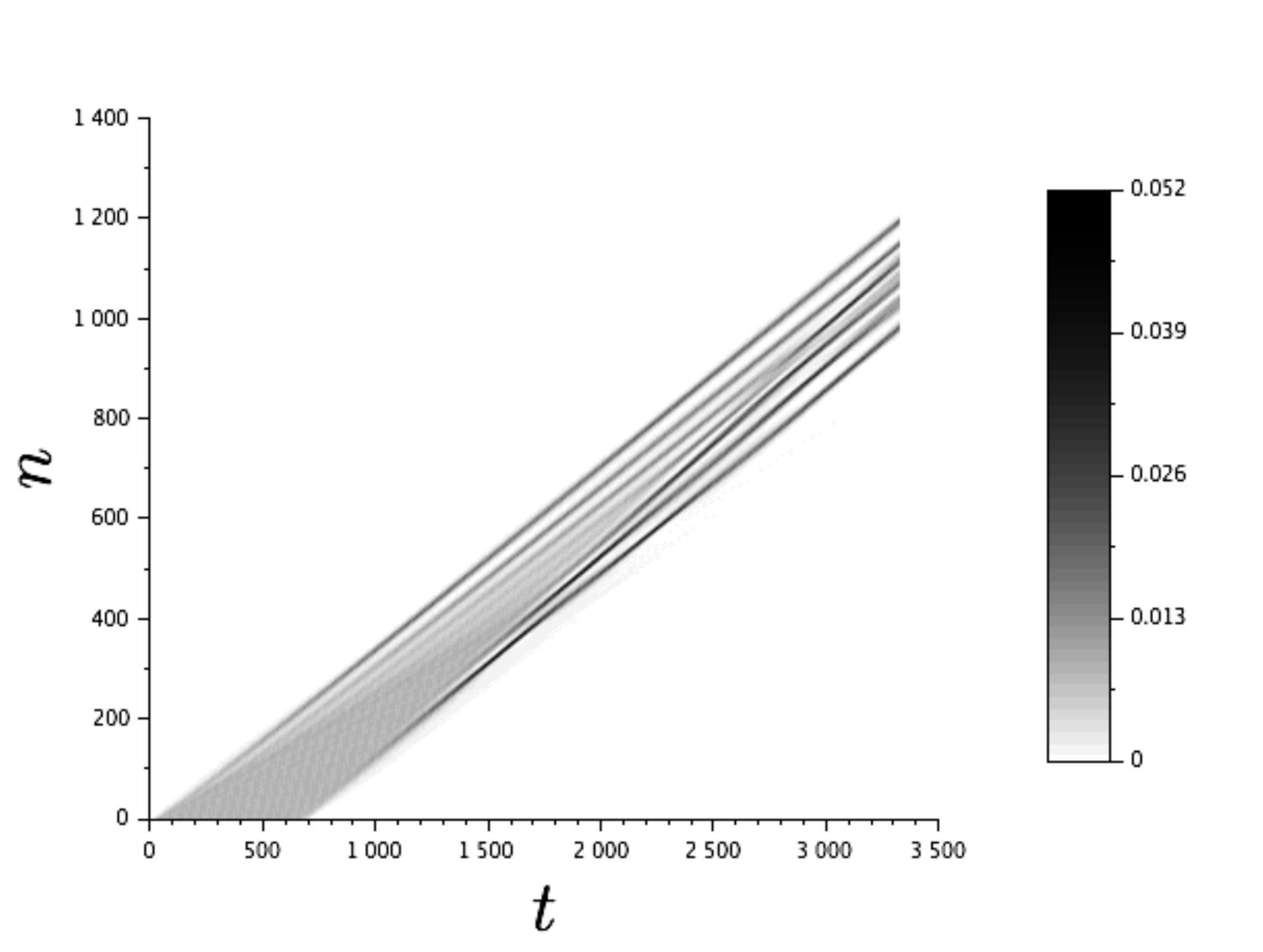}
\includegraphics[scale=0.37]{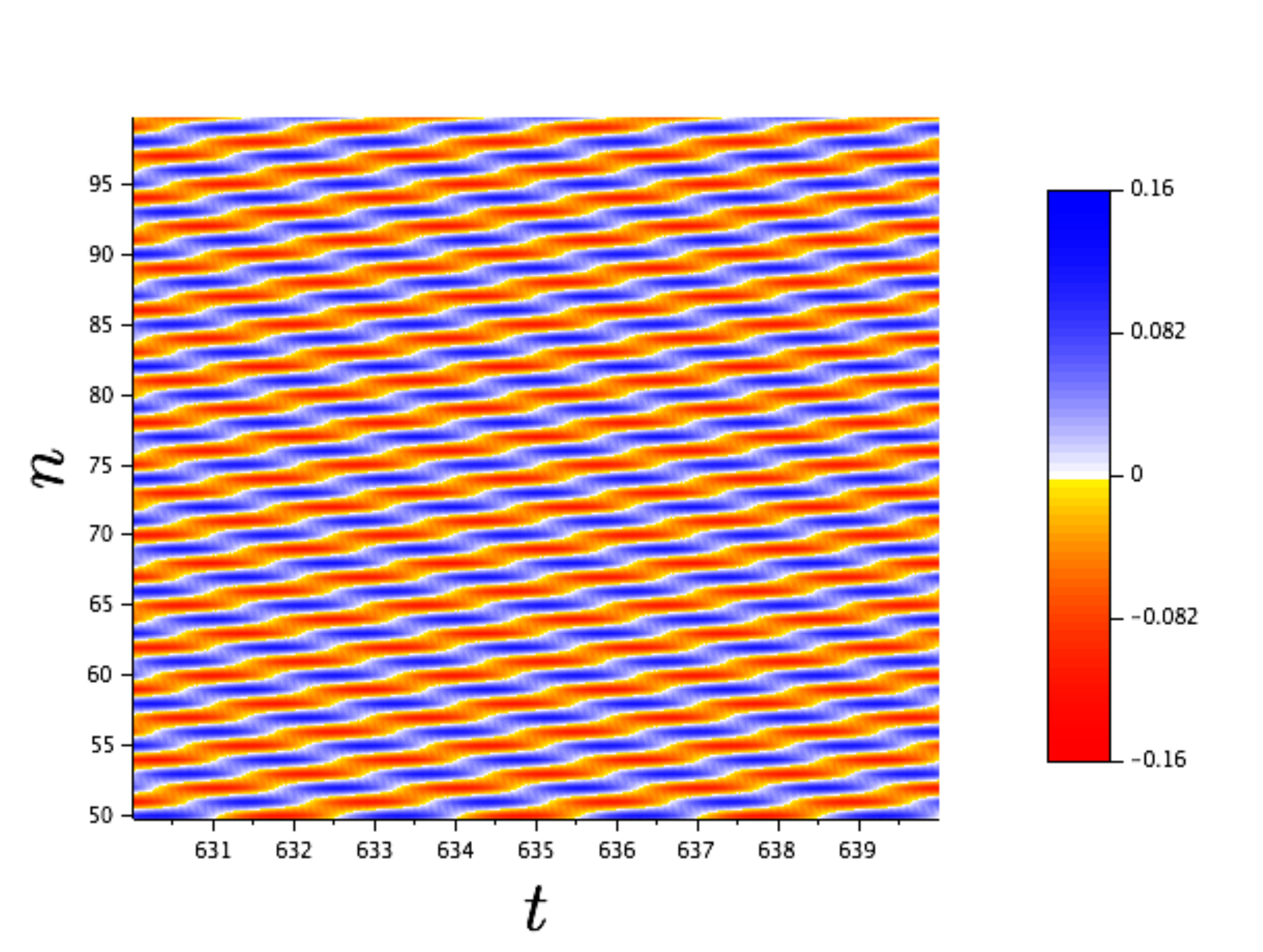}
\end{center}
\caption{\footnotesize
Space time diagrams illustrating the
system response to a periodic sinusoidal excitation applied at $n=0$
during some finite time interval (see text).
Left panel: energy density $e_n (t)$ defined by (\ref{edensity}).
Right panel: bead velocities $\dot{u}_n(t)$ for a space-time
zoom-in of the left panel.}
\label{fig:spatio}
\end{figure}

\begin{figure}[h]
\begin{center}
\includegraphics[scale=0.38]{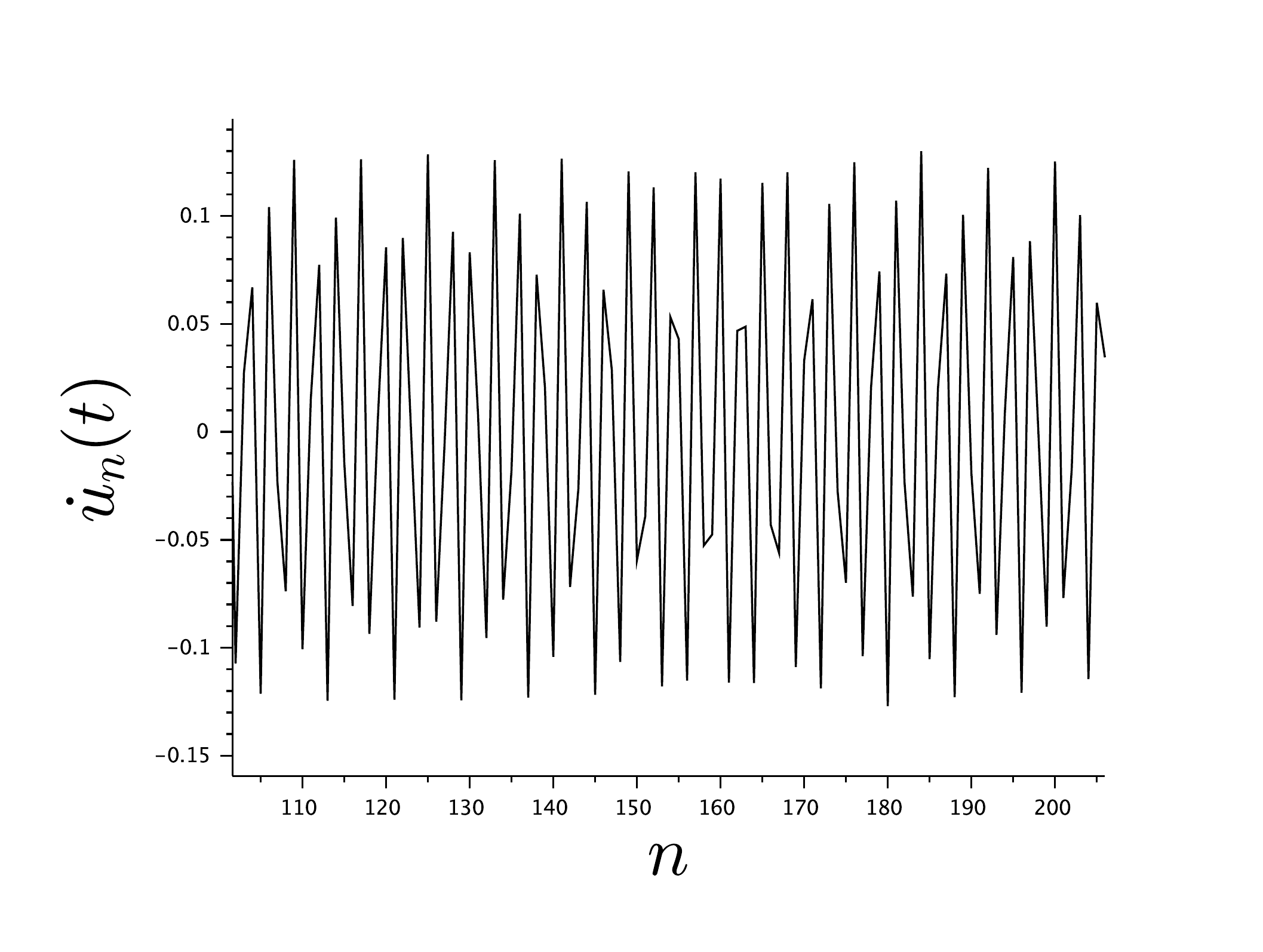}
\includegraphics[scale=0.37]{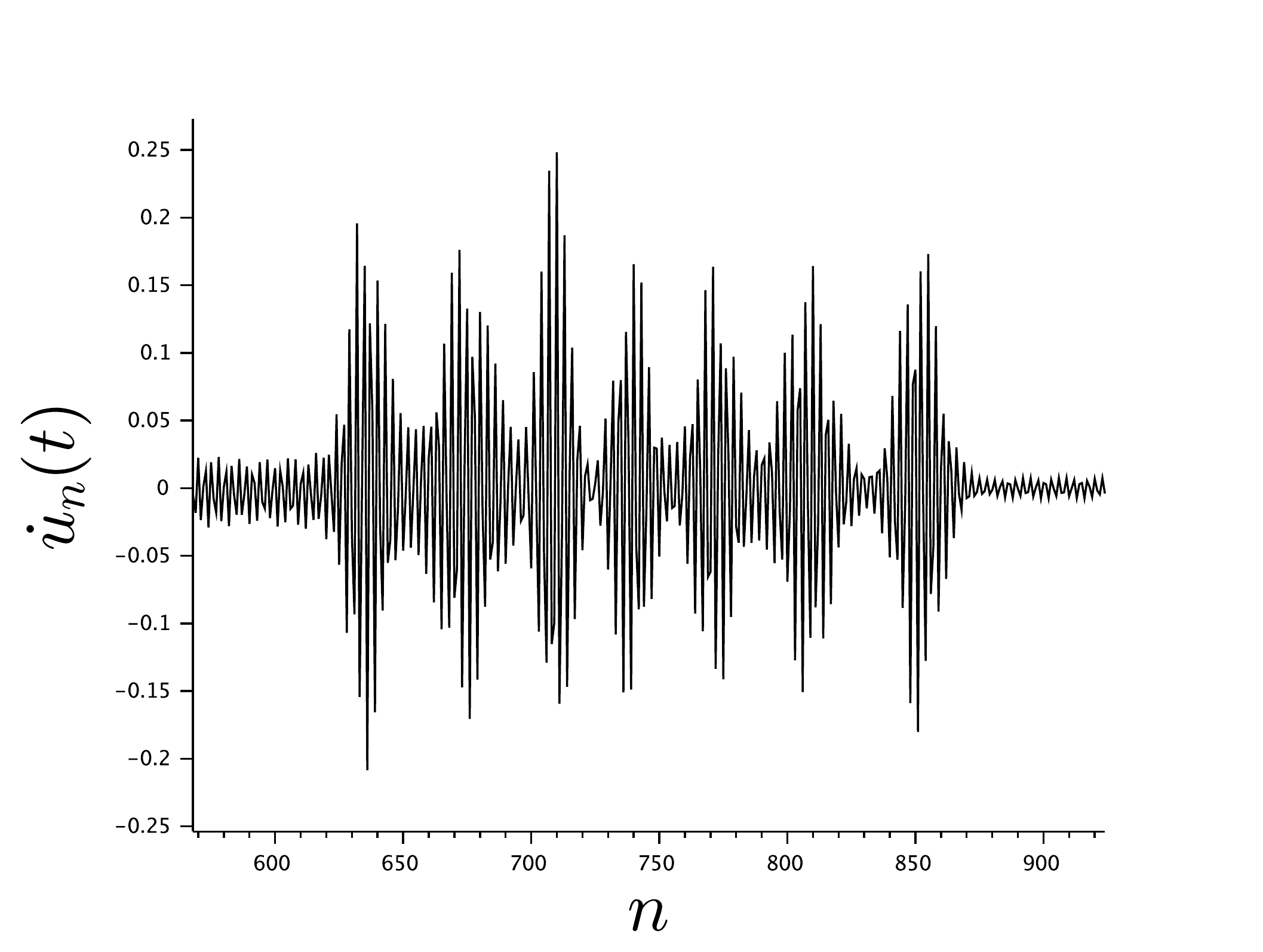}
\end{center}
\caption{\footnotesize
Snapshots of bead velocities $\dot{u}_n(t)$
at $t=779.9$ (left panel) and $t=2400$ (right panel)
in different parts of the chain, from the same simulation as Fig.~\ref{fig:spatio}.
}
\label{fig:spatio:stab1}
\end{figure}

In what follows, we illustrate the excitation of optical bright breathers for $\r = 10$
using the above approach.
We consider a chain of $2002$ elements, sufficiently long to ensure
that waves reflected from the ends remain negligible in the regions of interest
within the time frame of simulation. The chain is at rest at $t=0$.
We clamp the last bead and
impose a sinusoidal motion of the first bead at a frequency
within the focusing region of the optical band (see Fig.~\ref{fig:region}).
More precisely, we simulate equation
\eqref{eq:Hertz} with $n \in \{ 1,\ldots , 2000 \}$
and boundary conditions $u_{2001}(t)=0$,
$u_0 (t)=A_0\, \sin{(\omega_0\, t)}\, \chi (t)$ for $t \geq 0$,
where $\omega_0 = 2.1 = \omega_{+}(\theta )$ ($\theta \approx 2.34$) and
$\chi$ denotes a smooth plateau function, which is slowly varying compared to
the driving period $T_0 = 2 \pi / \omega_0 \approx 2.99$.
The envelope
$\chi (t)$ increases smoothly from $0$ to (near)-unity for $t \in [0, 30\, T_0]$,
is almost equal to unity for $t \in [30\, T_0, 200\, T_0]$,
decreases smoothly to $0$ for $t \in [200\, T_0, 260\, T_0]$
and almost vanishes for larger times.
This is achieved by fixing
$\chi (t) = \text{tanh}[r\, t]\, (1-\text{tanh}[r\, (t-t_{\text{m}})])/2$
with $r = 0.046$ and $t_{\text{m}}=700$.
We impose a moderate driving amplitude $A_0 = 0.06 = 0.135 \times \delta_0$.
Figures \ref{fig:spatio} and \ref{fig:spatio:stab1} describe the system response
to the above boundary excitation.

At the early stage of the simulation,
a periodic traveling wave close to the optical mode with frequency $\omega_0$
is established. This wave corresponds to the grey region in the left panel
of Fig.~\ref{fig:spatio} which displays the energy density $e_n$ in the chain.
This wave pattern is initially mainly confined between $n=0$ and $n=c_+\, t$,
where $c_+=\omega_+^\prime (\theta)$ is the group velocity of the carrier wave.
For $t \geq t_\text{m}$, the support of the traveling wave pattern stabilizes to roughly
$200$ lattice sites and is mainly transported with the group velocity $c_+$.
The right panel of Fig.~\ref{fig:spatio} shows the detail of the traveling wave pattern,
and a snapshot of bead velocities at a fixed time is shown in the left panel of Fig.~\ref{fig:spatio:stab1}.

In the second stage, the periodic traveling wave destabilizes leading to
a traveling multibreather state (i.e. a train of closely spaced traveling breathers).
The formation and propagation of these localized structures is visible in the
energy density plot of Fig.~\ref{fig:spatio} (left panel).
Breather profiles are shown in the right panel of Fig.~\ref{fig:spatio:stab1}
which provides a snapshot of bead velocities at a fixed time.

\section{\label{sec:defocusing} Dark breather solutions in the granular chain}

\subsection{\label{dedb}Dynamical excitation of dark breathers}

In this section, we show that unstable periodic traveling waves do not
necessarily lead to breather solutions (in contrast with the example of Figures \ref{fig:spatio}-\ref{fig:spatio:stab1})
and can generate in some cases long-lived dark breather solutions.
For this purpose, we simulate equation \eqref{eq:Hertz} with periodic
boundary conditions $u_{n+N}=u_{n}$, $v_{n+N}=v_{n}$ and $N=401$.
As previously we choose $\alpha = 3/2$, $\kappa =1$, and we now fix $\rho=0.32$, $\delta_0 = 0.27$.

\begin{figure}[h]
\begin{center}
\includegraphics[scale=0.34]{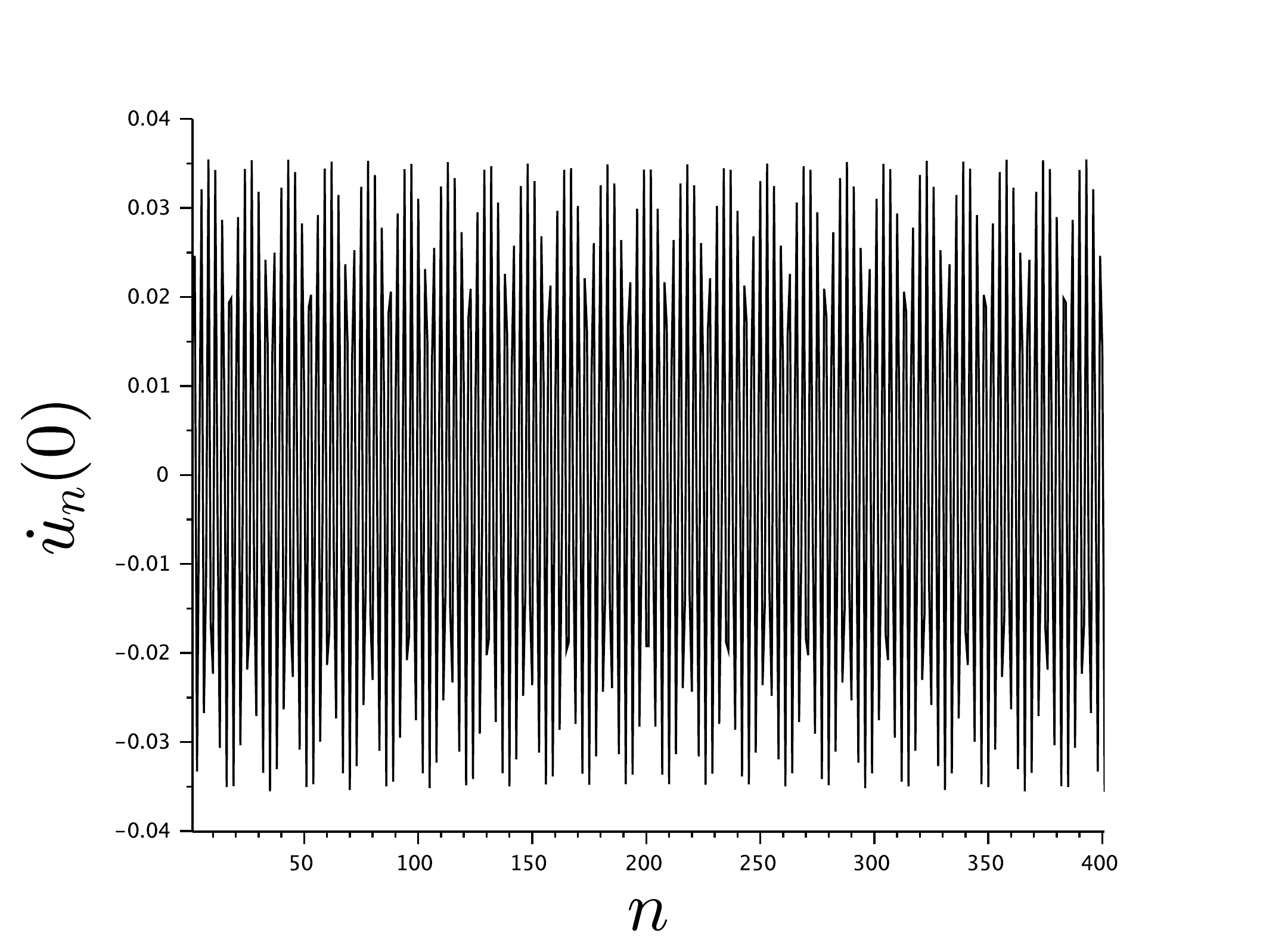}
\includegraphics[scale=0.34]{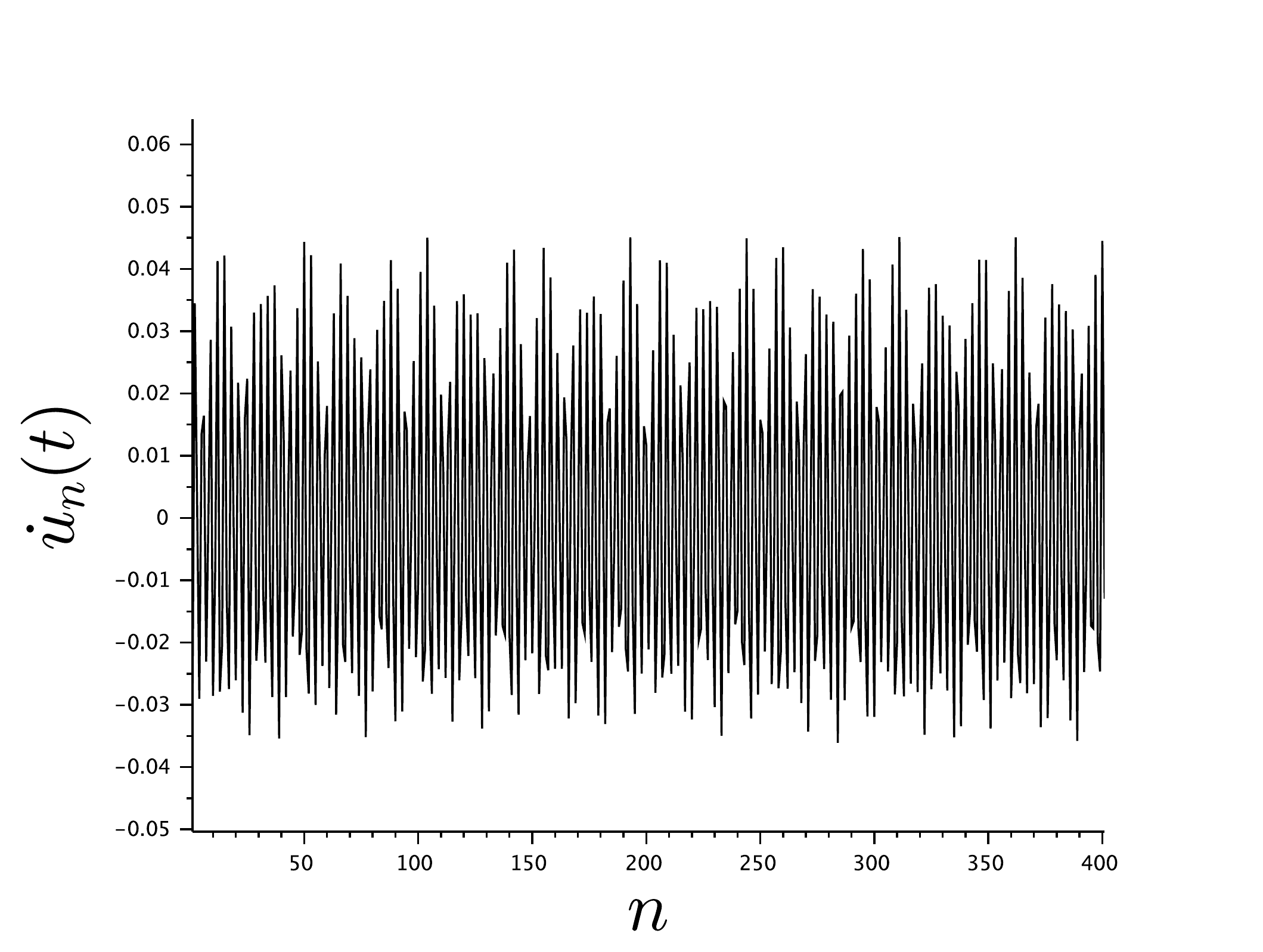}
\includegraphics[scale=0.34]{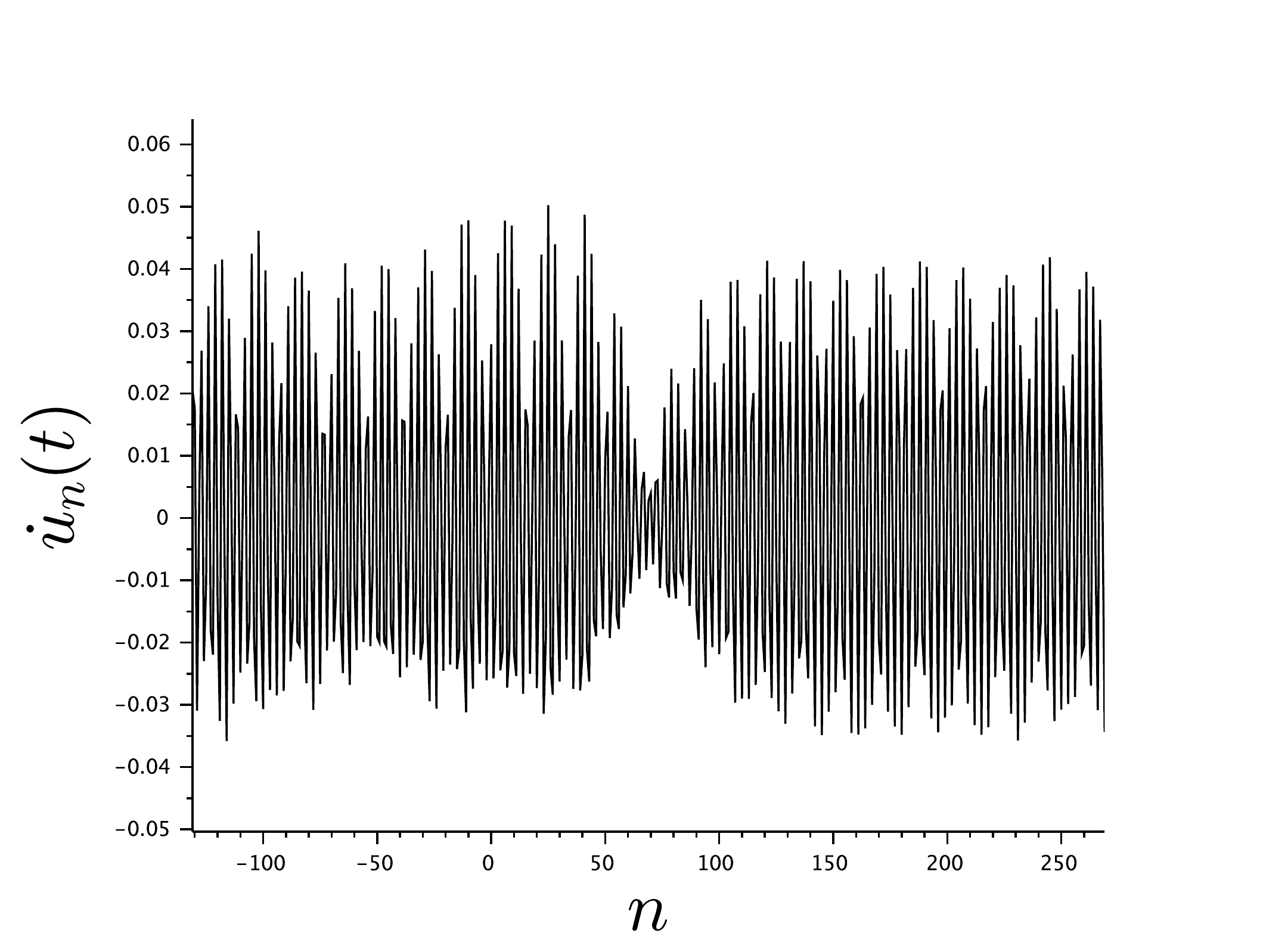}
\includegraphics[scale=0.34]{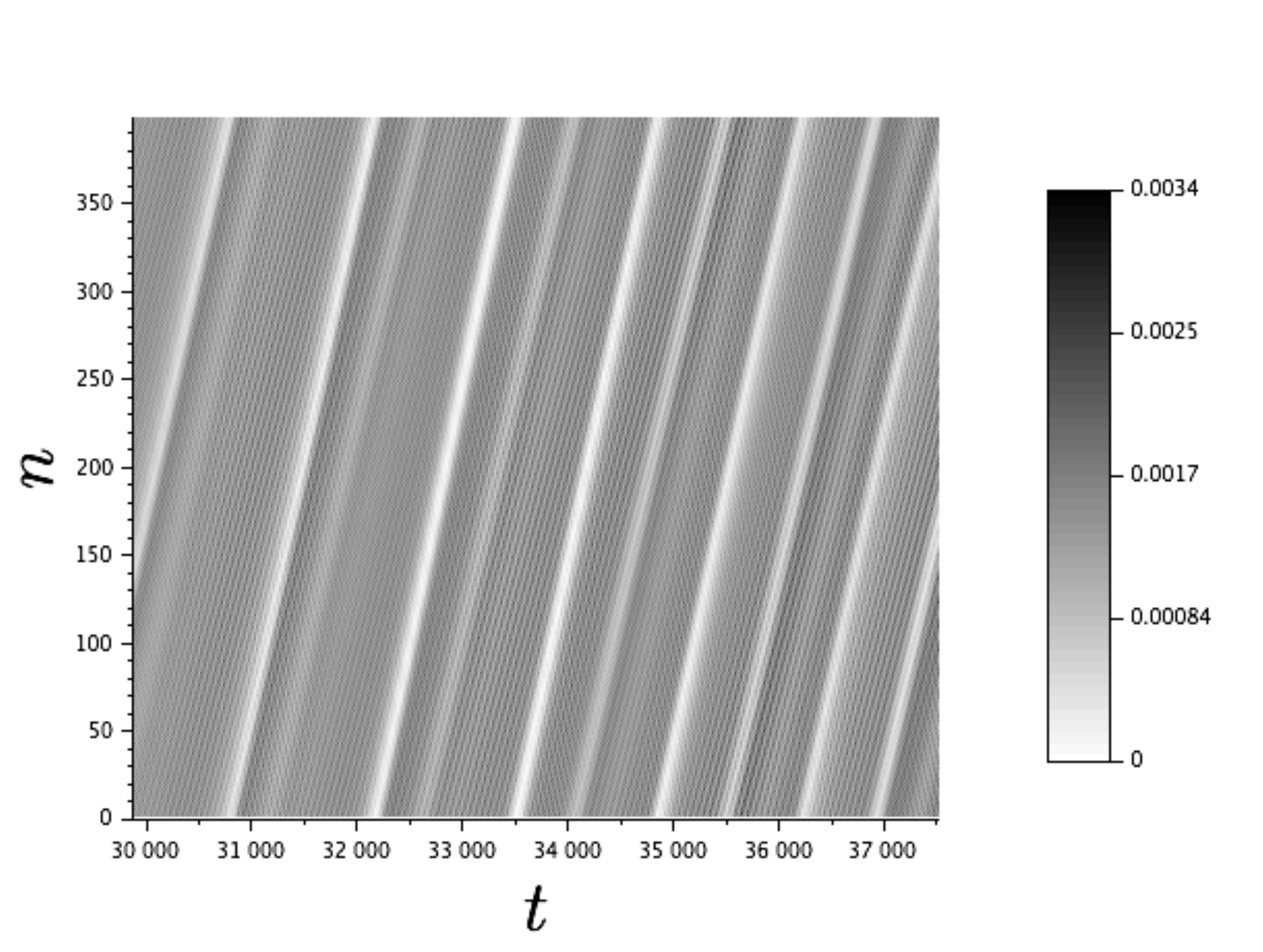}
\end{center}
\caption{\label{midarkb}
\footnotesize Evolution of the initial condition (\ref{cimi})
in system \eqref{eq:Hertz} with periodic boundary
conditions ($N=401$ particles).
Particle velocities $\dot{u}_n(t)$ are displayed at $t=0$
(top left panel), $t=17100$ (top right panel) and
$t=32400$ (bottom left panel).
The top right panel illustrates a short-wavelength oscillatory instability.
As shown in the bottom left panel, a traveling dark breather is generated at a later stage
(the profile of $\dot{v}_n(t)$ is qualitatively similar to $\dot{u}_n(t)$ shown in the figure).
The bottom right panel shows the
energy density $e_n(t)$ in grey levels (see definition (\ref{edensity})).}
\end{figure}

We integrate equation \eqref{eq:Hertz} numerically for the initial condition
\begin{eqnarray}
\label{cimi}
\nonumber
u_n(0)&=&\epsilon\, \sin{(\theta_0 n)}\, (1+b\, \cos{(2n\pi /N)}), \\
\nonumber
\dot{u}_n(0)&=&-\epsilon\, \omega\, \cos{(\theta_0 n)}\, (1+b\, \cos{(2n\pi /N)}), \\
v_n(0)&=&\frac{\kappa}{\kappa - \rho \omega^2}\, u_n(0), \\
\nonumber
\dot{v}_n(0)&=&\frac{\kappa}{\kappa - \rho \omega^2}\, \dot{u}_n(0)
\end{eqnarray}
corresponding to a slowly modulated acoustic mode, with amplitude
$\epsilon =0.03$, a wavenumber $\theta_0 = 252 \pi /N \approx 1.97$ in the band of unstable acoustic modes,
$\omega = \omega_{-}(\theta_0) \approx 1.17$ and $b=0.01$.
The initial velocity profile is shown in the top left panel of Fig.~\ref{midarkb}.

The time-evolution of the above initial condition is described in Fig.~\ref{midarkb}.
System \eqref{eq:Hertz}
generates a weakly modulated periodic traveling wave over a long transient
$t \in [0, 16300]$ (roughly $3000$ periods of the acoustic mode).
An oscillatory instability starts to develop around $t=16300$,
with a characteristic wavelength much shorter than
the initial modulation (the resulting perturbation of the
traveling wave is illustrated in the top right panel for $t=17100$).
For larger times around $t=30000$, one observes
the formation of a traveling dark breather
(the corresponding velocity profile at $t=32400$ is shown in the bottom left panel).
In Fig.~\ref{midarkb}, the space-time diagram
(bottom right panel) displays the energy density $e_n (t)$ in the chain.
The white line corresponds to the center of the traveling dark breather, where the energy
density decreases significantly. The dark breather propagates at a velocity $c_{\text{db}} \approx 0.3$
close to $\omega_{-}^\prime (\theta_0)$. One can notice that
a second dark breather appears to progressively detach from the
first one for $t > 33000$.
For $t \geq 43000$, dark breathers disappear and a turbulent regime
ensues.

Clearly, the focusing NLS equation does not capture the above dynamical features since it does not admit
dark breather solutions. As discussed in Sec.~\ref{focdef}, the band of focusing wavenumbers
is narrow for acoustic modes (it corresponds to $\theta \in (1.888,2.003)$ for the above parameter values)
and the range of validity of the focusing NLS equation is questionable (see Sec.~\ref{mmr}).
In contrast, as we shall see in Sections \ref{adb} and \ref{numdb}, dark breather solutions of
the defocusing NLS equation can be used to approximate (moving or stationary) dark breather solutions of the original lattice.
Near the limit of vanishing amplitude, dark breathers propagate at a velocity close to the group velocity of the carrier wave.
Near the lower edge of the defocusing band at $\theta \approx 2.003$,
the group velocity is close to $0.29$, and thus small-amplitude dark breathers propagate at velocities
compatible with the results of Fig.~\ref{midarkb} (bottom right panel).
Along this line, the excitation of a dark breather from the initial condition (\ref{cimi})
may be linked with the fact that the second harmonic with wavenumber
$\theta = - 2 \theta_0 + 2\pi \approx 2.3$ falls within the defocusing band.

\subsection{\label{adb}Approximate dark breather solutions}

We now turn our attention to the defocusing case of the NLS equation
($\o''\wt h >0$) and compute small-amplitude approximate dark breather solutions of
the original lattice (\ref{eq:Hertzd0S}).
Using \eqref{ansatz_uv}, \eqref{tanhA11} and \eqref{tanhA10}, we obtain
approximate solutions in terms of strain variables given by
\beq
\begin{split}
x_{n}^{A}(t) = \d^{-} u_n^{A}(t) &= \e\d^-\{M_3 + 2M_4\cos{[n\theta - \wt \o_{b}t + \beta F(\e^2t)]}\}\text{tanh}[\eta\e(n - n_0 - ct)] \\
&+\e^2\left\{f(\e^2t) - \frac{\l\m}{\wt h}\right\},\\
y_{n}^{A}(t) = \d^{-} v_n^{A}(t) &= \e\d^-\{M_3 + \frac{2\k M_4}{\k-\r\o^2}\cos{[n\theta - \wt \o_{b}t+\beta F(\e^2t)]}\}\text{tanh}[\eta\e(n - n_0 - ct)]\\
&+\e^2\left\{f(\e^2t)  - \frac{\l\m}{\wt h}\right\},
\label{eq:defocus_app}
\end{split}
\eeq
where we define
\beqs
M_3 = -\frac{\l\sqrt{-\o''\m}}{|\wt h|}, \q M_4 = \sqrt{\frac{-\m}{\wt h}}, \q  \text{and} \q \eta = \sqrt{- \frac{\m}{\o''}}.
\eeqs
Here $n_0$ corresponds to the spatial translation and $\wt\o_b = \o - \m\e^2$, where $\m$ is any real number such that $\m\o''<0$, $\m \tilde{h}<0$.
We recall that $c = \omega^\prime_{\pm}(\theta)$ is the group velocity of the carrier wave and
$f(\t)$ in the $O(\e^2)$ term is undetermined.

\vspace{1ex}

The special case $c=0$ of (\ref{eq:defocus_app}) corresponds to stationary dark breathers
consisting of time-periodic and spatially modulated standing waves.
This case occurs when the wave number is on the $\theta = \pi$ edges of the optical or acoustic branches, so that the group velocity $c$ vanishes
(the $\theta = 0$ edge of the optical branch where $\tilde{h}= 0$ cannot be included).
This yields in (\ref{eq:ppOmega}), (\ref{eq:lambda}) and (\ref{eq:h_tilde})
\beq
\beta =-4K_3\g , \q \wt h = (24K_4-32K_3^2/K_2)\g, \q \o'' = -K_2 \g, \q \l = -\frac{8K_3}{K_2}.
\eeq
Then $\o''\wt h = -(24K_2K_4 - 32K_3^2)\g^2 > 0$, which corresponds to the defocusing case. The same condition was obtained,
e.g., in \cite{James01,Chong13}; see also the discussion therein.
Since $f(\t)$ is an arbitrary function independent of $\xi$, we set
\beq
f(\e^2t) \equiv \frac{\l\m}{\wt h}
\eeq
to eliminate the $\e^2$-term in \eqref{eq:defocus_app}. The antiderivative of $f(\t)$ is now given by
$F(\t) = {\l\m \t}/{\wt h}$, which leads to
\beq\label{eq:Ftau}
\beta F(\t) = -\frac{2\m\t}{3}
\eeq
after evaluating $K_2 , K_3, K_4$ for $\alpha = 3/2$ (see (\ref{expv})).
Substituting \eqref{eq:Ftau} into the cosine terms of \eqref{eq:defocus_app} yields
\beqs
\cos{[n\theta - \wt\o_b t + \beta F(\e^2t)]} =\cos{(\o - \frac{1}{3}\mu\e^2) t}
\eeqs
and the leading order NLS approximation \eqref{eq:defocus_app} is now given by
\beq
\begin{split}
&x_{n}^{A}(t) \approx \d^-\left\{-\frac{2}{\sqrt{3}}+2(-1)^n\cos{(\o_{b}t)}\right\}\sqrt{\frac{3(\o_b - \o)}{\wt h}}\text{tanh}\left[\sqrt{\frac{3(\o_b - \o)}{\o''}}(n-n_0)\right] \\
&y_{n}^{A}(t) \approx \d^-\left\{-\frac{2}{\sqrt{3}} +\frac{2(-1)^n\k}{\k-\r\o^2}\cos{(\o_{b}t)}\right\}\sqrt{\frac{3(\o_b - \o)}{\wt h}}\text{tanh}\left[\sqrt{\frac{3(\o_b - \o)}{\o''}}(n-n_0)\right],
\label{eq:xy_app}
\end{split}
\eeq
where we have converted $\mu$ and $\e$ into a new single parameter $\o_b = \o - \frac{1}{3}\m\e^2$ measuring the breather frequency. Choosing $n_0 = 0$ in \eqref{eq:xy_app} results in a \emph{site-centered} solution, whereas the \emph{bond-centered} solution corresponds to $n_0 = 1/2$.
Notice that although the continuum
envelope approximation developed here permits an arbitrary selection
of $n_0$, discrete models typically only support such site- and bond-centered
solutions with the corresponding selection of $n_0$ discussed above;
see e.g.~\cite{Flach08}. The energetic differences between these
two solutions and the non-existence of solutions centered at other
values of $n_0$ (due to the absence of translational invariance in the
discrete problem) can only be detected beyond all algebraic
orders~\cite{hwang11}.

When $\o_b$ is close enough to $\o$, the ansatz \eqref{eq:xy_app} with $t = 0$ can be used as an initial seed for a Newton-type iteration to compute the numerically exact stationary
dark breather solutions of the discrete system \eqref{eq:Hertzd0S} of both
site-centered and bond-centered type.
This computation will be performed in Sec.~\ref{numdb}.
In Sec.~\ref{stabdb}, the stability properties of stationary
dark breathers will be analyzed, and traveling dark breathers
with $c\neq 0$ will be generated from unstable stationary dark breathers.

\subsection{\label{numdb}Numerical continuation of stationary dark breathers}\label{sec:db_exact}

We now use a Newton-type algorithm (Algorithm 2 in \cite{Aubry97}) with the initial seed \eqref{eq:xy_app} to compute numerically exact
stationary dark breather solutions of system \eqref{eq:Hertzd0S}. Computations are performed with periodic boundary conditions.
In what follows we set $\a=3/2$, $\k = 1$ and $\d_0 = 4/9$.

Let $x(t)$, $y(t)$, $\dot x(t)$ and $\dot y(t)$ denote the row vectors with component $x_n(t)$, $y_n(t)$, $\dot x_n(t)$ and $\dot y_n(t)$, respectively. Let $Z(t) := (x(t), y(t))$. We seek time-periodic solutions $(Z(t), \dot Z(t))$ of
the Hamiltonian system \eqref{eq:Hertzd0S}
for a prescribed period $T_b = 2\pi/\o_b$.
The problem is equivalent to finding the fixed points of the corresponding Poincar\'e map
$P_{T_b}[ (Z(0) ,\dot Z(0)) ] = ( Z(T_b), \dot Z(T_b) )$. We fix $\dot Z(0) = 0$ and use
the Newton-type algorithm to compute components $Z(0)$ of the fixed points.
The reader is referred to \cite{LJKV15} for more details on the numerical method.

To characterize the solution, we define the vertical centers of the solution
for the  $x$ and $y$ components \cite{Chong13},
\beq\label{eq:center}
C_x = \frac{\sup_{t\in[0,T_b]}x_1(t) + \inf_{t\in[0,T_b]}x_1(t) }{2},  \q C_y = \frac{\sup_{t\in[0,T_b]}y_1(t) + \inf_{t\in[0,T_b]}y_1(t) }{2}.
\eeq
The vertical center $C_x$ is set to be approximately zero
in our numerical computations (see \cite{LJKV15} for more details). We evaluate the dark breather amplitude through
\beq\label{eq:amp}
K_x = \frac{ \sup_{t\in[0,T_b]}x_1(t) - \inf_{t\in[0,T_b]}x_1(t) }{2}, \q  K_y = \frac{ \sup_{t\in[0,T_b]}y_1(t) - \inf_{t\in[0,T_b]}y_1(t) }{2}
\eeq
and the (squared) renormalized $\ell^2$ norm of $Z(0)$ can be defined as
\beq\label{eq:l2norm}
|| Z(0) ||^2_{\wt \ell^2} = \sum_{n} K_x^2 - |x_n(0) - C_x|^2 + \sum_{n} K_y^2 - |y_n(0) - C_y|^2.
\eeq
To evaluate the accuracy of the numerical solution, we also introduce the relative error
\beq\label{eq:relerr}
E_b(t) = ||Z(mT_b)- Z(0)||_{\infty} / ||Z(0)||_{\infty},
\eeq
where $m = \lfloor t/T_b \rfloor $ and $Z(mT_b) = (x(mT_b), y(mT_b))$ represents the strain profile after integrating \eqref{eq:Hertzd0S} over $m$ multiple of time periods, starting with the initial condition $Z(0)$.
Once the Newton-type solver converges to an exact dark breather solution, we use the method of continuation to obtain an entire family of dark breathers that corresponds to different values of $\o_{b}$.

\vspace{1ex}

\begin{figure}[h]
\centering
\centerline{\includegraphics[width=\textwidth]{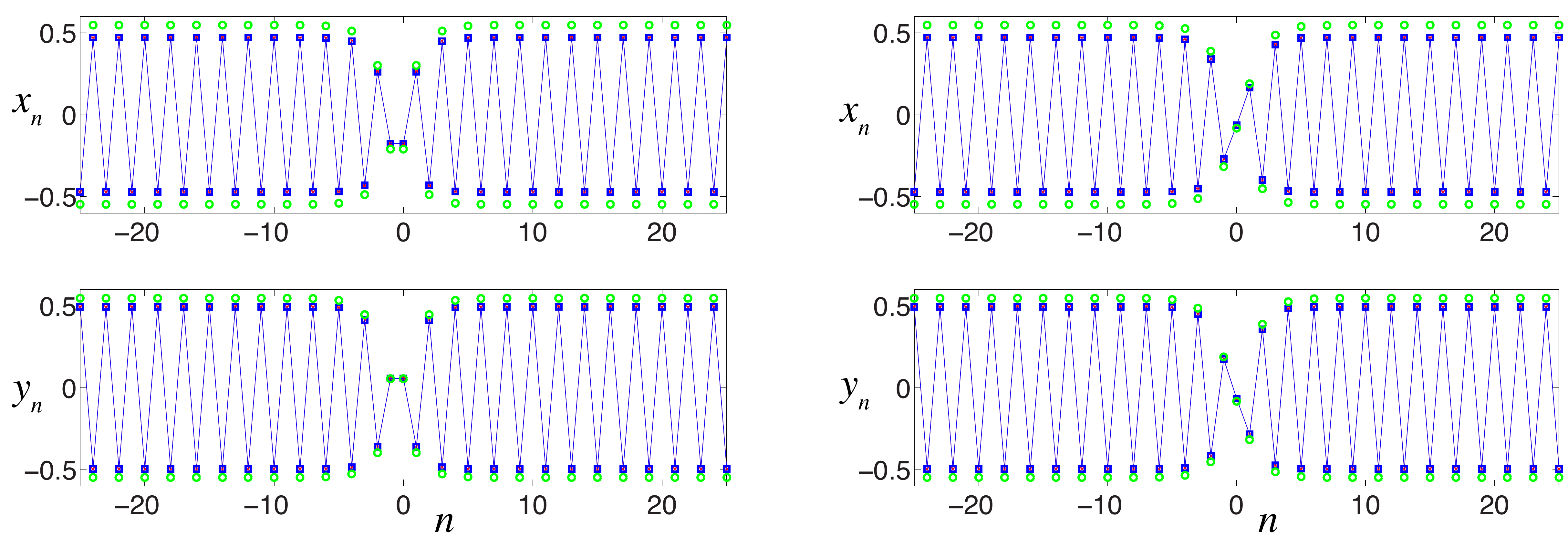}}
\caption{\footnotesize Left plot: a bond-centered dark breather solution (connected stars) with frequency $\o_{b} = 2.42$. Squares represent the strain profile after integration over $T_b$. Circles represent the initial profiles computed from the approximation \eqref{eq:xy_app}. Right plot: a site-centered solution.  Here $\a=3/2$, $\k =1$, $\d_0 = 4/9$ and $\r = 1/3$.}
\label{fig:opt_r03_wb242}
\end{figure}

To apply the above algorithm we first consider the mass ratio $\r = 1/3$. In that case
the linear frequencies of plane waves at the $\theta = \pi$ edge of optical and acoustic branches are $\o_{opt} = 2.4495$ and $\o_{acs} = 1.4142$, respectively. We choose a value of $\o_{b}$ that is slightly smaller than but close enough to $\o_{opt}$ or
$\o_{acs}$ to obtain a good initial seed with a small amplitude.
Sample profiles of both bond-centered and site-centered dark breather solutions with frequency $\o_{b} = 2.42$, along with their corresponding initial seeds from the NLS approximation \eqref{eq:xy_app} are shown in Fig.~\ref{fig:opt_r03_wb242}.
The relative errors $E_b(T_b)$ of both solutions are less than $3\times 10^{-9}$.

\begin{figure}[h]
\centering
\centerline{\includegraphics[width=\textwidth]{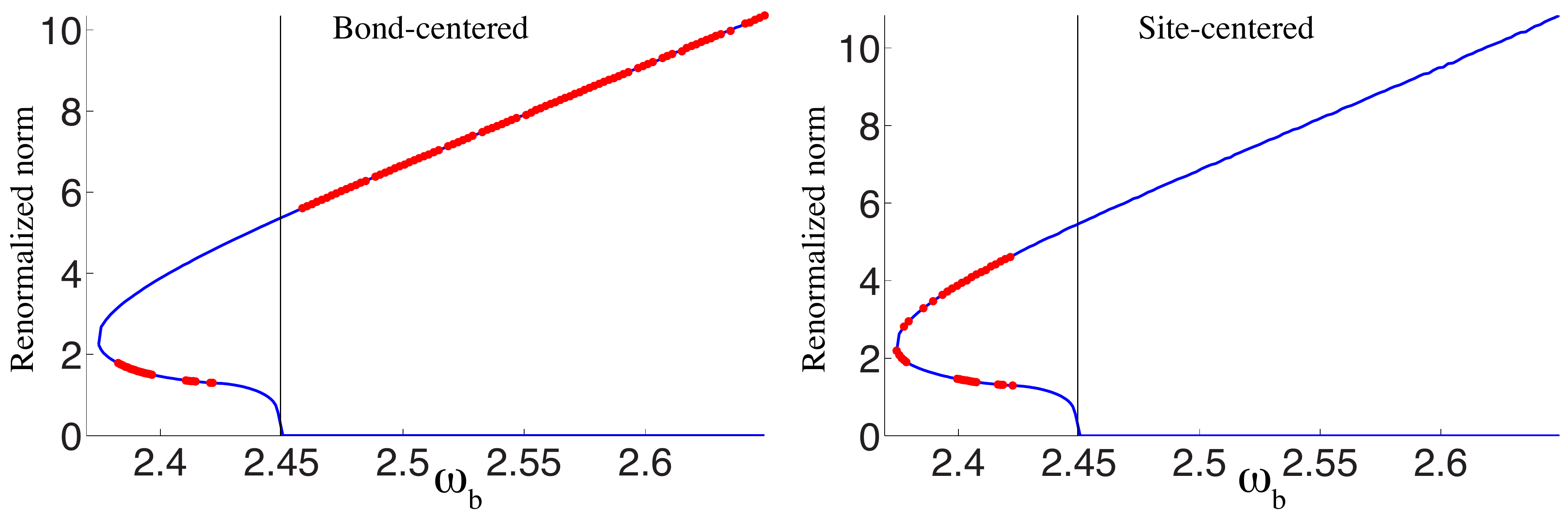}}
\caption{\footnotesize
Bifurcation diagram for stationary optical dark breathers in the case $\r = 1/3$ ($\k =1$, $\d_0 = 4/9$).
Left plot: renormalized $\ell^2$ norm of the bond-centered dark breather solution versus frequency $\o_{b}$ with the vertical center $C_x = 0$. The black vertical line shows the edge of the optical branch $\o_{opt} = 2.4495$. Right plot: same as the left, but for site-centered solution. Regions where a real instability is present are indicated by red dots.}
\label{fig:r03_norm}
\end{figure}

In the left plots in Fig.~\ref{fig:r03_norm}, we show the renormalized $\ell^2$ norm of the numerically exact bond-centered dark breathers that bifurcate from $\theta = \pi$ edge of the optical branch. The solutions do not exist for arbitrary values of $\o_b$; rather, there is a turning point at $\o_b \approx 2.37$.
We are able to continue the relevant solution branch past this turning point,
obtaining dark breather solutions making up the top part of the branch, with increasing amplitude as $\o_b$ increases. Therefore, dark breathers above the turning point can be regarded as \emph{strongly nonlinear} solutions, while solutions along the bottom part are \emph{weakly nonlinear}. As explained in the next subsection, solutions along the segments marked by red dots possess real instability, and the ones along the blue segments do not.

Using the same method for $\rho = 1/3$, we obtain qualitatively similar solution branches for acoustic
dark breathers bifurcating from $\omega_b = \o_{acs}$. The bifurcation diagram
for bond-centered solutions is shown in the left panel of Fig.~\ref{fig:acs_floq_bc_r03}.

\begin{figure}[htp]
\centering
\centerline{\includegraphics[width=\textwidth]{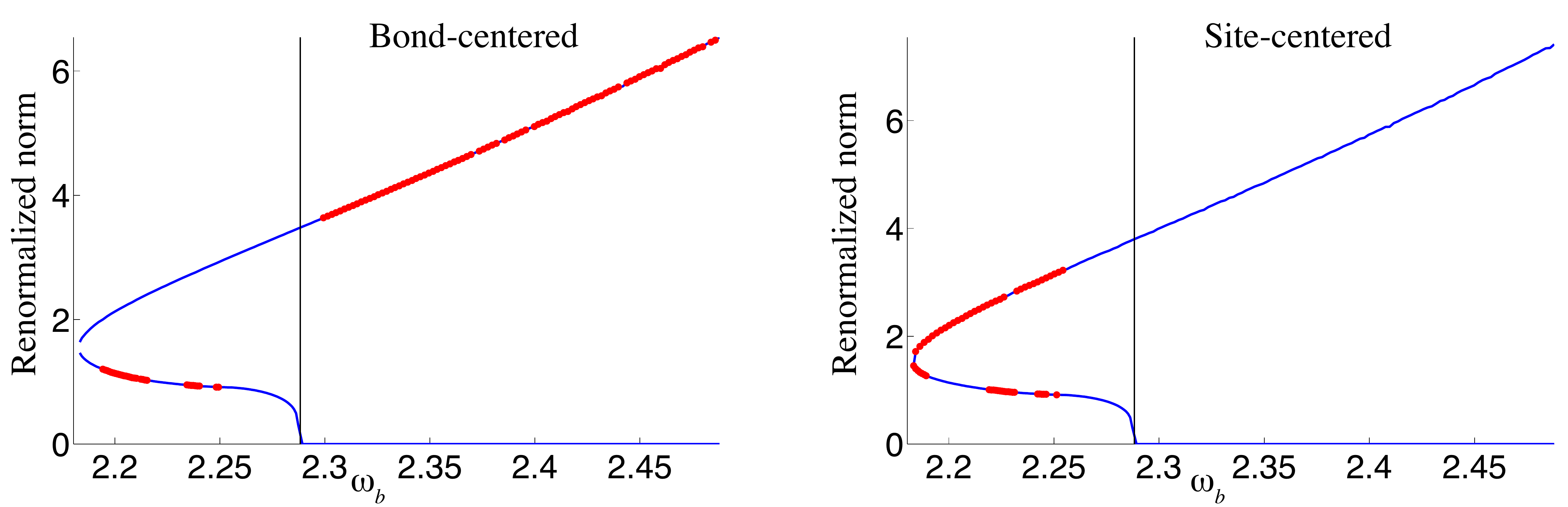}}
\caption{\footnotesize
Bifurcation diagram for stationary optical dark breathers in the case $\r = 1$ ($\k =1$, $\d_0 = 4/9$).
Left plot: renormalized $\ell^2$ norm of the bond-centered dark breather solution bifurcating from the optical branch versus frequency $\o_{b}$ with the vertical center $C_x = 0$. The black vertical line shows the edge of the optical branch $\o_{opt} = 2.2282$. Right plot: same as the left, but for site-centered solution. Portions of the curve where a real instability is present are indicated by red dots.}
\label{fig:r1_norm}
\end{figure}

We found the above bifurcations to be robust provided the mass ratio $\rho$ is not too small or too large.
For example, increasing mass ratio to $\rho = 1$, the continuation yields
a family of dark breathers bifurcating from the optical branch edge $\o_{opt} = 2.2882$.
The bifurcation diagrams of the renormalized $\ell^2$ norm of these solutions are
shown in Fig.~\ref{fig:r1_norm} and are similar to the $\r=1/3$ case.

The situation is different for large or small mass ratio.
For $\r = 10$, the continuation procedure worked quite well for dark breathers bifurcating from the optical branch but we encountered difficulty doing computations for the acoustic band due to the rapidly growing amplitude of those solutions as $\o_b$ decreases. As a result, only dark breathers with frequencies very close to $\o_{acs}$ were obtained.
For the case of a very small mass ratio, $\rho = 0.1$, we found that in contrast to the large mass ratio like $\r = 10$, we can only obtain dark breathers bifurcating from the acoustic branch and corresponding to a wide range of frequencies.

\subsection{\label{stabdb}Stability analysis for stationary dark breathers}\label{sec:stability_weak}

\begin{figure}[htp]
\centering
\centerline{\includegraphics[width=\textwidth]{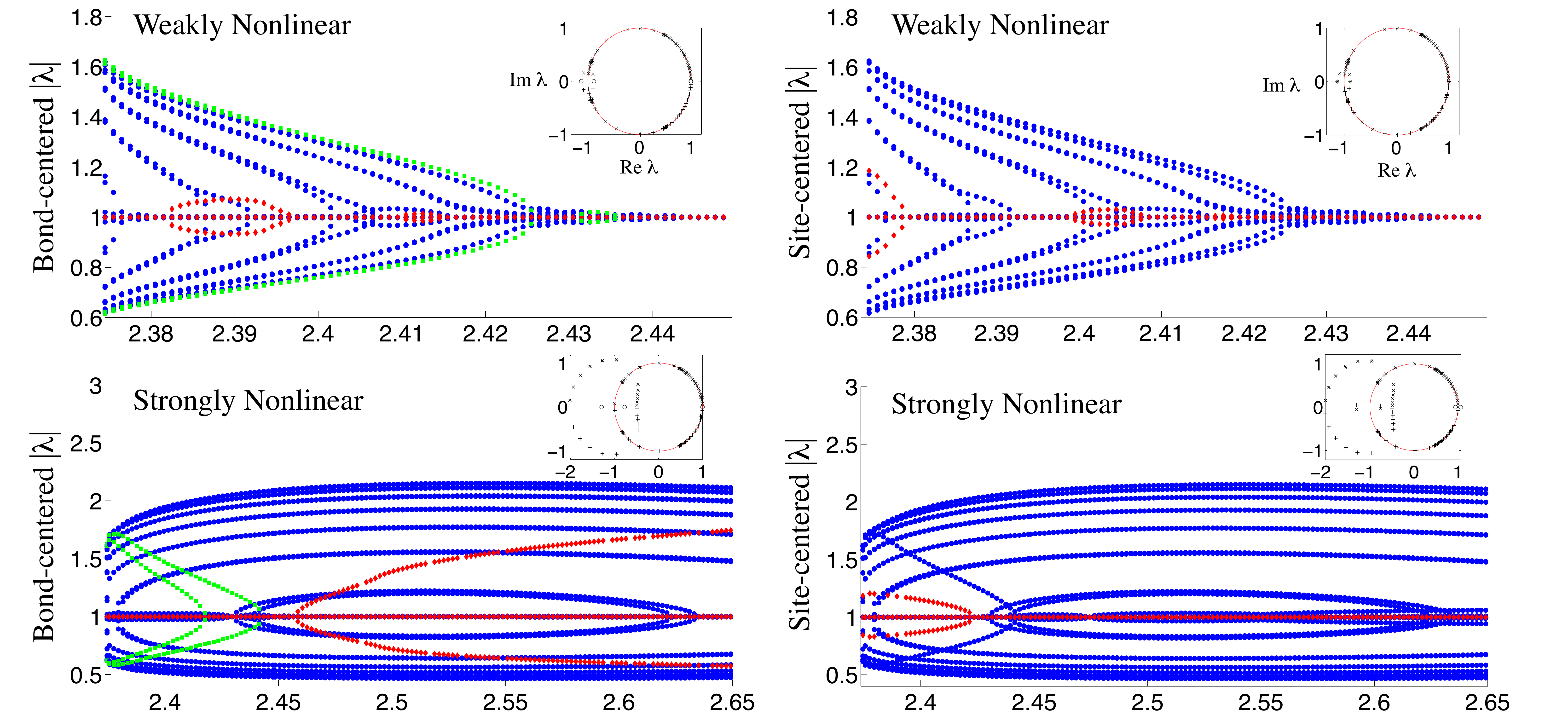}}
\caption{\footnotesize Moduli of the Floquet multipliers versus frequency $\o_{b}$ for weakly nonlinear (top) and strongly nonlinear (bottom) types of
optical dark breathers. Recall that these refer to solutions below and above the turning point in Fig.~\ref{fig:r03_norm}, respectively. Left and right plots correspond to bond- and site-centered dark breathers respectively.
Blue dots represent $\text{Im}(\l) \neq 0$, red diamonds represent $\mbox{Im}(\l) =0, \l > 0$ and green squares represent $\text{Im}(\l) = 0, \l < 0$. The Floquet multipliers for $\o_{b} = 2.42$ in the complex plane are shown in the respective insets. Here $\k =1$, $\d=4/9$ and $\r = 1/3$.}
\label{fig:opt_floq_r03}
\end{figure}

We now examine the linear stability of the obtained dark breather solutions using the standard Floquet analysis,
starting with the optical case and fixing $\rho = 1/3$.
The eigenvalues (Floquet multipliers) of the associated monodromy matrix of the variational equation of \eqref{eq:Hertzd0S} determine the linear stability of the breather solution; see also the relevant discussion e.g. of~\cite{Chong13}. The moduli of Floquet multipliers of the site-centered and bond-centered solutions, for both strongly and weakly nonlinear types, are shown in Fig.~\ref{fig:opt_floq_r03}, along with the numerically computed Floquet spectrum corresponding to the sample breather profile at $\o_{b}=2.42$. If any of these Floquet multipliers $\l_{i}$ satisfies $|\l_{i}|>1$, the corresponding breather is linearly unstable. We observed two types of instabilities in this Hamiltonian system. The \emph{real} instability takes place when there is a pair of real Floquet multipliers, one of which has magnitude greater than one. The second type is the \emph{oscillatory} instability  corresponding to a quartet of Floquet multipliers which do not lie on the unit circle and have nonzero imaginary parts.

\begin{figure}[htp]
\centering
\centerline{\includegraphics[width=\textwidth]{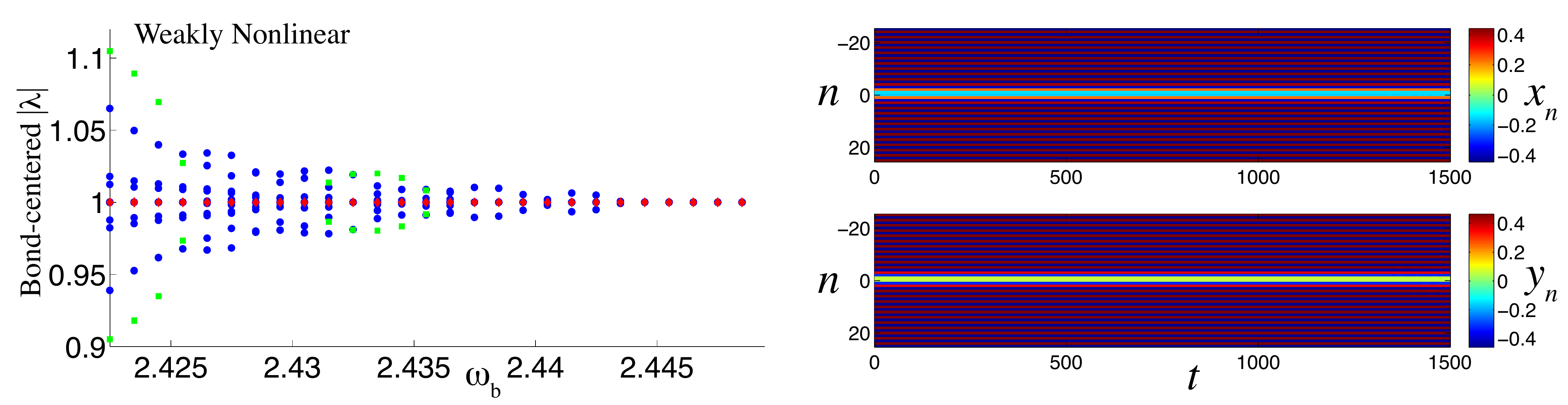}}
\caption{\footnotesize Left plot: moduli of Floquet multipliers of bond-centered
dark breathers for frequencies $2.4235 \le \o_{b}\le \o_{opt}$. Right plots: contour plot of the time evolution of the bond-centered solution for $\o_b = 2.4255$. The color bar corresponds to the magnitude of the strain $x_n$ (top) and $y_n$ (bottom). The dark breather appears to be very long-lived despite the instability suggested by the left
panel of the figure. Here $\k =1$, $\d=4/9$ and $\r = 1/3$. }
\label{fig:opt_bc_r03_wb24255}
\end{figure}

Numerical results suggest that both bond-centered and site-centered weakly nonlinear solutions are stable at the beginning of the continuation procedure, as shown in the top panels of Fig.~\ref{fig:opt_floq_r03}, similarly to what
was found in the homogeneous granular chain~\cite{Chong13}. However, as the corresponding frequency decreases, the amplitude of the breathers increases and they start to exhibit oscillatory instabilities. These marginally unstable modes remain weak until $\o_b$ reaches $\o_b \approx 2.4255$ in the left plot of Fig.~\ref{fig:opt_bc_r03_wb24255}.

To further investigate the long-term behavior of dark breather solutions, we examine
the relative error $E_b(t)$ defined in (\ref{eq:relerr}), with initial condition
$Z(0)$ given by the dark breather solution.
The relative error $E_b(1200)$ defined in \eqref{eq:relerr} stays below $2\times 10^{-4}$ when $2.4255\le \o_b \le \o_{opt}$ and increases dramatically at smaller frequencies. A representative space-time evolution diagram of bond-centered dark breather solution at $\o_{b} = 2.4255$ is shown in the right plots of Fig.~\ref{fig:opt_bc_r03_wb24255}, and the time evolution of site-centered solution is similar. This suggests that the dark breathers solutions of both types with frequency close to the linear frequency $\o_{opt}$ are long-lived and have marginal oscillatory instability.

As $\o_b$ further decreases, we observe the emergence of many new and stronger modes of oscillatory instability, with Floquet multipliers distributed symmetrically outside the unit circle around $-1$. In addition, for bond-centered dark breathers, pairs of real Floquet multipliers collide at $-1$ and move in the opposite directions as $\o_b$ decreases. This is associated with a period-doubling instability. Moduli of these multipliers are indicated by the green squares in Fig.~\ref{fig:opt_floq_r03} and in the left plot of Fig.~\ref{fig:opt_bc_r03_wb24255}.

\begin{figure}[htp]
\centering
\centerline{\includegraphics[width=\textwidth]{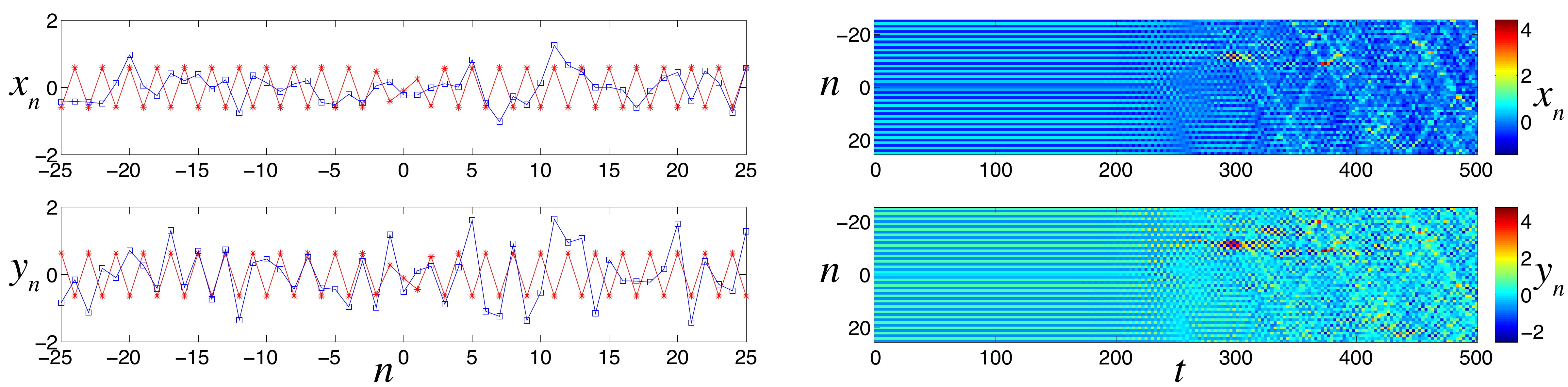}}
\caption{\footnotesize Left plot: snapshots of the strain profile of site-centered optical dark breather solution at time $t = 0$ (connected red stars) and $t = 500$ (connected blue squares). Right plots: contour plot of the time evolution of the site-centered solution for $\o_b = 2.40$. The color bar corresponds to the magnitude of the strain $x_n$ (top) and $y_n$ (bottom). Here $\k =1$, $\d=4/9$ and $\r = 1/3$.
Clearly, the dark breather structure gets destroyed as a result of
its spectral instability.}
\label{fig:opt_sc_r03_wb240}
\end{figure}

The emergence of many unstable quartets suggests that both bond-centered and site-centered solutions have strong modulational instability of the background, which leads to the breakdown of the dark breather structure, accompanied, as shown in Fig.~\ref{fig:opt_sc_r03_wb240}, by a chaotic evolution of both solutions after a short time. The same phenomenology (dismantling of the breather and chaotic evolution) also takes place for all strongly nonlinear dark breather solutions. For this reason, in the bifurcation diagram of Fig.~\ref{fig:r03_norm} only real instabilities are indicated. It is interesting that the strongly nonlinear bond-centered solutions typically exhibit real instability at frequencies greater than $\o_{opt}$, while the real instability emerges only when $\o_b$ is less than $\o_{opt}$ for the site-centered ones.

\vspace{1ex}

\begin{figure}[htp]
\centering
\centerline{\includegraphics[width=0.9\textwidth]{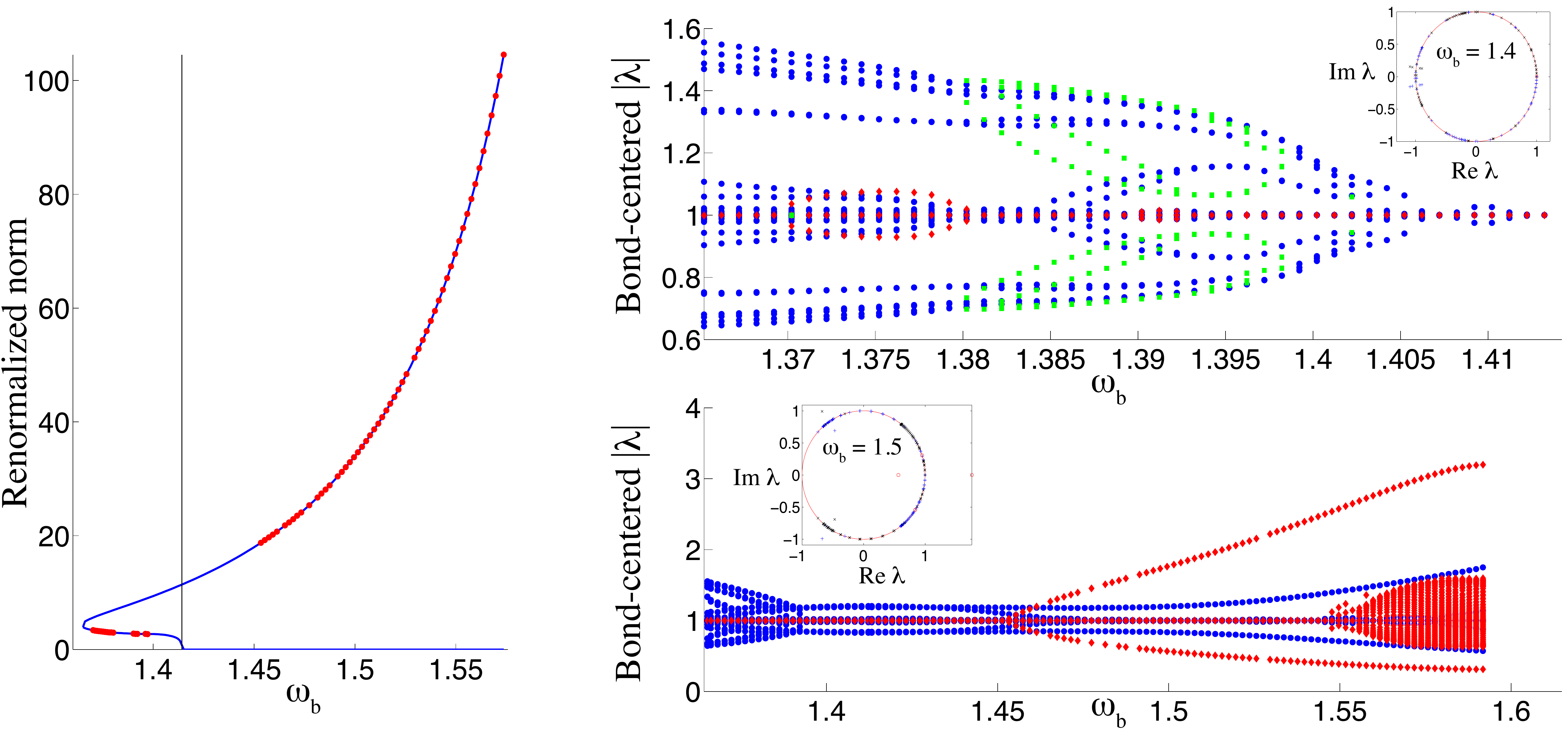}}
\caption{\footnotesize Left panel: renormalized $\ell^2$ norm of the bond-centered dark breather bifurcating from the acoustic branch versus $\o_b$ with the vertical center $C_x = 0$. The black vertical line shows the $\theta = \pi$ edge of the acoustic branch $\o_{acs}=1.414$. Right panel: moduli of the Floquet multipliers versus frequency $\o_{b}$ for weakly nonlinear (top) and strongly nonlinear (bottom) types. Insets: the Floquet multipliers in the complex plane. Here $\k =1$, $\d=4/9$ and $\r = 1/3$.}
\label{fig:acs_floq_bc_r03}
\end{figure}

We now  examine the stability properties of dark breathers that bifurcate from the $\theta = \pi$ edge of the acoustic branch, where the linear frequency is $\o_{acs} = 1.414$. The bifurcation diagram for solutions of the bond-centered type and the corresponding diagrams of the moduli of the Floquet multipliers versus frequency are shown in Fig.~\ref{fig:acs_floq_bc_r03}. We observe that the distribution of positive real Floquet multipliers in the top right plot (weakly nonlinear dark breathers) follows a pattern similar to the one for such breathers bifurcating from the optical branch (top left plot of Fig.~\ref{fig:opt_floq_r03}). Note also that the large arc for strongly nonlinear solutions in the bottom right plot of Fig.~\ref{fig:acs_floq_bc_r03} suggests that these dark breathers exhibit not only an oscillatory instability but also a real instability, which
possesses a considerably larger growth rate than the oscillatory one.

\begin{figure}[htp]
\centering
\centerline{\includegraphics[width=\textwidth]{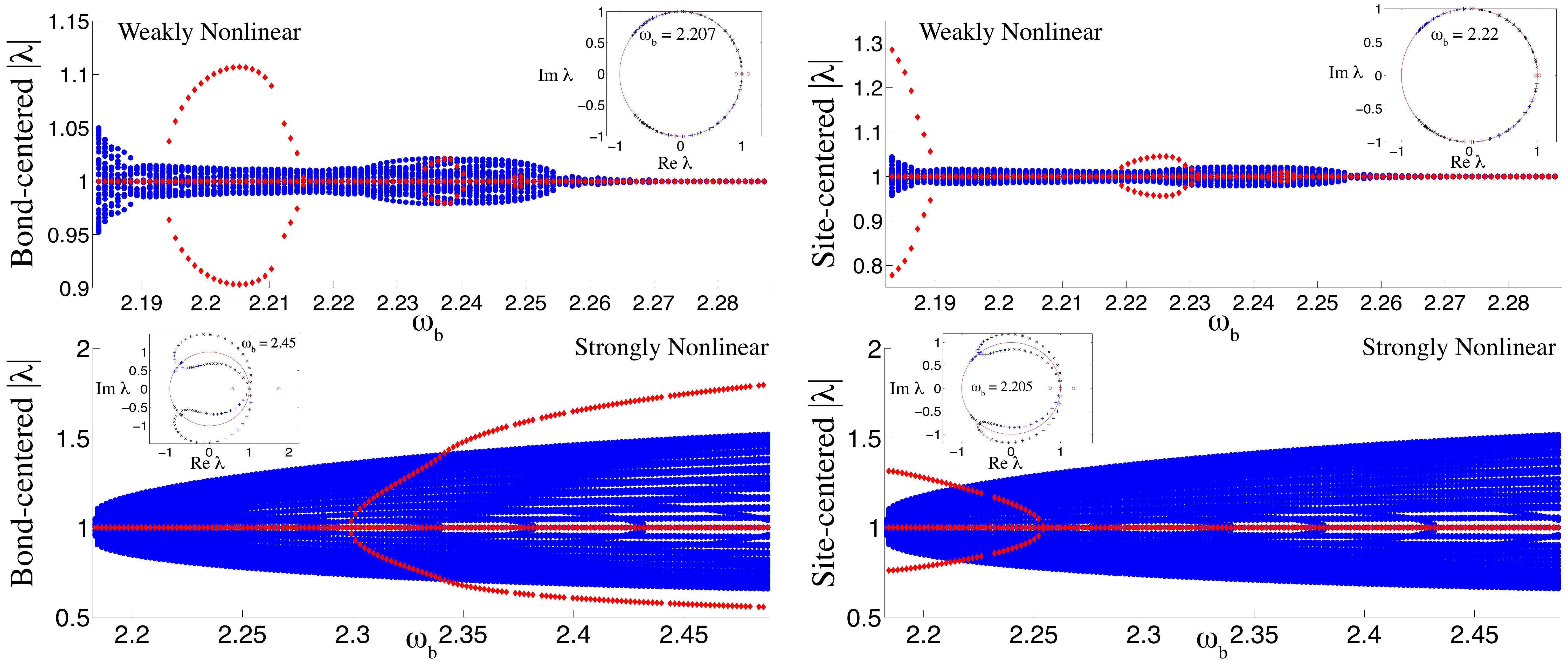}}
\caption{\footnotesize Results of the simulation with the same parameters as in Fig.~\ref{fig:opt_floq_r03} except for $\r = 1$. Here the corresponding dark breathers bifurcate from the optical branch.}
\label{fig:opt_floq_r1}
\end{figure}

\begin{figure}[htp]
\centering
\centerline{\includegraphics[width=\textwidth]{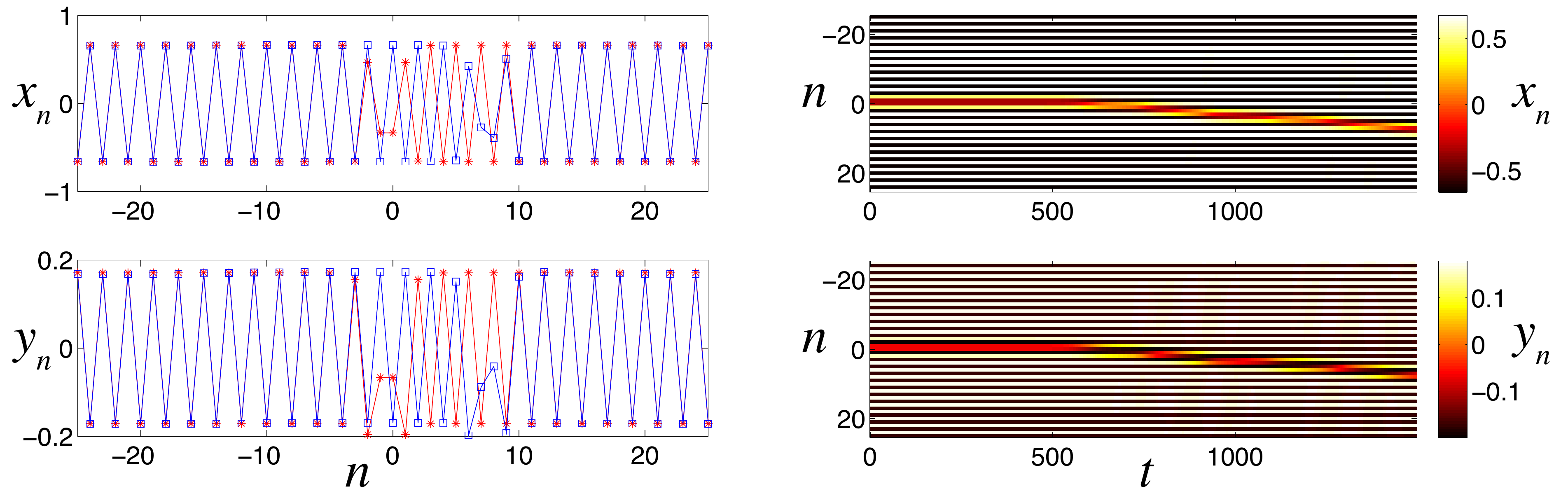}}
\caption{\footnotesize Left panel: sample profiles (connected stars) of bond-centered optical dark breathers at the frequency $\o_b = 2.207$. Connected squares represent strain profiles after integration over $526\, T_b \approx 1500$. Right panel: contour plots of the time evolution of the bond-centered solution for $\o_b = 2.207$. The color bar corresponds to the magnitude of the strain $x_n$ (top) and $y_n$ (bottom). Clearly, the instability of the stationary dark breather solutions sets it into motion. Here $\k =1$, $\d=4/9$ and $\r = 1$.}
\label{fig:weak_opt_bc_r1_evol}
\end{figure}

We now investigate the effect of the mass ratio on the linear stability of solutions. We first consider the mass ratio $\r = 1$ and keep all other parameters the same as in the previous simulations. We
have obtained in Sec.~\ref{numdb} a family of dark breathers bifurcating from the optical branch edge $\o_{opt} = 2.2882$,
with bifurcation diagrams similar to the $\r=1/3$ case (see Fig.~\ref{fig:r1_norm}).
The moduli of Floquet multipliers of the weakly and strongly nonlinear dark breathers solutions for both bond-centered and site-centered types are shown in Fig.~\ref{fig:opt_floq_r1}. We observe that the magnitude of the Floquet multipliers corresponding to the oscillatory instability is much weaker than in the $\r = 1/3$ case. In addition, the real instability becomes more significant than the oscillatory one at some frequencies, resulting in not only shorter lifetime of the solutions, but also setting the dark breathers in motion. A representative space-time evolution diagram for weakly nonlinear bond-centered dark breather solution of frequency $\o_b = 2.207$ is shown in Fig.~\ref{fig:weak_opt_bc_r1_evol}. Note that the positive Floquet multipliers again exhibit the same bifurcation structure as in the previous results.


Considering optical dark breathers for a large mass ratio $\r = 10$, we have
observed that the magnitude of Floquet multipliers corresponding to oscillatory instability is much larger than in the $\rho = 1$
case but smaller than in the $\rho = 1/3$ case, which suggests that the significance of oscillatory instability does not depend monotonically on the mass ratio. Furthermore, in this case, the oscillatory instability appears to set in
already in the immediate vicinity of the linear limit at the edge of the optical band.

As indicated in Sec.~\ref{numdb}, for a very small mass ratio $\rho = 0.1$ we have only obtained acoustic
dark breathers for a wide range of frequencies. In that case,
it is surprising that the real instability of some of the weakly nonlinear solutions is again more significant than the oscillatory instability, which is also a feature for the optical weakly nonlinear solutions at $\r = 1$. Again, similar patterns of the distribution of the positive real Floquet multipliers suggest that their bifurcation structure remains unaffected by different mass ratios.
%
%
 %
%
\section{Concluding remarks}
\label{sec:end_weak}
In this work, we investigated discrete breathers in a precompressed locally resonant granular chain. The precompression effectively suppresses the fully nonlinear character of the Hertzian interactions and leads to a weakly nonlinear system in the small-amplitude limit. Following the approach developed in \cite{Chong13, JKC13} for two limiting cases of the present model and adopting a multiscale asymptotic technique \cite{Gian04, Gian06, Schneider10}, we derived modulation equations that reduce to the NLS equation at finite mass ratio.

The focusing NLS equation was then used to investigate the moving bright breather solutions of the system at finite mass ratios. We showed numerically that breather solutions of the NLS equation can successfully approximate
small-amplitude moving optical bright breathers on a long but finite time scale at some wave numbers and various mass ratios.
In addition, we have shown the possibility to excite optical bright breathers by imposing a sinusoidal motion of the first element of the chain.

Contrary to the more standard dimer case \cite{Boechler10,Theo10}, the locally resonant granular chain
appears to possess bright breathers
in the neighborhood of $\theta=0$ wavenumbers of the optical
band. However, that very point is found to be singular, and breathers
in its immediate vicinity do not appear to be robust, while bright breathers
at larger wave numbers that are below a certain threshold are found to propagate nearly undistorted in the
resonator chain at any mass ratio.
At some other wave numbers for the optical branch and in the acoustic case,
perhaps especially so near the edges of the windows where
the focusing model is predicted via the multiscale method,
the NLS equation does not capture numerically observed phenomena, including eventual formation and steady motion of smaller breathers that detach from the initial breather and are associated with the other dispersion branch. In particular, robust propagation of nanoptera, bright breathers with small-amplitude tails, was observed in some cases.

In addition, we have found that dark breather solutions can be generated from instabilities of certain
periodic traveling waves (close to acoustic modes) in the locally resonant granular chain.
In order to analyze dark breather solutions, we used the analytical solutions of
the defocusing NLS equation to construct approximate dark breather solutions.
Using this approximation as an initial condition for a continuation procedure based on a Newton-type algorithm, we obtained both weakly and strongly nonlinear
stationary dark breathers of both the site-centered and bond-centered families for a wide range of frequencies. We then studied linear stability of the obtained solutions. The results revealed that only small-amplitude weakly nonlinear solutions with frequencies very close to the linear frequencies of the system are stable (for typical values
of the resonator mass $\rho$). In contrast, large-amplitude dark breather
solutions exhibit either real or oscillatory instabilities (or both). In particular, the strongly nonlinear dark breathers have a very unstable background, leading to the dismantling of their structure, accompanied by a chaotic evolution after a short time. We also observed long-lived traveling dark breathers, resulting
from the perturbation of stationary dark breathers subject to a real instability.
Finally, we showed that the mass ratio strongly affects the strength of the oscillatory instability of the dark breather solutions, but its influence on the distribution patterns of positive real Floquet multipliers is less pronounced.

Future theoretical challenges include the rigorous proof of the existence of long-lived or exact bright and dark breathers
in the case of stationary or propagating solutions, the analysis of their stability, and the comparison with the numerical results presented here.
A numerical study of the stability of periodic traveling and standing waves going beyond the NLS approximation will be also of interest, in order to
classify the parameter regimes leading to modulated periodic waves, localized structures or disordered regimes.
Along this line, the mechanism of the excitation of traveling dark breathers from unstable periodic waves illustrated in
Sec.~\ref{dedb} still needs to be explained.

On the experimental side, it will be interesting to investigate whether it is possible to generate either (approximate) moving bright breathers, perhaps actuating one boundary at a suitable, near-band-edge frequency, or stationary dark breathers in a slightly damped finite locally resonant chain driven at the ends in a way similar to \cite{Chong13, chong14} but using the woodpile setup \cite{kimyang15}.
In addition, as we observed in Sec.~\ref{sec:derivation_NLS}, the number of focusing regions within the
optical and acoustic bands can change when the main parameters, the mass ratio $\rho$
and precompression $\delta_0$, are varied. This opens up the possibility to change the system's response to external perturbations by varying precompression,
which constitutes a very interesting property for the design of adaptive metamaterials.\\

\noindent {\bf Acknowledgements.} G.J. acknowledges financial support from the Rh\^one-Alpes Complex Systems Institute (IXXI). The work of L.L. and A.V. was partially supported by the US NSF grants DMS-1007908 and DMS-1506904. P.G.K. gratefully acknowledges the support
of US AFOSR
through grant FA9550-12-1-0332. P.G.K.'s work at Los Alamos was supported
in part by the US Department of Energy. P.G.K. also acknowledges support from the ARO (W911NF-15-1-0604).
\appendix
\section{Appendix: Derivation of the modulation equations}
\label{App:mod}
In this appendix, we show the details of the derivation of the modulation equations \eqref{eq:mod1} and \eqref{eq:mod2}. In what follows, we will use the abbreviation $\sum_{k,j}$ for the summation over $k \in \hN_1$ and $j \in \hZ$, $|j| \le k$ in \eqref{eq:ansatz}.
The ansatz defined in \eqref{eq:ansatz} will be
denoted by $U_{n}^{A}=u_n$ and $V_{n}^{B}=v_n$.
After substituting \eqref{eq:ansatz} into \eqref{eq:Hertz}, we find that the right hand side of the second equation in \eqref{eq:Hertz} reads
\beq
-\frac{\k}{\r} \sum_{k,j}\e^k[B_{k,j}(\t,\xi) - A_{k,j}(\t, \xi)]E(t,n)^j,
\eeq
and its left hand side is
\bea
\ddot V_{n}^{B}(t) &:= \sum_{k,j}\e^k[-(j\o)^2B_{k,j} + 2\e ij\o c\partial_{\xi}B_{k,j} + \e^2(c^2\partial_{\xi}^2B_{k,j} - 2ij\o\partial_{\t}B_{k,j})\nonumber
\\
&-2\e^3c\partial_{\xi}\partial_{\t}B_{k,j} + \e^4\partial_{\t}^2B_{k,j}]E^{j}(t,n).
\eea
To match the coefficients of each $\e^kE^j(\t,\xi)$ term on both sides, we require that
\bea
&&\e^1E^0: \q 0 = -\frac{\k}{\r}(B_{1,0} - A_{1,0}) \q \Ra  \q B_{1,0} = A_{1,0}, \label{B10}\\
&&\e^1E^1: \q -\o^2B_{1,1} = -\frac{\k}{\r}(B_{1,1} - A_{1,1}) \q \Ra  \q (\k - \r\o^2)B_{1,1} = \k A_{1,1}, \label{B11}\\
&&\e^2E^1: \q -\o^2B_{2,1} + 2i\o c\partial_{\xi}B_{1,1} = - \frac{\k}{\r}(B_{2,1}-A_{2,1}),\label{B21}\\
&&\e^2E^2: \q -4\o^2B_{2,2} = -\frac{\k}{\r}(B_{2,2} - A_{2,2}),\label{B22}\\
&&\e^3E^0:\q c^2\partial_{\xi}^2B_{1,0}  = -\frac{\k}{\r}(B_{3,0} - A_{3,0}), \label{B30}\\
&&\e^3E^1: \q -\o^2B_{3,1} + 2i\o c\partial_{\xi}B_{2,1} + c^2\partial_{\xi}^{2}B_{1,1} - 2i\o\partial_{\t}B_{1,1} = - \frac{\k}{\r}(B_{3,1}-A_{3,1}).\label{B31}
\eea
Meanwhile, the right hand side of the first equation of \eqref{eq:Hertz} can be treated as a sum of linear part $L_n(U^A, V^B)$ and nonlinear part $N_n(U^A)$, where we define
\begin{align}
L_n(U^A, V^B) &:= \sum_{k,j}\e^k\{[2K_2(\cos{j\theta} -1) - \k)]A_{k,j} + \k B_{k,j} + 2iK_2\e\partial_{\xi}A_{k,j}\sin{j\theta} \nonumber\\
&+K_2\e^2\partial_{\xi}^2A_{k,j}\cos{j\theta} + O(\e^3)\}E^{j}
\end{align}
and
\begin{align}
N_{n}(U^A ) &:= - \e^2(K_3 s D_1 A_{1,1}^2E^2 + c.c.) \nonumber\\
&+ \e^3\{2K_3 D_1 \bar A_{1,1}\partial_{\xi}A_{1,1} + (2K_3 s D_1\bar A_{1,1}A_{2,2} - 3K_4 D_1^2|A_{1,1}|^2A_{1,1}-2K_3 D_1\partial_{\xi}A_{1,0}A_{1,1})E\nonumber\\
&+ [2K_3 D_1(D_1-3)A_{1,1}\partial_{\xi}A_{1,1} - 2K_3 s D_1 A_{1,1}A_{2,1}]E^2\nonumber\\
&+ [2K_3 s(D_1+s^2)A_{1,1}A_{2,2} + K_4 D_1^2(3 - D_1)A_{1,1}^3]E^3 + c.c\} + h.o.t.,
\end{align}
with the abbreviation h.o.t. meaning higher order terms.
Here $s = 2i\sin{\theta}$ and $D_1 = 4\sin^2(\theta/2)$. The left hand side of the first equation of \eqref{eq:Hertz} reads
\begin{align}
\ddot U_{n}^{A}(t) &:= \sum_{k,j}\e^k[-(j\o)^2A_{k,j} + 2\e ij\o c\partial_{\xi}A_{k,j} + \e^2(c^2\partial_{\xi}^2A_{k,j} - 2ij\o\partial_{\t}A_{k,j})\nonumber\\
&-2\e^3c\partial_{\xi}\partial_{\t}A_{k,j} + \e^4\partial_{\t}^2A_{k,j}]E^{j}(t,n).
\end{align}
To match the coefficients of each $\e^kE^j(\t,\xi)$ term on both sides, one needs
\begin{align}
&\e^1E^0: 0 = -\k A_{1,0} + \k B_{1,0} \q \Ra  \q A_{1,0} = B_{1,0}, \label{A10}\\
&\e^1E^1: -\o^2A_{1,1} = (-D - \k)A_{1,1} + \k B_{1,1}, \label{A11}\\
&\e^2E^1:  -\o^2A_{2,1} + 2i\o c\partial_{\xi}A_{1,1} = (-D-\k)A_{2,1} +\k B_{2,1}+2iK_2\partial_{\xi}A_{1,1}\sin{\theta},\label{A21}\\
&\e^2E^2: -4\o^2A_{2,2} = (-4K_2\sin^2\theta- \k)A_{2,2} + \k B_{2,2} - K_3 s D_1A_{1,1}^2,\label{A22}\\
&\e^3E^0: c^2\partial_{\xi}^2A_{1,0}  = -\k A_{3,0} + \k B_{3,0} + K_2\partial_{\xi}^2A_{1,0} + 2K_3D_1\bar A_{1,1}\partial_{\xi}A_{1,1} + c.c.,\label{A30}\\
&\e^3E^1: -\o^2A_{3,1} + 2i\o c\partial_{\xi}A_{2,1} + c^2\partial_{\xi}^{2}A_{1,1} - 2i\o\partial_{\t}A_{1,1} =(-D-\k)A_{3,1}+\k B_{3,1}\label{A31}\\
&+2iK_2\partial_{\xi}A_{2,1}\sin{\theta}+K_2\partial_{\xi}^2A_{1,1}\cos{\theta}+2K_3 s D_1\bar A_{1,1}A_{2,2} - 3K_4D_{1}^2|A_{1,1}|^2A_{1,1}
-2K_3D_1\partial_{\xi}A_{1,0}A_{1,1}.\nonumber
\end{align}
Note that both (\ref{B10}) and (\ref{A10}) yield $A_{1,0} = B_{1,0}$ but the two coefficients are not zero, in contrast with the non-resonant homogeneous chain problem ($\rho=0$). We now derive an equation to determine them. Using \eqref{B30} and \eqref{A30}, we obtain
\beq\label{A10A11}
[c^2(1 + \r)-K_2]\partial_{\xi}^{2}A_{1,0} = 2K_3D_1\bar A_{1,1}\partial_{\xi}A_{1,1} + c.c. ,
\eeq
where we assume the non-resonance condition \eqref{eq:nonresonance2}.

Now combining (\ref{B11}) and (\ref{A11}) yields $\textbf{M}\,(A_{1,1}, B_{1,1})^{T} = 0$, where the matrix $\textbf{M}$ is given by \eqref{M}. This yields that $(A_{1,1}, B_{1,1})^{T}$ is an eigenvector of $\bf{M}$ corresponding to zero eigenvalue and $A_{1,1}$ and $B_{1,1}$ are thus connected by
\beq\label{B11A11}
B_{1,1} = \frac{\k}{\k - \r\o^2}A_{1,1}.
\eeq
Note that $\k - \r\o^2 \neq 0$, since one can check that
$\o_-^2 (\theta ) \leq \o_-^2 (\pi ) < \k / \r$ and
$\o_+^2 (\theta ) \geq \o_+^2 (0 ) > \k / \r$.
In addition, (\ref{B21}), (\ref{A21}) and \eqref{B11A11} yield
\beas
\textbf{M} \left( \begin{array}{c}A_{2,1} \\ B_{2,1} \end{array}\right) =  2i\partial_{\xi}A_{1,1}
\, W,
\eeas
with
\beq
W =  \left( \begin{array}{c} \o c - K_2\sin\theta\\ \o c\k/(\k - \r\o^2) \end{array}\right)
\eeq
and the matrix $\mathbf{M}$ defined in (\ref{M}).
The range of $\mathbf{M}$ being orthogonal to
$$
W^\ast =  \left( \begin{array}{c} 1-\rho \, \omega^2 / \kappa \\ \rho \end{array}\right),
$$
we further obtain the compatibility condition ${W^\ast}\cdot W=0$, which reads
\beq\label{wc1}
\frac{\r\k}{\r\o^2 - \k} = \frac{\o c - K_2\sin\theta}{\o c} \; \frac{\k - \r\o^2}{\k} .
\eeq
This yields
\beq\label{eq:c}
c = \frac{K_2 \sin{\theta}\, (\k - \r\o^2)^2}{\o [\r\k^2 + (\k - \r\o^2)^2]},
\eeq
and one can check that $c = \o'(\theta)$ by differentiating the dispersion equation $\text{det}\, \mathbf{M}=0$, or
\beq
\label{eq:disp}
\o^4 - (D + \k + \k/\r)\o^2 + D\k/\r =0
\eeq
with respect to $\theta$.

Consider now the equations (\ref{B22}) and (\ref{A22}) for $\e^2E^2$, which yield
\bea
&&B_{2,2} = \frac{\k}{\k - 4\r\o^2}A_{2,2},\\
&&A_{2,2} = \frac{K_3sD_1(\k - 4\o^2\r)}{(4\o^2 - 4K_2\sin^2\theta-\k)(\k-4\o^2\r) + \k^2}A_{1,1}^2\label{A22A11}.
\eea
This solution exists under the non-resonance condition \eqref{eq:nonresonance1}, which is equivalent to $\o(2\theta) \pm 2\o(\theta) \neq 0$, as can be easily verified by substituting $\o \rightarrow \pm 2\o$ and $\theta \rightarrow 2\theta$ in \eqref{eq:disp}.

In the same manner, \eqref{B21}, \eqref{B31} and \eqref{A31} yield the following linear system:
\beas
\textbf{M} \left( \begin{array}{c}A_{3,1} \\ B_{3,1} \end{array}\right) =  2i\partial_{\xi}A_{2,1}\, W + \left( \begin{array}{c} (c^2 - K_{2}\cos\theta)\partial^2_{\xi}A_{1,1}\\ c^2(3\r\o^2+\k)/(\k-\r\o^2)\partial^2_{\xi}B_{1,1} \end{array}\right)-2i\o \left( \begin{array}{c} \partial_{\t}A_{1,1}\\ \partial_{\t}B_{1,1} \end{array}\right)\\
 + \left( \begin{array}{c} -2K_3sD_1\bar A_{1,1}A_{2,2} + 3K_4D_1^2|A_{1,1}|^2A_{1,1}+2K_3D_1\partial_{\xi}A_{1,0}A_{1,1}\\ 0\end{array}\right).
\eeas
In order for the right hand side to lie in
$\mbox{range}\, \mathbf{M} = (W^\ast )^\perp$, and in view of \eqref{wc1},
the following compatibility condition must be satisfied
\beas
(c^2 - K_2\cos\theta)\partial^2_{\xi}A_{1,1} - 2i\o\partial_{\t}A_{1,1} -2K_3sD_1\bar A_{1,1}A_{2,2} + 3K_4D_1^2|A_{1,1}|^2A_{1,1}+2K_3D_1\partial_{\xi}A_{1,0}A_{1,1}\\
=\frac{\r\k}{\r\o^2-\k}\{\frac{c^2(3\r\o^2+\k)}{(\k-\r\o^2)}\partial_{\xi}^2B_{1,1} - 2i\o\partial_{\t}B_{1,1}\}.
\eeas
Using (\ref{B11A11}), (\ref{A22A11}) and substituting $B_{1,1}$ and $A_{2,2}$ into the above identity yields the following modulation equation in terms of $A_{1,1}$ and $A_{1,0}$~:
\beq
\begin{split}
-2i\o\frac{\r\k^2 + (\r\o^2-\k)^2}{(\r\o^2 - \k)^2}\partial_{\t}A_{1,1} = \left\{K_2\cos\theta - c^2 [1-\frac{3\o^2\r^2\k^2+\r\k^3}{(\r\o^2-\k)^3}]\right\}\partial_{\xi}^2A_{1,1}\\
+ \left\{\frac{2K_3^2s^2D_1^2(\k-4\o^2\r)}{(4\o^2+K_2s^2-\k)(\k-4\o^2\r)+\k^2} - 3K_4D_1^2\right\}|A_{1,1}|^2A_{1,1}-2K_3D_1\partial_{\xi}A_{1,0}A_{1,1},
\end{split}
\label{mod1_old}
\eeq
which is coupled to (\ref{A10A11}).
Introducing
\beqs
\gamma = \frac{(\r\o^2 - \k)^2}{\o[\r\k^2 + (\r\o^2-\k)^2]},
\eeqs
one can show that the curvature is given by
\bea\label{PPomg}
\o'' = \{K_2\cos\theta - c^2[1 - \frac{3\o^2\r^2\k^2+\r\k^3}{(\r\o^2-\k)^3}]\} \g .
\eea
This completes the derivation of the coupled modulation equations \eqref{eq:mod1} and \eqref{eq:mod2}.

\bibliography{refs_notes}
\end{document}